\documentclass[notitlepage, final, 10pt, pra, aps, twocolumn, floatfix, showpacs, letter]{revtex4-1}
\pdfoutput=1

\usepackage{lmodern}
\usepackage[english]{babel} 
\usepackage{graphicx}
\usepackage{color}  
\usepackage[final]{changes}
\setremarkmarkup{\ \textcolor{orange}{ (#2)}}   
\usepackage{subcaption}
\usepackage[space]{grffile}
\definecolor{navyblue}{rgb}{0.0, 0.0, 0.5} 
\usepackage{amsmath, amssymb, amsfonts,bm}
\usepackage{latexsym}   
\usepackage{bbm} 
\usepackage{mathtools}  
\usepackage{placeins}   
\usepackage[protrusion=true, expansion=true]{microtype}
\usepackage[colorlinks=true, breaklinks=false, linkcolor=navyblue, urlcolor=navyblue, citecolor=navyblue]{hyperref}

\newcommand{\ket}[1]{\mathinner{|{#1}\rangle}}

\newcommand{\expe}[1]{\left\langle #1 \right\rangle}

\begin{document}

\title{From the moving piston to the dynamical Casimir effect: \\ explorations with shaken condensates}

\author{Marios H.~Michael$^1$}\email{marios\_michael@g.harvard.edu}
\author{J\"org Schmiedmayer$^2$}
\author{Eugene Demler$^1$}
\affiliation{$^1$Departments of Physics, Harvard University, Cambridge, Massachusetts 02138, USA}
\affiliation{$^2$Institute of Atomic and Subatomic Physics, Technische Universität Wien, Vienna 1020, Austria}

\date{\today}
\selectlanguage{english}
\begin{abstract}
Recent experimental realizations of uniform confining potentials for ultracold atoms make it possible to create quantum acoustic resonators and explore nonequilibrium dynamics of quantum field theories. These systems offer a promising new platform for studying the dynamical Casimir effect, since they allow to achieve relativistic, i.e. near sonic, velocities of the boundaries. In comparison to previously studied optical and classical hydrodynamic systems ultracold atoms allow to realize a broader class of dynamical experiments combining both classical driving and vacuum squeezing. In this paper we discuss theoretically two types of experiments with interacting one dimensional condensates with moving boundaries. Our analysis is based on the Luttinger liquid model which utilizes the emergent conformal symmetry of the low energy sector of the Lieb-Liniger model. The first system we consider is a variable length interferometer with two Y-junctions connected back to back. We demonstrate that dynamics of the relative phase between the two arms of the interferometer can be analyzed using the formalism developed by Moore in the problem of electro-magnetic vacuum squeezing in a cavity with moving mirrors. The second system we discuss is a single condensate in a box potential with periodically moving walls. This system exhibits classical excitation of the mode resonant with the drive as well as nonlinear generation of off-resonant modes. In addition we find strong parametric multimode squeezing between modes whose energy difference matches integer multiples of the drive frequency.
\end{abstract}

\maketitle

\section{Introduction}
Understanding quantum dynamics of many-body systems with time-dependent size and geometry is at the heart of many
fundamental  problems in physics. In cosmology, inflation and subsequent expansion of the universe  
resulted in strong quantum fluctuations of the inflanton field which then froze into classical inhomogeneities\cite{Dodelson03}. 
This led to the formation of the lumpy nature of the currently observed universe with its
galaxies and cosmic voids. In nuclear physics, emission of neutrons and 
$\alpha$-particles during the fission processes can be studied from the perspective of particles moving 
in the presence of time-dependent boundaries\cite{David10}. In field theory and quantum gravity Hawking radiation has a direct 
analogue in the phenomenon of photon creation by a relativistically accelerating mirror\cite{Fulling75}. 

Many common properties of systems with time dependent boundaries can be 
understood from the perspective of the dynamical Casimir effect\cite{Moore70} (see Ref. \cite{Dodonov10} for a review).
In this paradigmatic problem one considers an optical cavity with mirrors moving relative to each other. 
While classically changing the length of the cavity should not change its vacuum state, quantum mechanically 
mirrors moving at velocities comparable to the speed of light produce a "squeezed vacuum" state. 
In the ground state of a cavity one finds only virtual 
photons arising from the zero point motion of harmonic oscillators describing eigenmodes of the cavity.
When mirrors move, they can turn these virtual photons into real particles. Moreover one finds that
photons created in such process appear as entangled pairs, which is why the dynamical Casimir effect has 
been suggested as a resource for 
several applications in quantum information\cite{Giuliano14}. 
What makes 
dynamical Casimir effect particularly subtle is that the entire Hilbert space changes in time.
Thus one faces the problem of not just understanding evolution of many degrees of freedom in Hilbert space
but of combining the "old" and the "newly created" ones. 

While dynamical Casimir effect has been originally formulated for optical cavities, it is rather challenging to realize
with electromagnetic fields. The main obstacle is the requirement of mirrors moving at speeds comparable to the speed of light.
This motivated the search for alternative realizations of this fundamental phenomenon. Important milestones in this direction have been
achieved in recent experiments with one dimensional arrays of superconducting Josephson junctions\cite{Wilson11} and Bose-Einstein Condensates of 
ultracold atoms\cite{Jaksula12}. In Josephson Junction arrays parameters of
the SQUIDs could be changed in time thus modifying the optical path length. This resulted in excitation of photons analogous 
to dynamical Casimir effect although the geometric size of the system remained fixed. In experiments with Bose-Einstein Condensates of 
ultracold atoms modulation of the transverse confining potential was shown to lead to the production of pairs of excitations at half of the modulation frequency. 
These experiments provided the first demonstration of coherent particle production in systems with time-dependent parameters and gave the first indications of the correlated nature of the dynamical Casimir effect. In particular, in experiments of Jaskula et al. it was shown that modulation excited particles at opposite momenta. However, the full extent of the coherent nature of the dynamical Casimir effect remains largely unexplored. Absence of the moving walls in these experiments implied that there was no mixing between different modes although there was squeezing of individual modes.


Motivated by recent progress in creating flat box potentials for ultracold atoms \cite{Zoran17}, we discuss theoretically a new approach for exploring the dynamical Casimir effect in experiments with one dimensional condensates. 
Weakly interacting bosons in 1d exhibit emergent Lorentz symmetry because their low energy dynamics can be described using the Luttinger liquid formalism of sound-like excitations\cite{Giamarchi04}. 
A system of finite size with a uniform density can then be interpreted as an acoustic resonator. 

We consider in this project two distinct systems for exploring the intrguing physics of the dynamical Casimir effect, both of which can be realized using currently available experimental techniques.

\begin{figure}
\includegraphics[trim={2.5cm 2cm 4cm 2cm},clip, scale=0.7]{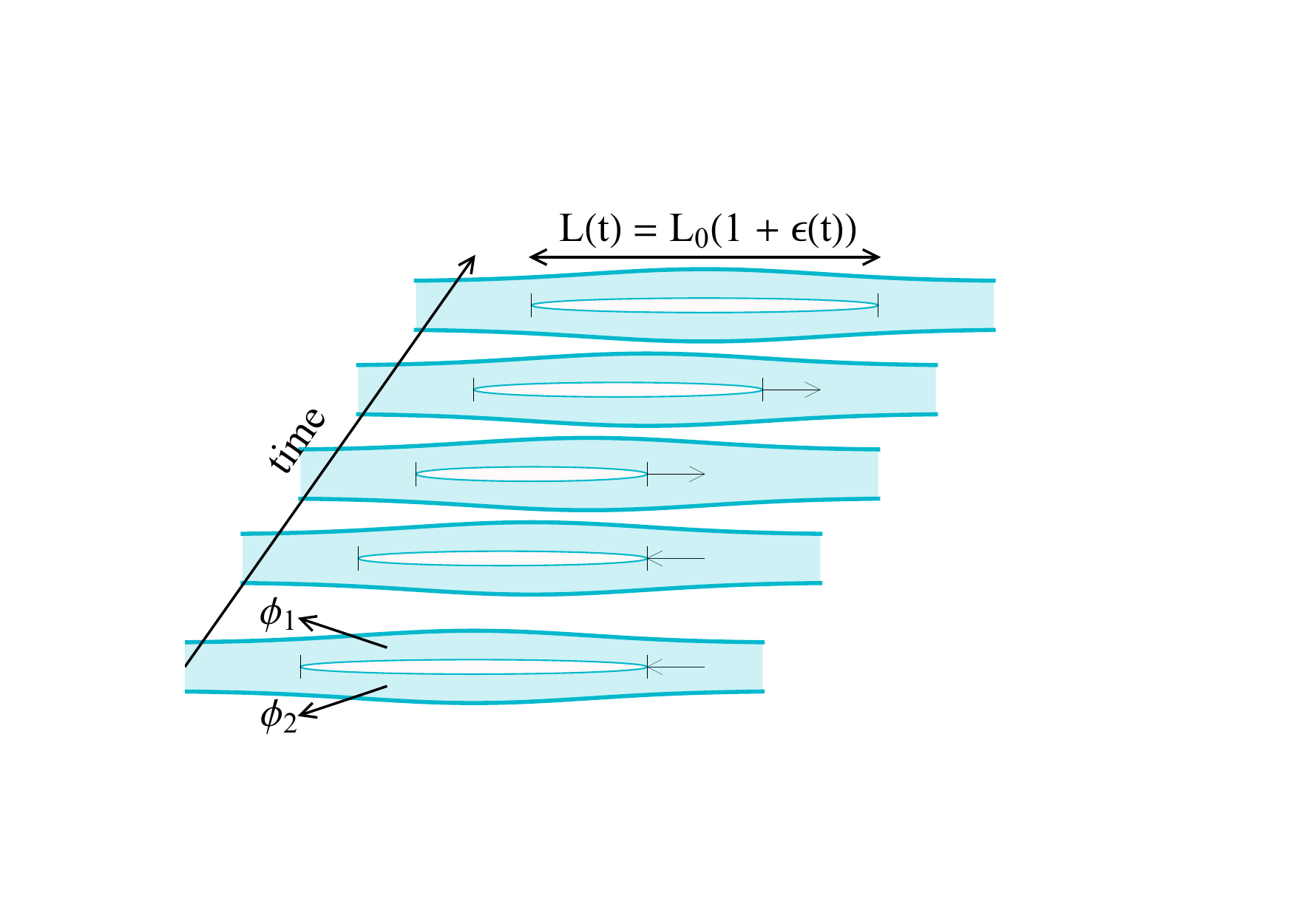}
\caption{Quantum zipper. One of our proposed experiments is a variable length interferometer made up of two Y-junctions back to back. The relative phase between the BECs on the two arms of the intereferometer, $\phi_{rel}$, obeys Dirichlet boundary conditions at the junctions, $\phi_1(0,t) = \phi_2(0,t)$ and $\phi_1(L(t),t) = \phi_2(L(t),t)$. Oscillating one of the Y-junctions, simulates the moving mirror in an electromagnetic cavity with the relative phase being identified with the QED gauge field, $\phi_{rel} \leftrightarrow A$. }
\label{fig:relphi}
\end{figure}

\paragraph{The quantum zipper:}
The first system we propose consists of two Y-junctions\cite{Stenholm03} connected back to back. It can also be understood as a variable length one dimensional atomic interferometer (see Fig.~\ref{fig:relphi}). The relevant degrees of freedom come from the relative phase between the two arms of the interferometer, $\phi_{rel} = \phi_1 - \phi_2$. In the connecting point of the Y junctions the relative phase must be equal to zero, which provides the analogue of the mirror boundary conditions of electromagnetic cavities. The time dependent geometry arises from changing the length of the "split" part of the condensate, which can be achieved,for example, with RF potentials. As we discuss below it offers a faithful representation of the original dynamical Casimir effect considered in the context of cavity QED (for details see Sec.~\ref{sec:zipper}). The sound velocity in a BEC is of the order of a few millimeters per second, allowing near sonic velocities to be achieved by the boundary, required for resonant effects to be significant. We note however, that one should avoid supersonic motion where the Luttinger liquid approach breaks down. The primary experimental observable, the relative phase between the two condensates, $\phi_{rel}$, can be measured by observing the interference after a transverse expansion of the atoms \cite{Schmiedmayer14}\cite{Schumm05}\cite{Hoff08}. 

In our analysis we primarily address the problem of periodic modulation of the atomic interferometer, $L(t) = L_0 \left(1 + \epsilon(1 - \cos\left( \omega t\right)\right)$, although the same approach can be used to study other types of dynamical excitations. The main consequences of the time dependent change of the boundary is the generation of a squeezed state and the associated particle creation(phonons). These effects are encoded in the quantum fluctuations of $\phi_{rel}$ and the energy stored in the relative degrees of freedom. The most interesting dynamics occurs when the driving frequency matches an integer multiple of the fundamental energy difference of the Luttinger liquid acoustic resonator. At short times the moving boundary excites multiple modes in the system with the number of phonons in every mode growing quadratically in time. In this regime we observe strong correlations between modes whose frequencies differ by the drive frequency. This results in a checker-board type correlation matrix, $\expe{\phi_{rel}(n,t) \phi_{rel} (m,t) }$, shown in Fig.~\ref{fig:pert}. At late times parametric pair production process is dominant in which pairs of identical phonons are produced, each at half the frequency of the drive. At this stage the number of phonons in non-resonant modes starts to decrease, whereas in the parametrically resonant mode the number of phonons increases linearly in time (see Fig.~\ref{fig:cross}).

\begin{figure}
\includegraphics[trim={2cm 2cm 2cm 2cm},clip,scale=0.6]{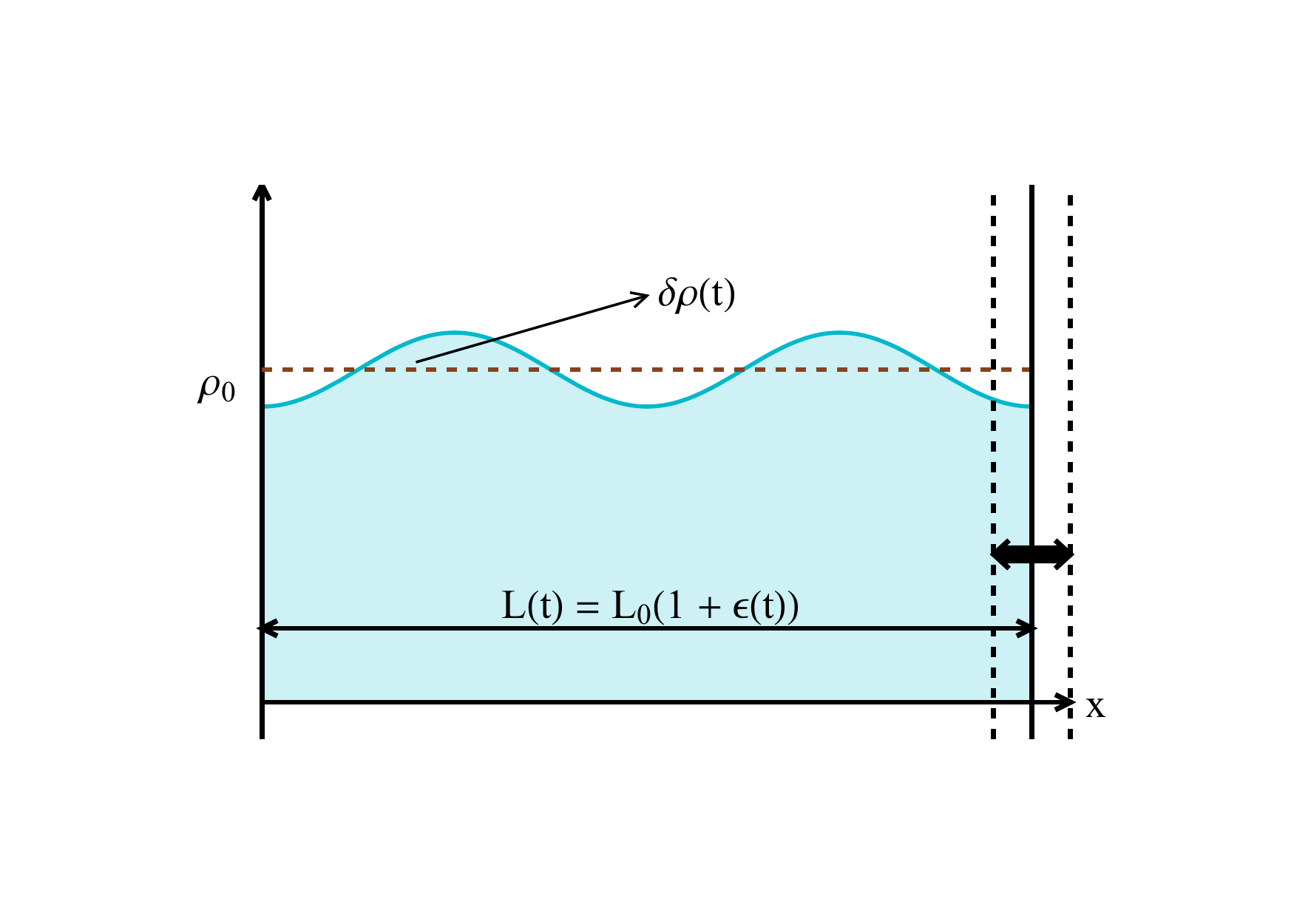}
\caption{1d quantum fluid shaken by a moving wall. The moving wall excites both classical expectations as well as quantum fluctuations of the density and current density of the fluid.}
\label{fig:shakebox}
\end{figure}

\paragraph{Shaking a box:} In the second set-up, we consider a single 1d box containing interacting ultracold bosons. We discuss a shaking type protocol, in which one of the confining walls of the trap oscillates at a fixed frequency, resulting again in the periodic modulation of the condensate length. Similar results are obtained when both walls are moving. The primary experimental observable that we discuss in this set-up is the local density profile (see Fig.~\ref{fig:shakebox}). 

It is useful to put our analysis in perspective of earlier work on related systems. In the area of classical acoustics and hydrodynamics the problem of a pipe with a moving piston has received considerable attention (see refs.\cite{Rudenko09}\cite{Enflo02} and references therein). The main limitation of these studies is the restriction to classical expectation values of operators and disregard of correlations. Furthermore, most of the papers used simplified boundary conditions in which the flow velocity was set to match the wall's velocity not at the actual moving position of the walls but at some fixed point. In the linear response regime such an approximation is justified, however in many cases we are interested in dynamics beyond linear response. In this case changing the boundary conditions leads to dramatic changes of the long time dynamics. On the other hand, the advantage of the classical treatment is that it allows to include effects of non-linearities, which have been shown to lead to the proliferation of shock waves. 

In the area of non-equilibrium quantum field theories the most closely related problem to what we discuss in this paper is the dynamical Casimir effect originally studied by Moore in the context of electromagnetic cavities\cite{Moore70}. In this class of problems one only considers time evolution of correlation functions, since classical expectation values of the fields are forbidden by symmetry ( $\overrightarrow{A} \rightarrow - \overrightarrow{A}$ is the symmetry of both the Hamiltonian and the Dirichlet boundary conditions, $\left. \overrightarrow{A}\right|_{boundary} = 0$, on the moving mirrors). In our analysis of the cold atoms in a periodically squeezed box we consider both the classical expectation values of operators and their correlation functions. At first sight it may appear that the problem of a quantum fluid in a box of variable length is very different from the canonical Casimir effect. In the former case the boundary conditions correspond to the flow velocity at the position of the moving walls to equal the wall's velocity. In the case of the optical cavity of variable length the boundary conditions for the vector potential is to vanish at the boundary \cite{Moore70}. 

We begin our analysis of the shaken box problem by introducing an exact transformation that allows to convert the hydrodynamic boundary conditions for this system into Dirichlet type conditions with $j = 0$ on the moving walls, where $j$ is the current of the atoms. As a consequence of this transformation an additional time-dependent quadratic potential is introduced, which is responsible for producing the classical expectation values of the fields. Experimentally, it may be useful to have separate controls over the classical and quantum components of the drive. This can be achieved by combining the motion of the walls with a compensating external time dependent parabolic potential, which is readily available in experiments with ultracold atoms. Motivated by this consideration we include both types of driving in our analysis. Correlation functions are closely related to the Greens functions introduced for solving classical linear differential equations. Hence, the transformation that we introduced and correlation functions that we found are useful for the classical solution of the moving piston problem. They allow to treat the boundary conditions in the linearized dynamics exactly. 

We find that results of this improved analysis of the boundary conditions is rather dramatic at long modulation times. This difference arises from the fact that including moving walls introduces mixing between different modes even in linear hydrodynamics. The physical reason for the mixing is the Doppler shift of waves reflected from the moving walls. One of the most dramatic results of the mode mixing is the non-trivial evolution of the resonantly driven mode. 

At short times the amplitude of this mode grows linearly in time, similarly to the case of a resonantly driven single harmonic oscillator. At the same time, non-resonant modes become populated due to mixing with the resonantly driven one. At longer times, the increase of the amplitude of the resonantly driven mode saturates and eventually this mode becomes suppressed. We attribute this suppression to the destructive interference between the drive and the previously excited non-resonant modes. This should be contrasted to having simple damping in a system without mode coupling. In the latter case, when an eigenmode is driven resonantly, its amplitude first increases and then saturates to some finite value when the
energy deposited by the drive balances the energy lost to dissipation. The effect of mode squeezing coming from the
dynamical Casimir effect is much more dramatic: it suppresses the resonant mode amplitude(see Fig.~\ref{fig:supp}). This behavior is also to be contrasted with the usual non-linear steady states. In the latter, as the amplitude increases the resonant mode becomes detuned from the drive. In such cases, we have again an increasing resonant amplitude that eventually saturates when it goes off-resonance.

We arranged our paper so that the quantum zipper and shaken box systems are discussed separately. When analyzing the zipper system we focus on correlation functions and quantum fluctuations. In the case of the shaken box our emphasis is on the classical expectation values of the density and velocity, which corresponds to the coherent part of the wavefunction for individual modes.

This paper is organized as follows. We begin with a brief discussion of hydrodynamics of interacting bosons in 1d which provides the basis of the Luttinger liquid model for this system. We show that for the interferometer shown in Fig.~\ref{fig:relphi} the relative phase dynamics can be described by the wave equation with the Dirichlet boundary conditions at the points of the Y junctions. We present concrete predicitions for the quantum zipper with periodically modulated length at the end of Sec.~\ref{sec:fluc}. 

In Sec.~\ref{sec:shake} we switch to the shaken box problem and discuss how conservation of mass leads to inhomogeneous boundary conditions in the shaken box experiment. Through a series of transformations we derive a mathematical formulation of the model constructed in such a way that it obeys Dirichlet type boundary conditions $j=0$ at the positions of the walls at the expense of having an additional time-dependent potential. A naive perturbative treatment of the problem is given in Sec.~\ref{sec:classical} that helps readers develop an intuition for the classical behavior of the system. This is followed by introducing mode quantization and analyzing correlation functions. This formalism provides a natural way of including the changing geometry of the system and allows us to solve the classical problem of a moving piston. We note that within linear hydrodynamics our solution is exact and goes beyond previous papers that treated the boundary conditions only approximately.

In linear hydrodynamics, dynamics of the classical expectation values of operators and of their second order correlations are decoupled. Hence, one may be tempted to conclude that quantum fluctuations in the shaken box should be identical to those of the quantum zipper case (i.e. vacuum squeezing in optical cavity). We point out that there is a subtle difference in the character of commutation relations between the two types of problems. In Sec.~\ref{sec:val} we discuss effects of non-linear dispersion, non-linearities and thermal fluctuations, which should be relevant for actual experiments with shaken condensates and in Sec.~\ref{sec:beyond} we give an alternative Hamiltonian prescription to go beyond Luttinger liquid. Finally, in Sec.~\ref{sec:conc} we provide a summary and a brief discussion of interesting open issues.

\section{The Quantum Zipper}
\label{sec:zipper}

\subsection{Luttinger liquid formalism for interacting 1d systems}

A hydrodynamic approach is a general framework that describes the long wavelength and small frequency limit of dynamics in a large variety of systems. To make our analysis more transparent, in this section we show explicitly how the hydrodynamic equations of motion can be obtained for a cold gas of weakly interacting bosons. In the hydrodynamic limit, a superfluid is described by its low lying "gapless" modes which correspond to current-density fluctuations. The variables obey generically the continuity equation and the Josephson relation.
\begin{subequations}
\begin{align}
\partial_t \rho(x,t) &= - \frac{\hbar \rho_0 }{m}\partial_x \left(  \partial_x \phi(x,t) \right),\label{eq:continuity} \\
\hbar \partial_t \phi(x,t) &= -U_0 \rho(x,t),\label{eq:phieq}\\
\left[\rho(x,t) , \phi(x',t) \right] &= i \delta(x - x')\label{eq:comrel}
\end{align}
\label{eq:EofM}
\end{subequations}
where higher order derivative terms have been ignored and we limited ourselves to equations of motion around the equilibrium uniform state. The canonically conjugate variables are the density, $\rho(x,t)$ and the phase $\phi(x,t)$ of the superfluid, where the quantum nature is manifested through their commutation relations (Eq.~\ref{eq:comrel}). $\rho_0$ is the average density of the superfluid, $m$ is the mass of the bosons and $U_0$ the effective interaction strength of the point interaction between the bosons.

More generally for gapless systems, low lying excitations of interacting 1d systems, in the case of both bosonic and fermionic particles, have a unified description given by the standard Tomonaga-Luttinger liquid hamiltonian\cite{Voit95}:
\begin{subequations}
\begin{align}
H &= \int \frac{dx}{2 \pi} v_s \left( K (\pi j)^2 + \frac{1}{K} \left(\partial_x \theta \right)^2 \right), \\
\left[ j(x,t) , \theta(x,t) \right] &= i\delta(x-x')
\end{align}
\end{subequations}
Here $j$ is the current density and $\theta$ is its conjugate variable. Where in this expression variables are switched following the conventional form found in the literature of Luttinger liquids\cite{Voit95}. For general interacting systems, the speed of sound, $u$, and parameter $K$ should be found numerically. $K$ does not influence the dynamics but affects the overall scale of the fluctuations. What is essential for our discussion is that the Luttinger liquid model gives rise to a linear wave equation of the type:
\begin{equation}
\left( \partial_t^2 - v_s^2 \partial_x^2\right) \phi(x,t) = 0 
\end{equation}
In the case of weakly interacting bosons (Eq.~\ref{eq:EofM}), the speed of sound is given by $v_s^2 = \frac{U_0 \rho_0}{m}$. The other scale of interest needed to define our theory is the healing length, $\xi_h$. The corresponding momentum, $\Lambda = \frac{1}{\xi_h}$, presents a cut-off at which the dispersion relation is no longer linear. For weakly interacting bosons this cut-off is given by $\xi_h = \frac{\hbar}{mv_s}$. In more general scenarios, the cut-off corresponds to the momentum whose energy is of the order of the chemical potential.

Therefore, low energy excitations of a broad class of gapless 1d interacting systems can be analyzed from the perspective of sound modes in an acoustic resonator. 

\subsection{Mathematical formulation of the quantum zipper problem}

We consider a 1d interferometer of ultracold atoms shown in Fig.~\ref{fig:relphi}. In the two arms of the interferometer we have quantum fluids, which can be described with quantum fields $\{\rho_1, \phi_1\}$ and $\{\rho_2, \phi_2\}$ respectively. We assume that the average densities are equal so that the sound velocities are the same. In this case we can separate the symmetric variables $\rho_+ = \rho_1 + \rho_2$, $\phi_+ = \phi_1 + \phi_2$ from the relative degrees of freedom $\rho_{rel} = \rho_1 - \rho_2 $, $\phi_{rel} = \phi_1 - \phi_2$ \cite{Kitagawa11}. We will only analyze dynamics of the relative degrees of freedom which can be measured using interference experiments. These obey\cite{Kitagawa11}:

\begin{subequations}
\begin{align}
\partial_t \rho_{rel}(x,t) &= - \frac{\hbar }{m}\partial_x \left( \rho \partial_x \phi_{rel}(x,t) \right),\\
\hbar \partial_t \phi_{rel}(x,t) &= -U_0 \rho_{rel}(x,t) ,\\
\left[\rho_{rel}(x,t) , \phi_{rel}(x',t) \right] &= i \delta(x - x')
\end{align}
\label{eq:EofMring}
\end{subequations}

Note that any potential that acts the same way on both arms of the interferometer does not affect the relative phase $\phi_{rel}$.
The boundary conditions obeyed by the relative phase $\phi_{rel}$ are Dirichlet since at the point where the two BECs meet they must have the same phase. Combining the equations of motion then leads to the wave equation:
\begin{subequations}
\begin{align}
\left(\partial_t^2 - v_s^2 \partial_x^2 \right) \phi_{rel}(x,t) &= 0 , \\
\left[\phi(x,t) , \partial_t \phi_{rel}(x',t) \right] &= i \delta(x - x'),\\
\phi_{rel}(0,t) = \phi_{rel}(L(t),t) &= 0 
\end{align}
\label{eq:casimiranalogue}
\end{subequations}
where $v_s$ is the speed of sound. As promised Eq.~\ref{eq:casimiranalogue} offers a direct analogue to cavity QED (see below).

\section{Dynamical Casimir Effect}
\label{sec:quantum}

Discussion in this section is arranged as follows. We begin by obtaining the quantum time-dependent eigenstates of the quantum zipper problem following the formalism developed by Moore. These modes are used to compute both the retarded and Keldysh Greens functions of this system. Quantum fluctuations in the system can be obtained from the equal time values of the Keldysh Greens functions. Retarded Greens functions will be used in the discussion of the shaken box to analyze the classical part of the response.

\subsection{Generalized Formalism of the dynamical Casimir effect}
\label{subsec:formal}
Moore pointed out that quantum resonators with time-dependent length do not have a Hamiltonian description\cite{Moore70}. Instead the fields should be quantized using the equations of motion. Here, we follow closely the discussion in the original paper by Moore\cite{Moore70}, and represent the main field by the letter $A(x,t)$ making the analogy between vector potential in cavity QED:
\begin{subequations}
\begin{align}
\left( \partial_t^2 - \partial_x^2 \right) A (x,t) &= 0, \\
A(0,t) &= A(L(t),t) =0  
\end{align}
\label{eq:kleingordon}
\end{subequations}
To simplify the notations we set $v_s =1$ in the discussion of this section. We will recover proper dimensions later in the paper.
The general strategy is to find a set of orthonormal solutions of the Klein-Gordon equation with a moving boundary, Eq.~\ref{eq:kleingordon}, and expand the field, $A(x,t)$, in terms of orthonormal solutions with respect to the Klein-Gordon inner product:
\begin{equation}
\{f | g \} = -i \int^{L(t)}_0 dx \left( f(x,t) \partial_t g(x,t) - g(x,t) \partial_t f(x,t) \right)
\label{eq:kginner}
\end{equation}
Provided that $f$ and $g$ satisfy the Klein-Gordon equation, Eq.~\ref{eq:kleingordon}, the inner product defined in Eq.~\ref{eq:kginner} is time-independent. An orthonormal basis consists of a set of solutions $\{f_n\}$ such that $\{f_n , f_n^*\}$ span the space of solutions of Eq.~\ref{eq:kleingordon} and obey:
\begin{subequations}
\begin{align}
\{f_n | f_m^*\} &= \delta_{n,m} , \\
\{f_n | f_m\} = \{f_n^* | f_m^* \} &= 0 
\end{align}
\label{eq:orthonorm}
\end{subequations}
Using such a basis, any solution can be expanded as linear combination of the basis functions:
\begin{equation}
A = \sum_n(c_n f_n + c_n^* f_n^* ) 
\label{eq:cexp}
\end{equation}
where the fact that $A$ is real was used. Finally, quantization of this theory is achieved by promoting the time-independent coefficients to creation/annihilation operators:
\begin{subequations}
\begin{align}
c_m &\rightarrow \hat{c}_m , \quad \qquad c_m^* \rightarrow \hat{c}_m^\dag, \\
\hat{c}_m^\dag &= \{ f_m | \hat{A} \}, \quad \hat{c}_m = -\{f_m^* | \hat{A} \}  \\
[ \hat{c}_m, \hat{c}_n^\dag ] &= \delta_{n,m}
\end{align}
\label{eq:cdef}
\end{subequations}
The step of promoting coefficients to creation/annihilation operators using a basis with a particular normalization, corresponds to a particular choice of commutation relations obeyed by the field $A(x,t)$.

It is important to note that the creation/annihilation operators defined in this way are time-independent and since we are working in the Heisenberg picture the states are also time-independent. As a result, if the initial ground state is a vacuum for a set of annihilation operators $c_m$, then the state of the system will remain the vacuum state of those operators while all the time-dependence of observables is taken into account by the basis functions. Furthermore, notice that the definition of creation/annihilation operators and the vacuum state depends on the choice of basis functions we choose to expand our fields in. 

In the fixed box case, $L(t)= L_0$, expanding the field in the basis functions amounts to a fourier decomposition of the field:
\begin{equation}
f_n = \frac{1}{\sqrt{n \pi} }e^{- i \frac{n \pi }{L_0} t} \sin\left( \frac{n \pi}{L_0} x \right)
\label{eq:eigenfixed}
\end{equation}
With the positive frequency solutions corresponding to creation operators and the negative frequency solutions to annihilation operators.

Finding the eigenmodes while the wall is moving is achieved by performing a conformal transformation which preserves the equations of motion but fixes the boundary:
\begin{subequations}
\begin{align}
s + w &= R( t + x ) , \\
s - w &= R( t - x ) , \\
\left( \partial_s^2 - \partial_w^2 \right) A(s,w) &=0 , \\
A(s,0) = A (s, 1) &= 0 , \\
\Rightarrow R(t+L(T)) - R(t-L(t)) &=2
\end{align}
\end{subequations}
In the new frame, the solutions have the form of the fixed box, Eq.~\ref{eq:eigenfixed}, while the transformation function is found by requiring that it fixes the boundary. In the original coordinates the eigenmodes are given by:
\begin{subequations}
\begin{align}
f_n(x,t) &= \frac{i}{\sqrt{n \pi} } \frac{ e^{-i n \pi R(t+x) } - e^{-i n \pi R(t-x)} }{2}, \\
R(t + L(t) ) &= R( t -L(t) ) + 2 \label{eq:recR}
\end{align}
\label{eq:eigenmov}
\end{subequations}
Note that in order to define the transformation function $R(z)$ uniquely, one needs to set the value of $R(z)$ in an interval $z \in \left(t -L(t), t + L(t) \right)$  for some fixed value of t, which can then be used to exactly evaluate $R(z)$ numerically \cite{Cole95}. The complete evolution of the system is then given by realizing that for $t \leq 0$ the box size was constant and the system was in the ground state. The system must then be in the vacuum state of annihilation operators, $\{c_m(0)\}$, defined such that their basis functions take the fixed box form at $t=0$:
\begin{equation}
f_n(x,t = 0) = \frac{i}{\sqrt{n \pi} }\frac{\left( e^{-i n \pi \frac{( t + x)}{L_0} } - e^{ - i n \pi \frac{( t - x)}{L_0} } \right)}{2}
\end{equation} 
The modes should be identical to the fixed box modes through-out the box, $x \in \left( 0, L(t) \right)$, which translates to the following initial condition for $R(z)$:
\begin{equation}
R(z) = \frac{z}{L_0} , \quad z \in \left( - L_0, L_0 \right)
\label{eq:transt0}
\end{equation}
To recap: the set of modes in Eq.~\ref{eq:eigenmov}-\ref{eq:transt0} define a set of annihilation operators, $\{c_m(0) \}$, for which the ground state at $t=0$ corresponds to their vacuum state. However, since both the state and the operators are time-independent the system remains in the vacuum of these operators while all the time dependence is taken into account by the eigenfunctions, $f_n(x,t)$.

Having successfully quantized the theory we are now in a position to calculate the non-equilibrium Greens functions of the system. We will be interested in both the symmetric and antisymmetric correlation functions:
\begin{widetext}
\begin{align}
D^K(x,t;x',t') &= -i \expe{ \left\{ A(x,t), A(x',t')\right\}},\\ 
&= -  2 i \mbox{Re}\left[ \sum_n f_n(x,t) f_n^*(x',t') \right],\\
D^R(x,t;x',t') &= -i \theta(t - t') \expe{ \left[ A(x,t) , A(x',t') \right]}, \\ \label{eq:DR}
&= 2 \theta(t - t') \mbox{Im}\left[\sum_n f_n(x,t) f_n^*(x',t') \right]  
\end{align}
\label{eq:Greens}
\end{widetext} 

While most of the discussion in this section addresses the problem described by the homogeneous equations (\ref{eq:EofMring}), let us make a detour and consider the inhomogeneous version of this problem:
\begin{equation}
\left( \partial_t^2 - \partial_x^2 \right) A(x,t) = - V(x,t)
\label{eq:inhomEofM}
\end{equation}

As we will see a problem of this type appears in our discussion of the single shaken box protocol. We observe that solution to Eq.~\ref{eq:inhomEofM} can be easily obtained using the Greens function for the homogeneous problem (Eq.~\ref{eq:DR}).We know that the retarded Greens function, $D^R$, satisfies the equation (see Appendix~\ref{app:Gr} for details):
\begin{equation}
\left(\partial_t^2 - \partial_x^2 \right) D^R(x,t; x',t') = - \delta( x-x') \delta(t -t') 
\label{eq:RetGreen}
\end{equation}
with the boundary condition $D^R(t <t') =0$. The solution of Eq.~\ref{eq:inhomEofM} can then be written as:
\begin{equation}
\expe{A}(x,t) = \int_0^t dt'\int_0^{L(t')} dx' D^R(x,t;x',t') V(x',t') 
\end{equation}

Returning to the homogeneous problem, Eq.~\ref{eq:EofMring}, we observe that the covariance matrix, $C(x,x',t)$, encodes the fluctuations of the variable $\hat{A}$  through:
\begin{subequations}
\begin{align}
C(x,t;x',t') &= \frac{1}{2} \expe{ \left\{ A(x,t) A(x',t) \right\}}, \\
&= \frac{i}{2}D^K(x,t; x',t)
\end{align}
\end{subequations}
Therefore, the retarded and Keldysh Greens functions can be used to analyze the classical response and quantum fluctuations respectively.

In order, to make concepts such as squeezing and parametric resonance more transparent, it is convenient to relate the eigenmodes to the instantaneous Fourier components of the field and their fluctuations. This is achieved by introducing the notion of instantaneously at rest modes. Instead of using Eq.~\ref{eq:transt0} as initial conditions for $R(z)$ so that the eigenmodes have the fixed box eigenmodes' form at $t=0$, we could have chosen the initial condition:
\begin{equation}
R(z) = \frac{z}{L(t')} , \quad z \in \left( t' - L(t') , t' + L(t') \right) 
\end{equation}
This would correspond to a different set of eigenmodes that have the fixed box solution's form at $t = t'$:
\begin{equation}
f_{n,t'} (x,t = t') = \frac{i}{\sqrt{n \pi} }\frac{\left( e^{-i n \pi \frac{( t + x)}{L(t')} } - e^{ - i n \pi \frac{( t - x)}{L(t')} } \right)}{2}
\label{eq:restj}
\end{equation} 
In fact, we are always free to define fixed box initial conditions for $R(z)$ throughout the box for any time $t'$, as long as the wall's motion is not supersonic, $ \left|\frac{d L(t)}{dt}\right| < v_s$, as shown in Appendix~\ref{app:inst}. Working with the instantaneously at rest modes is convenient since they correspond to the Fourier transform of the signal at $t=t'$. Different modes define different set of creation/annihilation operators that define different vacua. A set of instantaneously at rest operators at time $t$ is related to the set of instantaneously at rest operators at time $t = 0$, whose vacuum defines the state of the system, via a bogoliubov transformation that is found using Eq.~\ref{eq:cexp},\ref{eq:cdef}:
\begin{subequations}
\begin{align}
\underline{c}(t) &= \underline{\underline{U}}(t) \cdot \underline{c}(0) + \underline{\underline{V}}(t) \cdot \underline{c}^\dag (0), \\  
 U_{n,m} &= -\{f^*_{n,t} | f_{m,0}\} = \{f_{m,0} | f^*_{n,t} \} ,\\
 V_{n,m} &= -\{f^*_{n,t} | f^*_{m,0}\} = \{f_{m,0}^* | f^*_{n,t} \}
\end{align}
\end{subequations}
Two basis that are related to each other via a Bogoliubov transformation, generically, see each other's vacuum as a squeezed state. As a result, the instantaneously at rest operators, $c(t)$, will see the vacuum of the $c(0)$ operators as a squeezed state as long as $V \neq 0$. In other words, the ground state is continuously squeezed with respect to the instantaneously at rest phonon operators while the boundary is moving. In particular, if we turn off the drive at $t=t'$, the system will remain in the squeezed state seen by the operators, $\{ c_m (t') \}$.

\section{Quantum Fluctuations}
\label{sec:fluc}

We now use results of the previous section to discuss quantum fluctuations of the phase, $\phi_{rel}$, in the quantum zipper problem. We use the standard description of the fluctuations in terms of the symmetric correlator:
\begin{subequations}
\begin{align}
C(x,t;x',t') &= \frac{1}{2} \expe{ \left\{\phi_{rel} (x,t) \phi_{rel} (x',t) \right\}}, \\
&= \frac{i}{2}D^K(x,t; x',t)
\end{align}
\label{eq:fluct0temp}
\end{subequations}

We define instantaneous spatial eigenmodes using the Fourier transform:
\begin{equation}
\begin{split}
\phi_{rel}(m, t) &= \frac{2}{L(t)} \int_0^{L(t)} \sin\left( \frac{ n \pi}{L(t)} x \right) \phi_{rel}(x,t)dx , \\
&= \frac{1}{\sqrt{n \pi}} \left( c_n(t) e^{-i n\pi t/L(t)} + c^\dag_n(t)e^{i n\pi t/L(t)}\right) 
\end{split}
\end{equation}
Quantum fluctuations due to squeezing can then be calculated in the Fourier basis by the following formula, as shown in Appendix~\ref{app:comp}:
\begin{equation}
\begin{split}
&\expe{\phi_{rel}(n,t) \phi_{rel}(m,t) } = \frac{1}{\pi \sqrt{n m}}\times \\
&\sum_l \bigg(V^\dag_{n,l}V_{l,m}e^{i (n - m) \pi t/L(t)} +  \\
&U_{n,l} U^\dag_{l,m} e^{ - i ( n - m ) \pi t/L(t)} + \\
&V^\dag_{n,l}U^\dag_{l,m}e^{i ( n + m) \pi t/L(t)} + \\
&U_{n,l}V_{l,m} e^{ - i \pi ( n + m ) t/L(t) }  \bigg)
\end{split}
\label{eq:fluctT0}
\end{equation}
Where the covariance matrix in the fourier representation takes the form:
\begin{equation}
C(n,m;t) = \mbox{Re}\left[ \expe{ \phi_{rel}(n,t) \phi_{rel}(m,t) } \right]
\end{equation}

The overlaps, $U_{n,m}(t), V_{n,m}(t)$, are given by
\begin{subequations}
\begin{align}
&U_{n,m} = -\{j^*_{n,t} | j_{m,0} \} = \{j_{m,0}|j^*_{n,t}\} \\
&=\sqrt{\frac{n}{m} } \frac{1}{2 L(t)} \times \int_{-L(t)}^{L(t)} dx e^{ i \pi \left( n\frac{t+x}{L(t)} - m R_0(t+x) \right)}, \\
&V_{n,m} = -\{j^*_{n,t} | j^*_{m,0} \} = \{j^*_{m,0}|j^*_{n,t} \} \\
&= - \sqrt{\frac{n}{m} } \frac{1}{2 L(t)} \times \int_{-L(t)}^{L(t)} dx e^{ i \pi \left( n\frac{t+x}{L(t)} + m R_0(t+x) \right)}
\end{align}
\label{eq:bogmat}
\end{subequations}




\subsection{Perturbative regime: Mode mixing}
First, we study the problem using perturbation theory valid only at  short times. $R(z)$ is expanded in powers of the perturbing function $\epsilon(t)$ presented here up to 2nd order:
\begin{equation}
\begin{split}
&R(z) = \frac{z}{L_0} - 2n \epsilon(z) + n^2 L_0 \frac{d \epsilon^2(z)}{dz},\\
&z \in ( -L_0 + 2 n L_0, L_0 + 2 n L_0) 
\end{split}
\label{eq:pertR}
\end{equation}
Where $n$, in this situation is an integer labeling time intervals of $2 L_0$ i.e. $n = t \mbox{mod} 2L_0$. This expansion was first suggested by Ref.\cite{Dalvit98}, however for completeness we provide an alternative derivation in Appendix~\ref{app:pert}.

Because of the secular terms, this expansion is only valid for:
\begin{equation}
t_{pert.} < \frac{1}{\epsilon \omega} 
\end{equation}
at $t = \frac{1}{\epsilon \omega}$ the 2nd order term becomes comparable to the first order term and the expansion breaks down. Expanding the Bogoliubov matrices to linear order in $\epsilon$ gives:
\begin{subequations}
\begin{align}
U_{k,l}(t = n 2 L_0 ) &= e^{i k \pi 2n \epsilon} \left( \delta_{k,l} -i \pi \sqrt{ k l }  n \epsilon (\delta_{k-l , \omega} + \delta_{k - l, -\omega}) \right), \\
V_{k,l}(t= n 2 L_0) &= e^{- i k \pi 2 n \epsilon} \left( i \pi \sqrt{k l} n \epsilon \right) \left( \delta_{k + l ,\omega}+ \delta_{k+l, - \omega}\right)
\end{align}
\label{eq:bogmatpert}
\end{subequations}

\begin{figure*}
\captionsetup[subfigure]{justification=centering}
    \centering
    \begin{subfigure}[b]{0.3\textwidth}
        \includegraphics[trim = {1cm, 1cm, 2cm, 2cm},clip,scale=0.3]{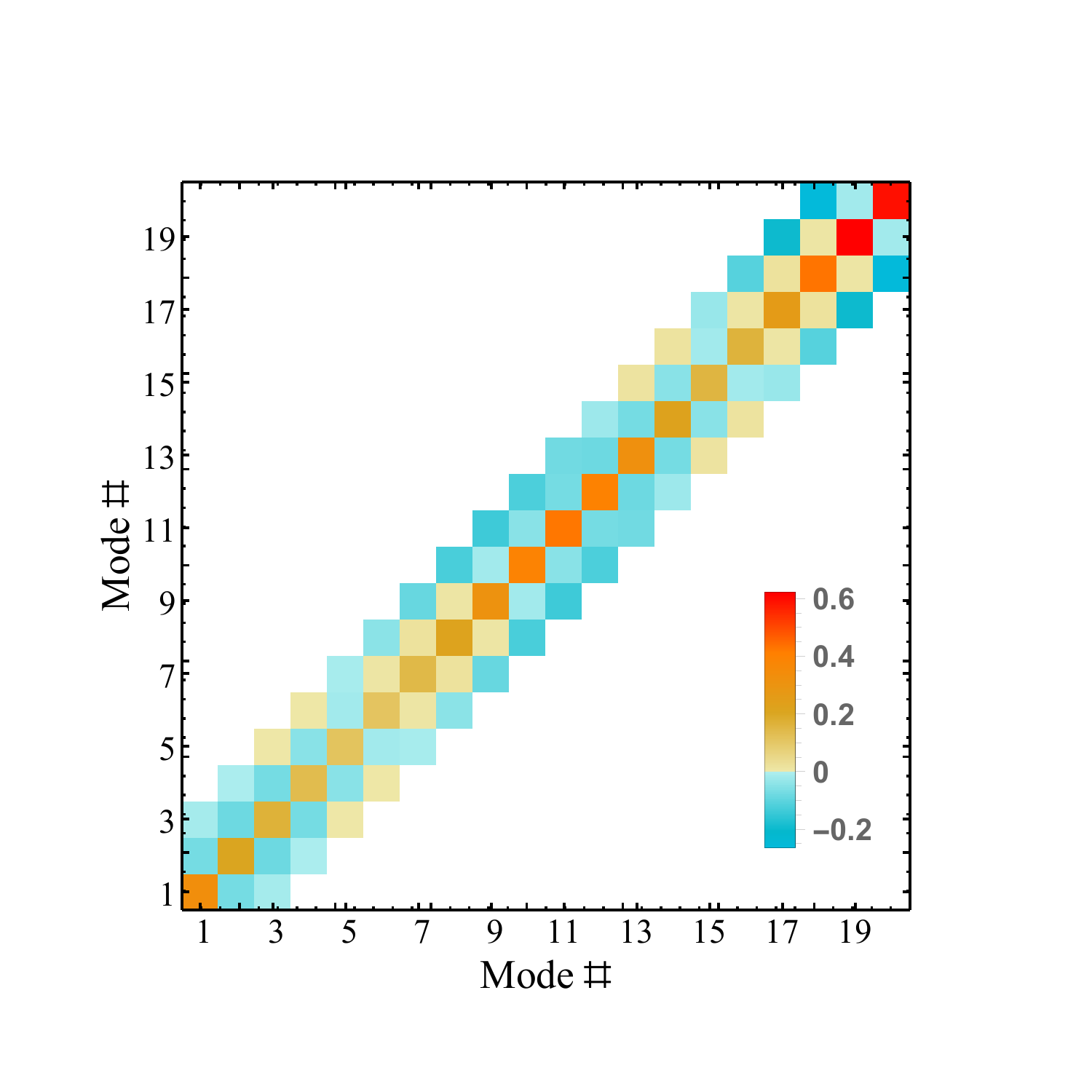}
        \caption{n = 1}
        \label{fig:pert1}
    \end{subfigure}
    \begin{subfigure}[b]{0.3\textwidth}
        \includegraphics[trim = {1cm, 1cm, 2cm, 2cm},clip,scale = 0.3]{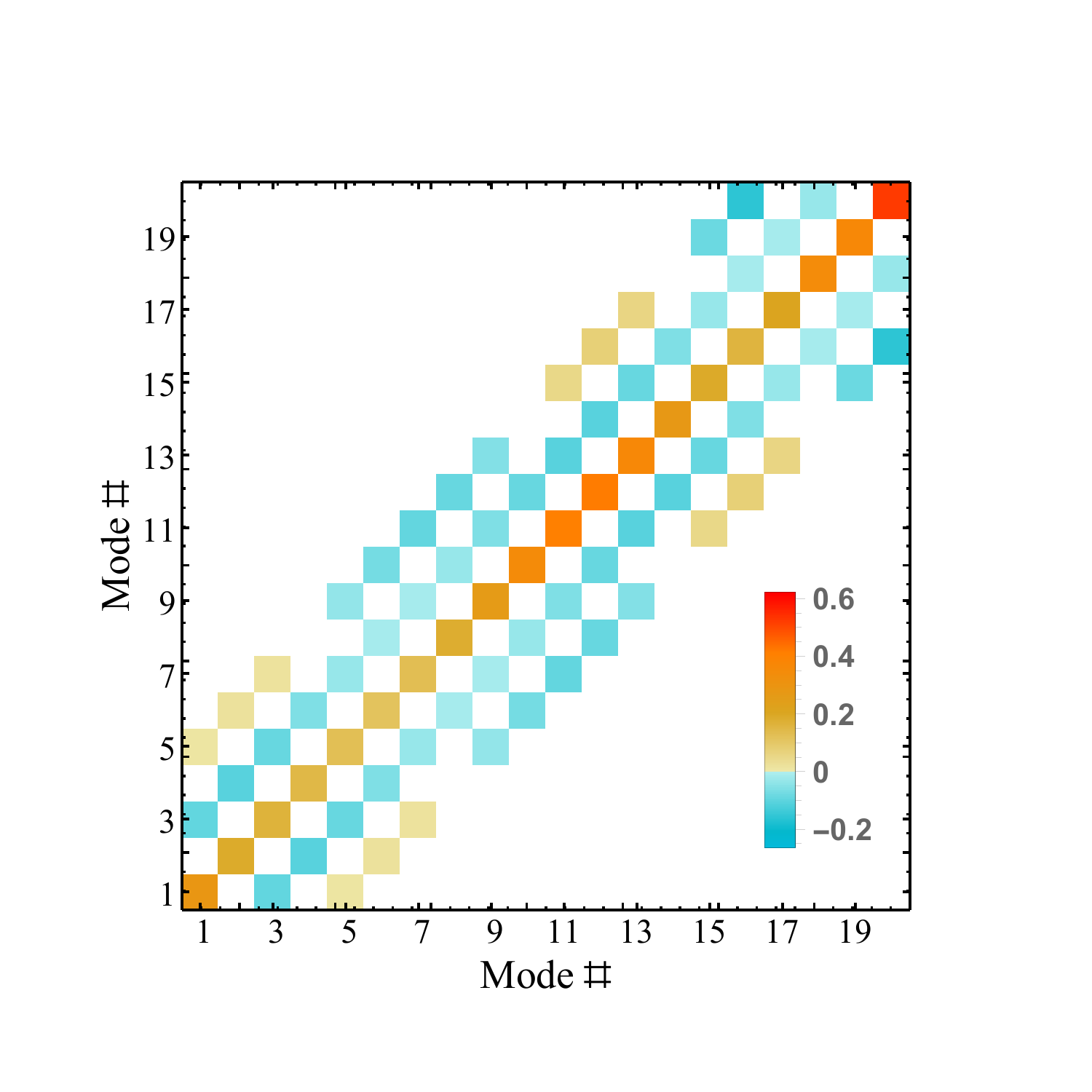}
        \caption{n=2}
        \label{fig:pert2}
    \end{subfigure}
    \begin{subfigure}[b]{0.3\textwidth}
        \includegraphics[trim = {1cm, 1cm, 2cm, 2cm},clip,scale=0.3]{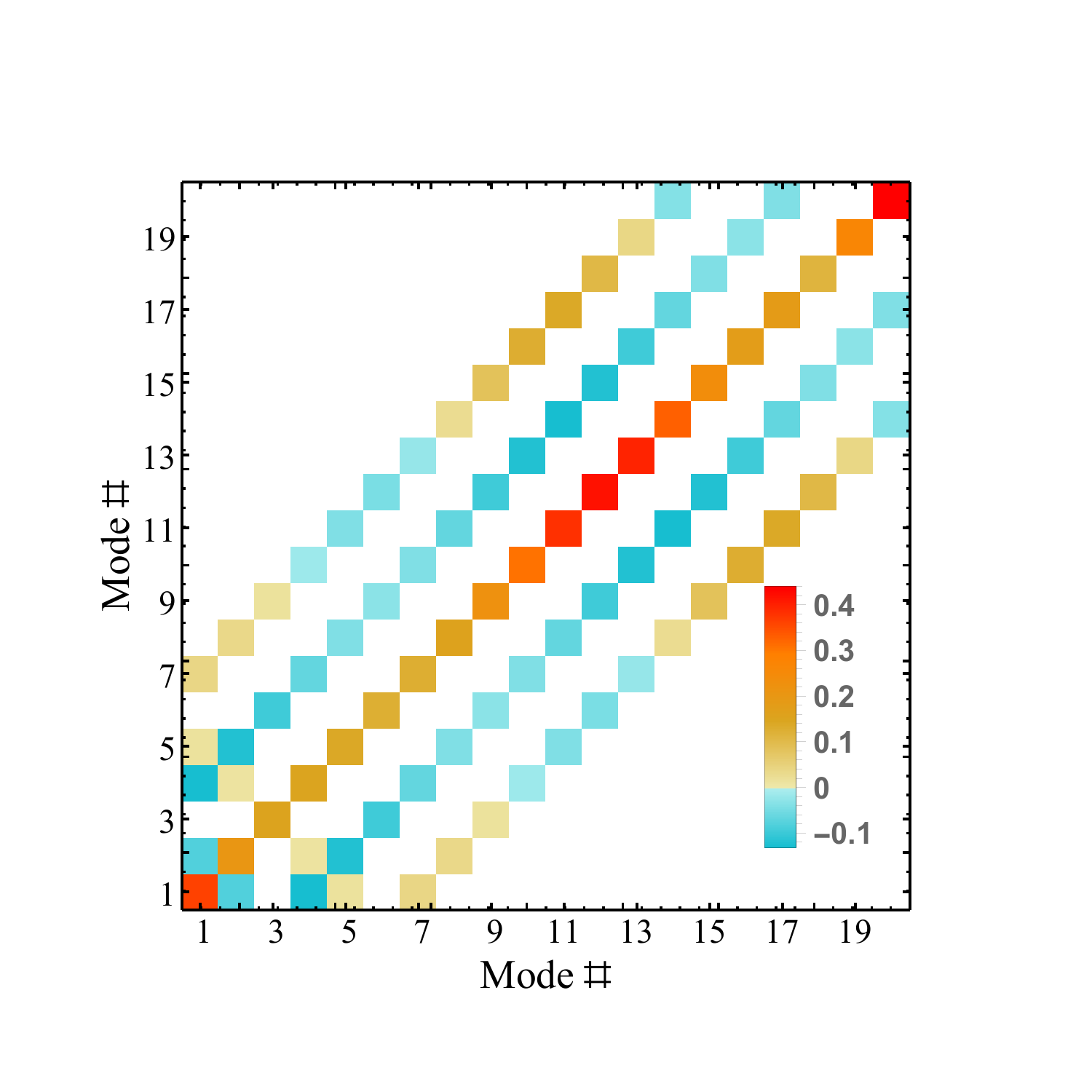}
        \caption{n=3}
        \label{fig:pert3}
    \end{subfigure}
    \begin{subfigure}[b]{0.3 \textwidth}
        \includegraphics[trim = {1cm, 1cm, 2cm, 2cm},clip,scale=0.3]{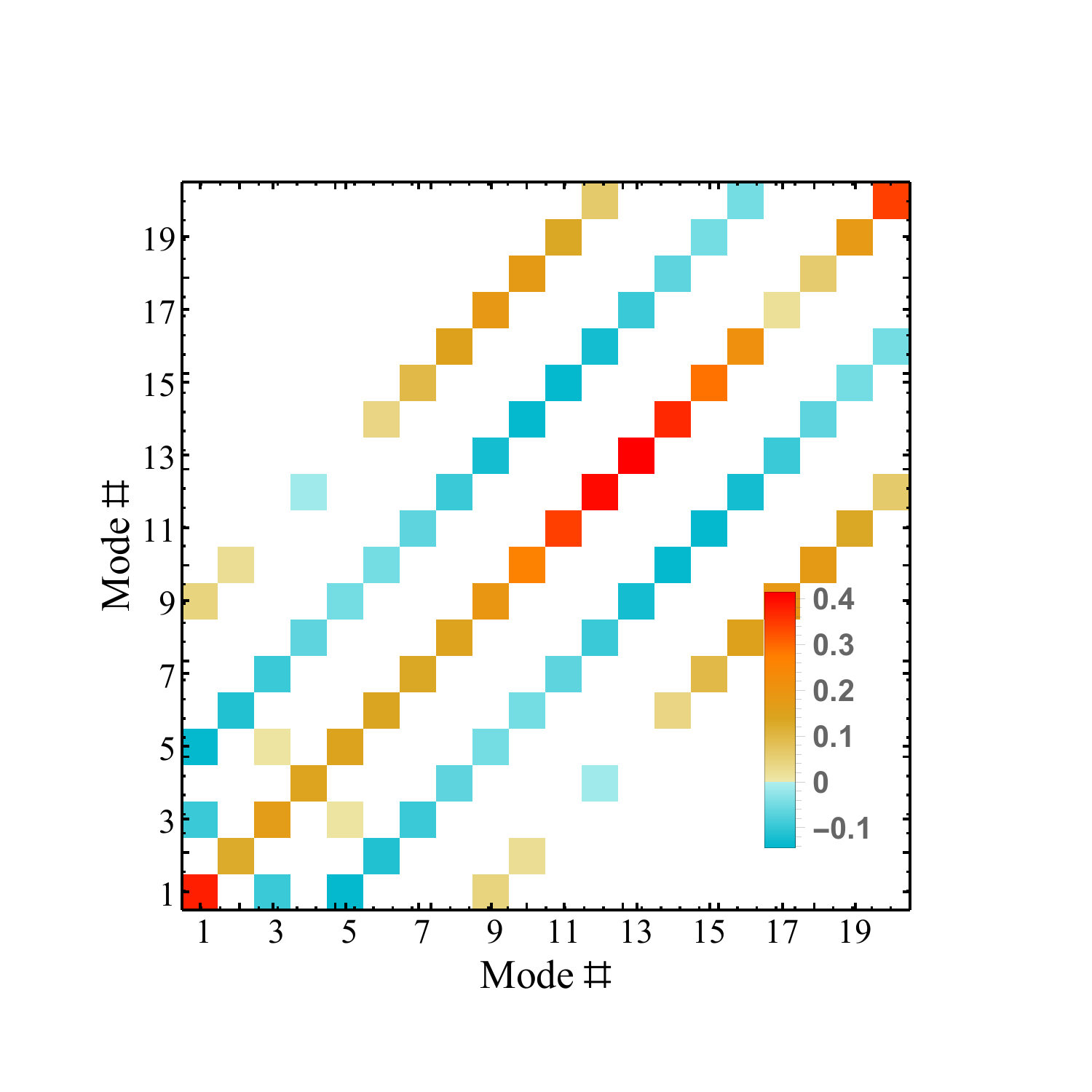}
        \caption{n=4}
        \label{fig:pert4}
    \end{subfigure}
    \begin{subfigure}[b]{0.3 \textwidth}
        \includegraphics[trim = {1cm, 1cm, 2cm, 2cm},clip,scale=0.3]{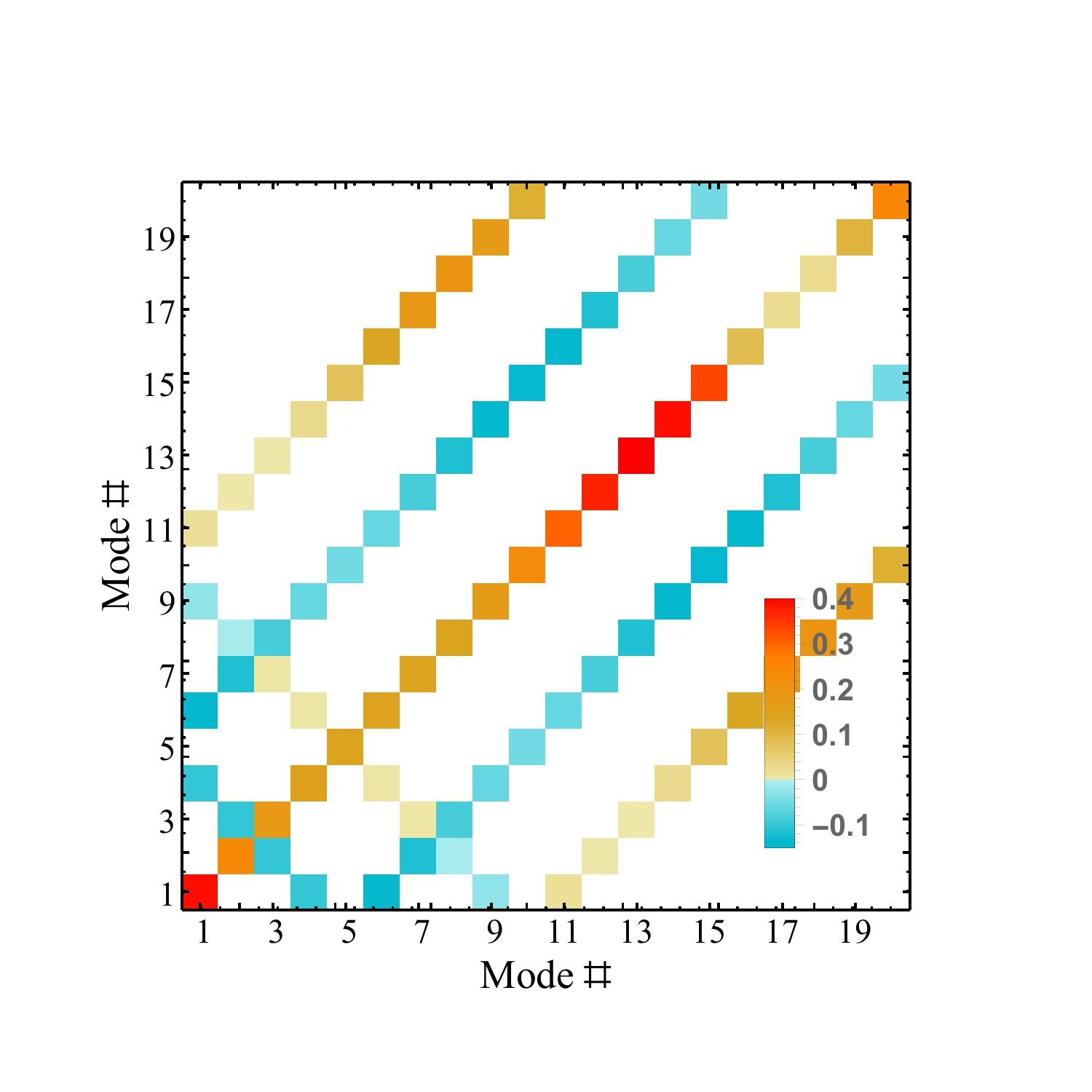}
        \caption{n=5}
        \label{fig:pert5}
    \end{subfigure}   
    \caption{Matrix plots of the correlation matrix $\expe{\phi_{res}(n,t) \phi_{res}(m,t)}$ for short times. At short time the weight is transferred from the diagonal to off-diagonal elements that differ by the frequency of the drive. Units of the plots are the phase quantum, while the plots where taken for $\epsilon =0.01$ and $t = 10 L_0/v_s$.}
\label{fig:pert}
\end{figure*}

Perturbation theory shows that at short times the drive couples only modes that are related to each other by adding or subtracting the drive's frequency. However, the non-zero $V$ Bogoliubov matrix already demonstrates phonon creation since the phonon number, $P_n$ is given by
\begin{equation}
P_n = \left( V^\dag V \right)_{n,n}
\end{equation}
This quantity grows quadratically with time since $V$ grows linearly in perturbation theory. Fig.~\ref{fig:pert} shows the fluctuations of $\phi_{rel}$ when the system is driven on resonance with the first 5 eigenmodes at stroboscopic times. In the plots, we see the emergence of a checker-board pattern where modes that differ by the driving frequency become correlated.

\subsection{Asymptotic state: Parametric pattern}
\begin{figure}
\includegraphics[scale=0.4]{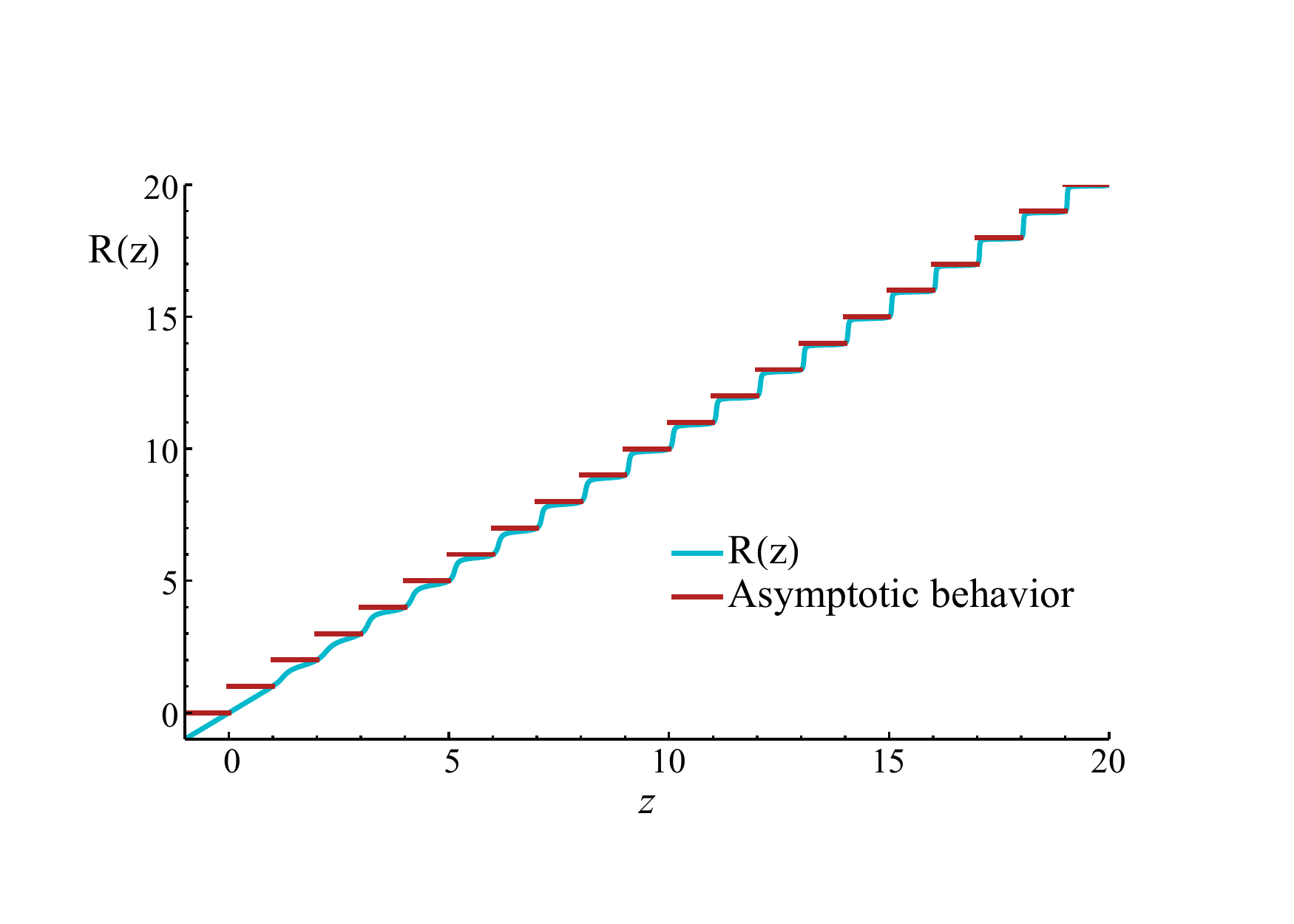}

\caption{Example of the transformation function, $R(z)$, plotted against its asymptotic behavior for a wall that oscillates at the second harmonic frequency, $L(t) = L_0 \left( 1 + \epsilon\left( 1 - \cos\left( \frac{2 \pi}{L_0} t \right) \right) \right)$. At late times $R(z)$ approaches the staircase function. How quickly this occurs depends on the driving amplitude, however in this example for $\epsilon = 5 \%$ the asymptotic behavior is established after a few oscillations.}
\label{fig:asympt}
\end{figure}

\begin{figure*}
\captionsetup[subfigure]{justification=centering}
    \centering
    \begin{subfigure}[b]{0.3\textwidth}
        \includegraphics[trim = {1cm, 1cm, 2cm, 2cm},clip,scale=0.3]{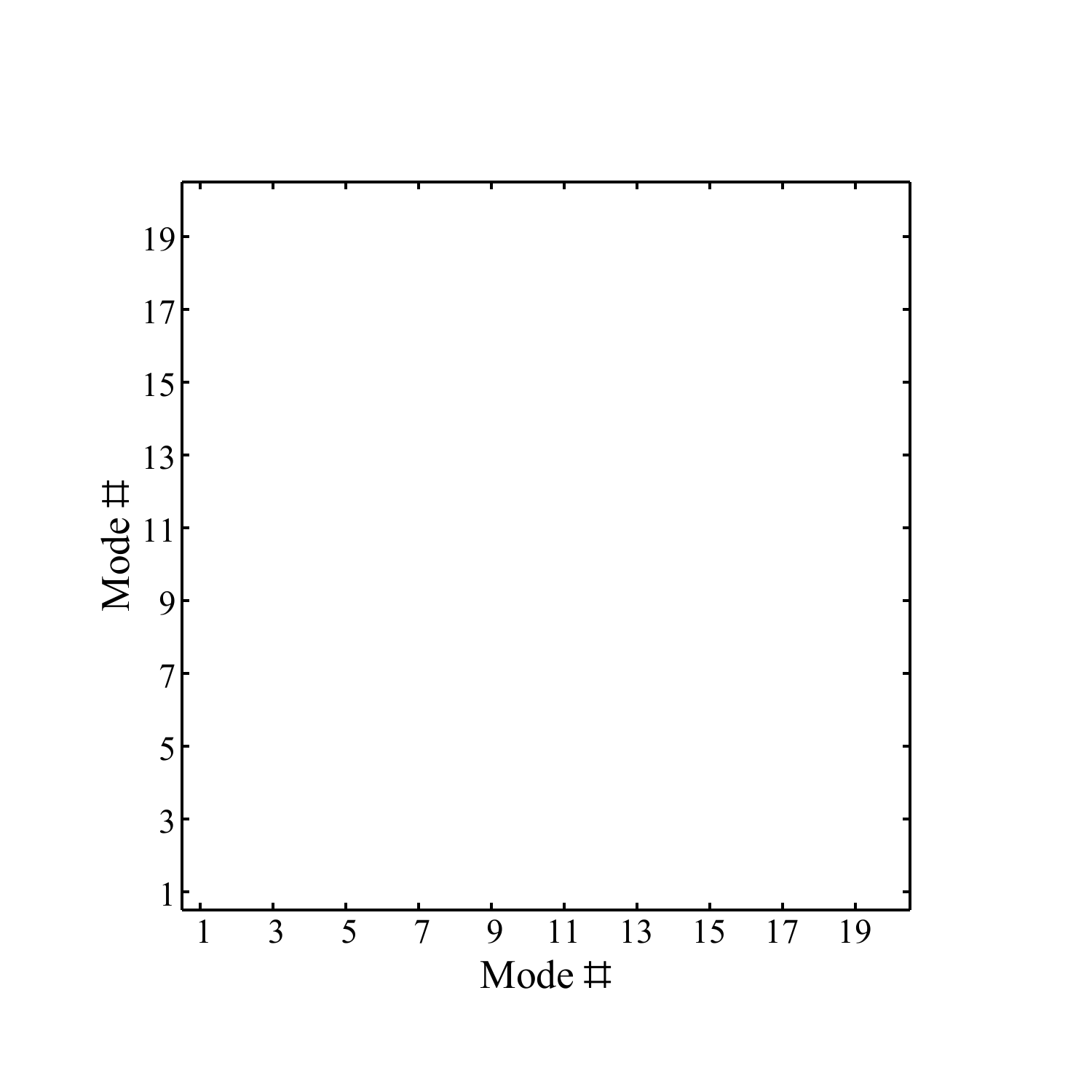}
        \caption{n = 1}
        \label{fig:Cmn1}
    \end{subfigure}
    \begin{subfigure}[b]{0.3\textwidth}
        \includegraphics[trim = {1cm, 1cm, 2cm, 2cm},clip,scale = 0.3]{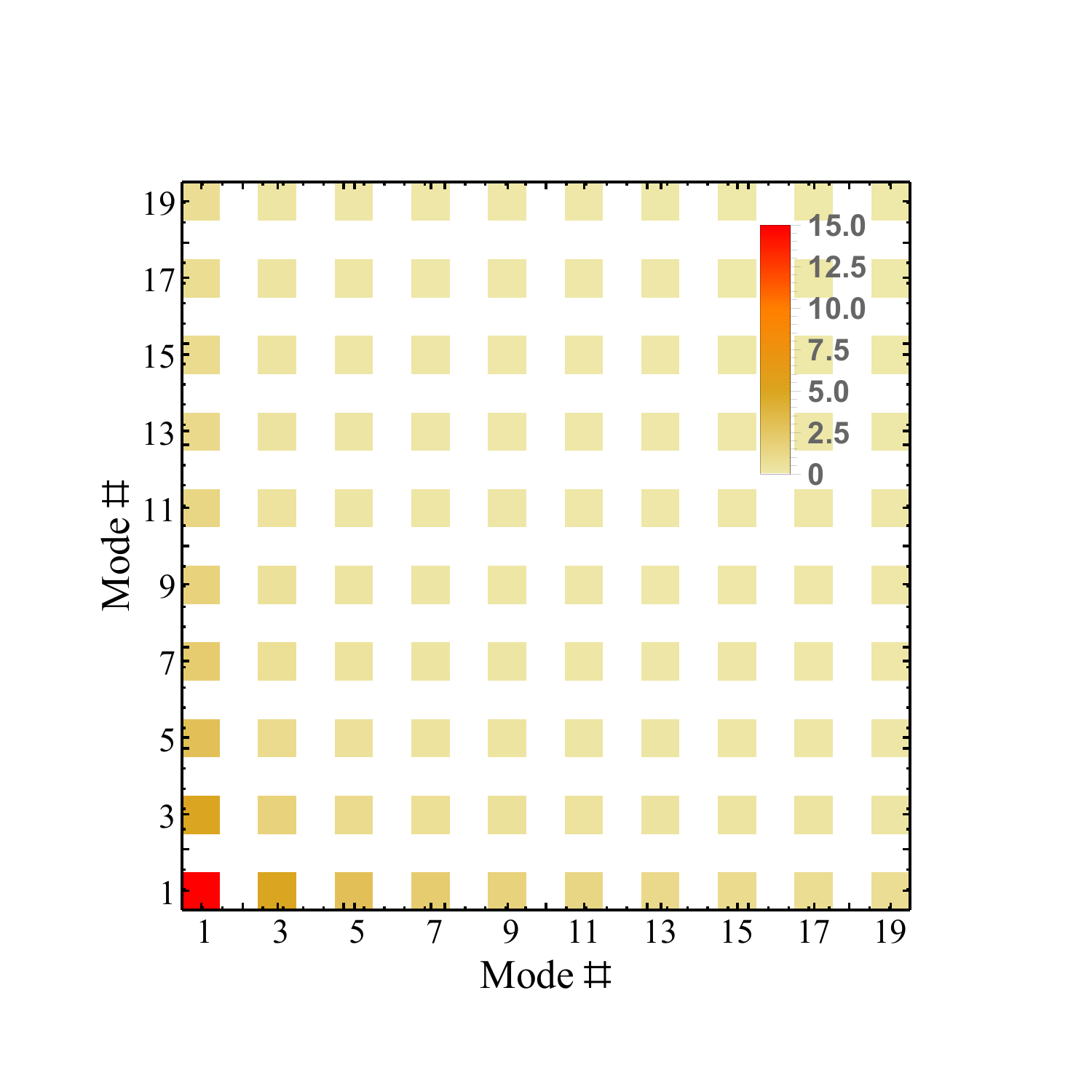}
        \caption{n=2}
        \label{fig:Cmn2}
    \end{subfigure}
    \begin{subfigure}[b]{0.3\textwidth}
        \includegraphics[trim = {1cm, 1cm, 2cm, 2cm},clip,scale=0.3]{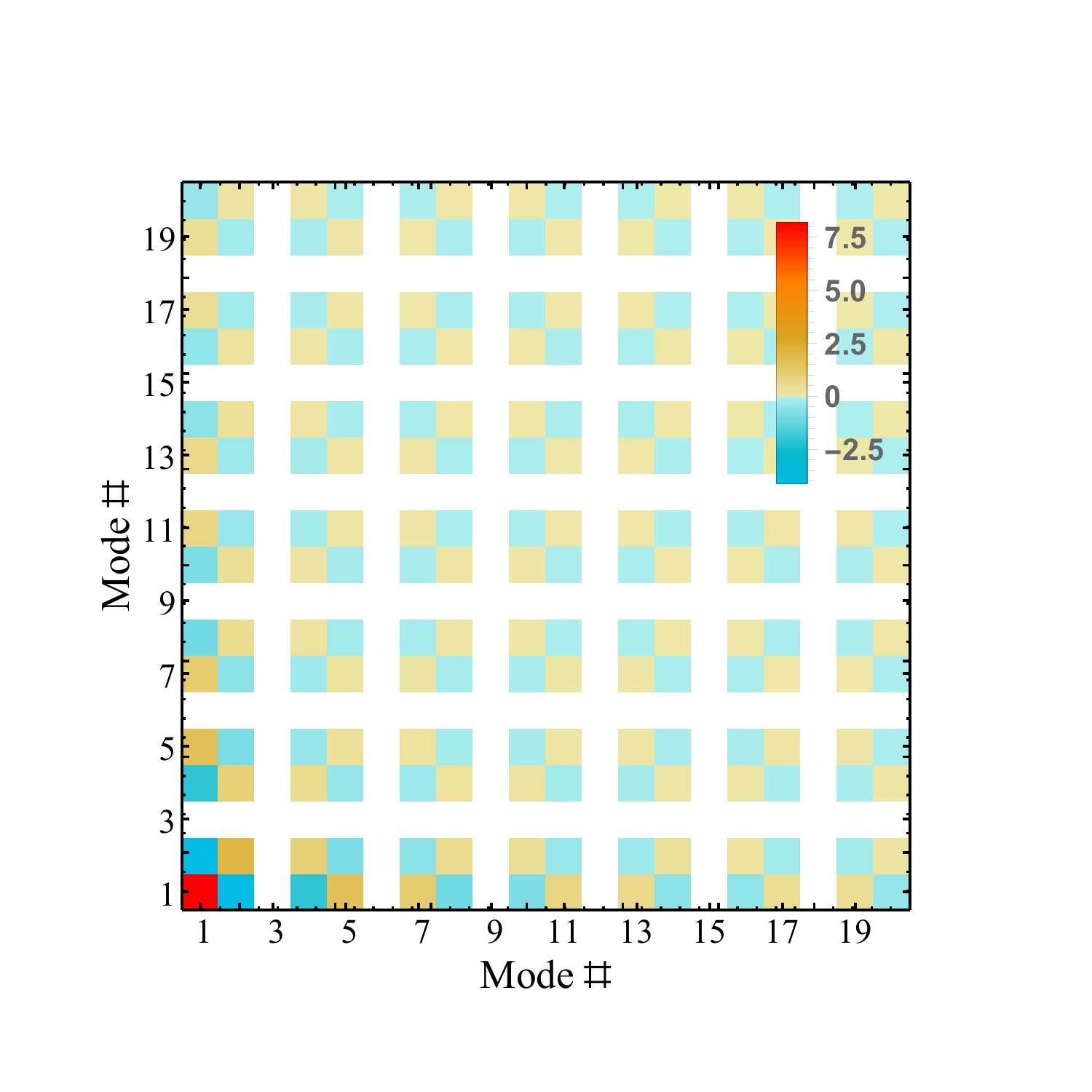}
        \caption{n=3}
        \label{fig:Cmn3}
    \end{subfigure}
    \begin{subfigure}[b]{0.3 \textwidth}
        \includegraphics[trim = {1cm, 1cm, 2cm, 2cm},clip,scale=0.3]{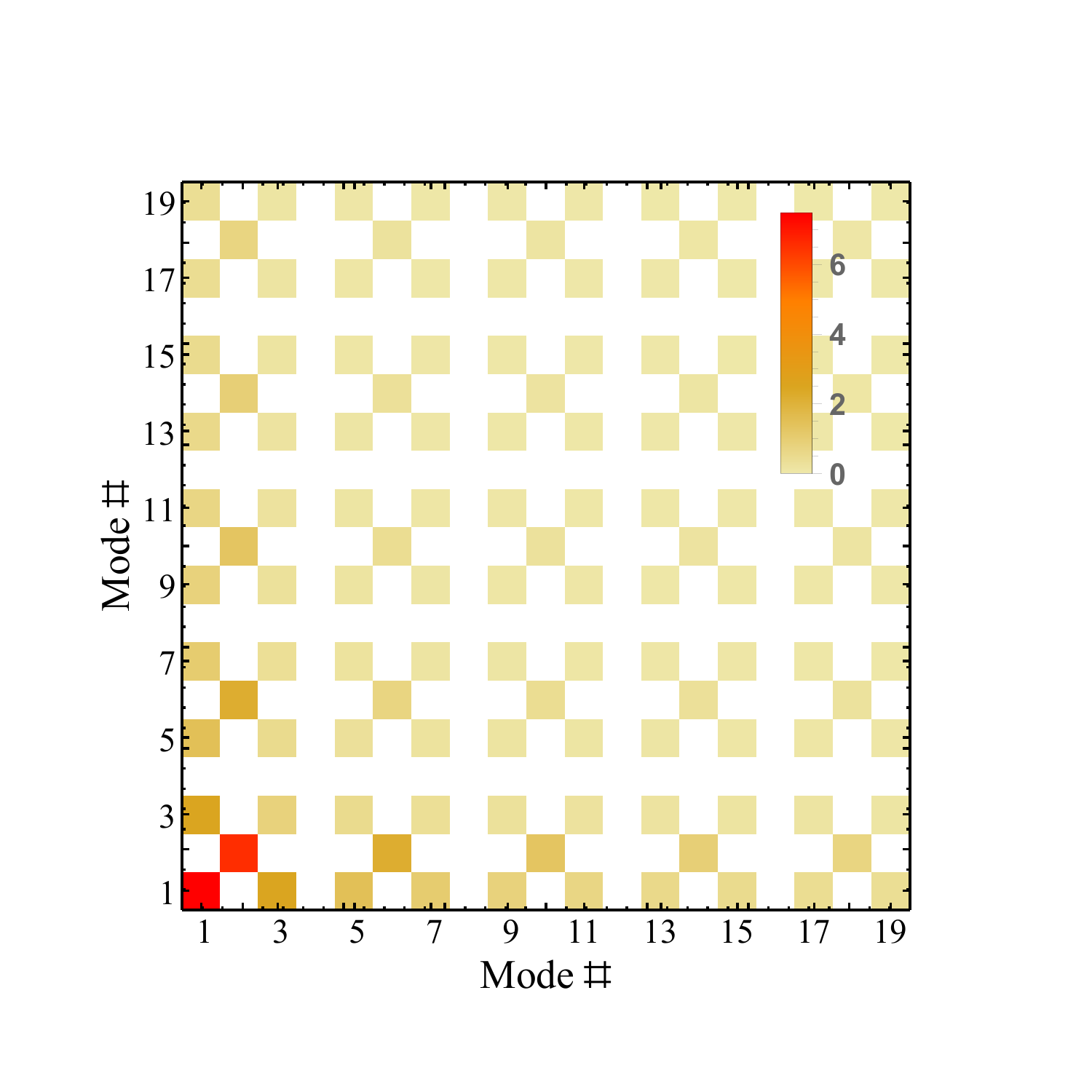}
        \caption{n=4}
        \label{fig:Cmn4}
    \end{subfigure}
    \begin{subfigure}[b]{0.3 \textwidth}
        \includegraphics[trim = {1cm, 1cm, 2cm, 2cm},clip,scale=0.3]{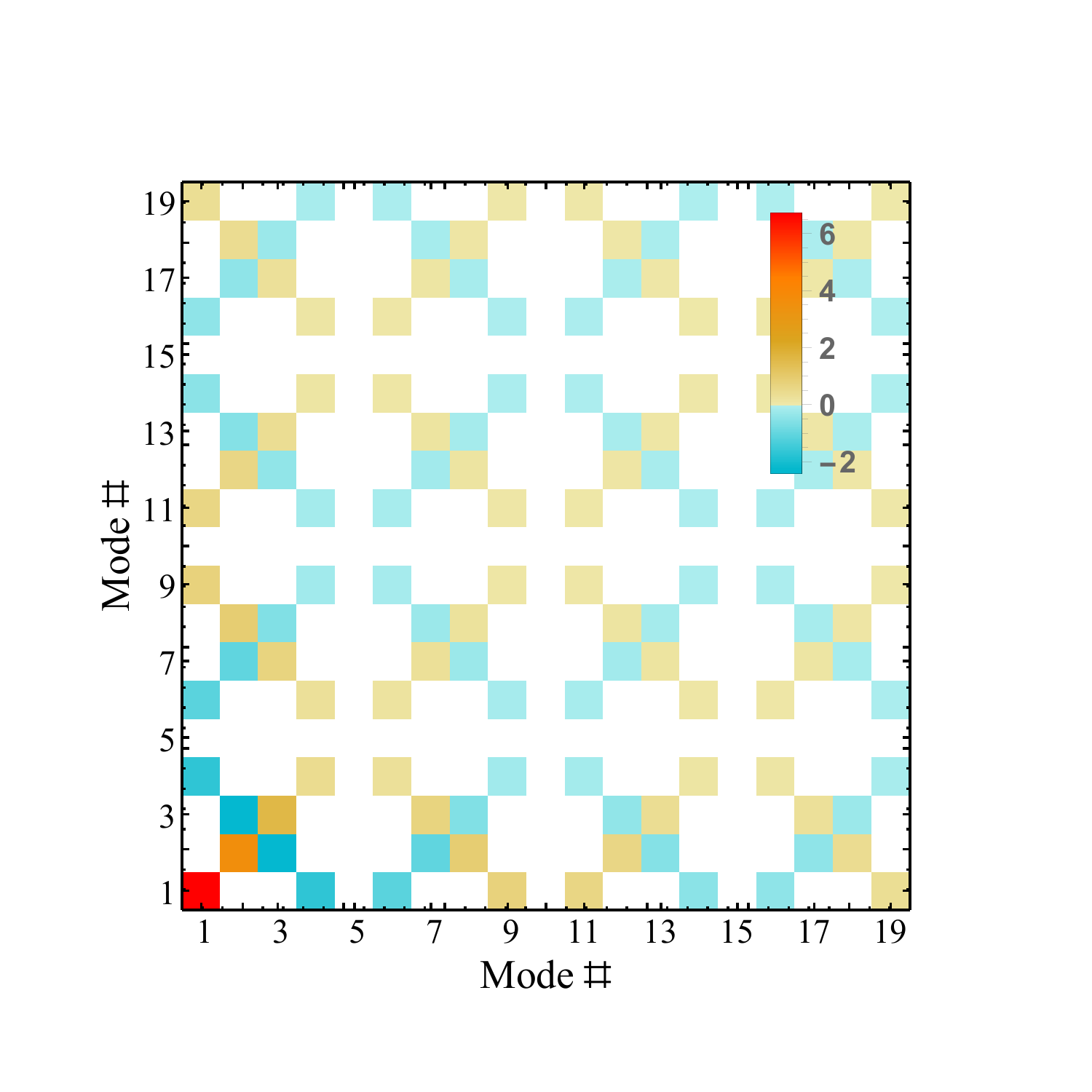}
        \caption{n=5}
        \label{fig:Cmn5}
    \end{subfigure}   
    \caption{Matrix plots of $\expe{\phi_{res}(n,t) \phi_{res}(m,t)} - \expe{\phi_{res}(n,0) \phi_{res}(m,0)} $ as $t \rightarrow \infty$, evaluated at $t =$ multiples of the period of the drive for different drive frequencies, $\omega_n = v_s \frac{n \pi}{L_0}$. The values are obtained using up to 100 modes, corresponding to a momentum cut-off, $\Lambda \sim 100\frac{\pi}{L_0}$. As expected for parametric resonance, we get resonances for $n \geq 2$. The numerical values of the plots are of order $\mathcal{O} (10)$ as anticipated from Eq.~\ref{eq:asymexpe}.}
\label{fig:parares}
\end{figure*}

We now discuss the driven quantum zipper in the limit of long modulation time. When the boundary oscillates at one of the resonant frequencies, the system is expected to display parametric resonance, where the dominant process is pair production of phonons whose individual frequency adds up to the drive frequency. The signatures of such resonances are enhanced correlations (divergent in the limit at $t \rightarrow \infty$ and in the absence of a cut-off) between modes involved in this dominant process as well as between all other modes connected to them by integer multiples of the frequency leading to patterns shown in Fig.~\ref{fig:parares}. This can be demonstrated analytically using the formalism developed by Cole and Schieve \cite{Cole95}, which states that in the case of on resonant driving frequencies, $\omega = \frac{n \pi}{L_0}$, the asymptotic behavior of $R(z)$ is a staircase function independent of the amplitude of the drive.

An example of the asymptotic behavior of $R(z)$ for a boundary motion of the form $L(t) = L_0 (1 - \epsilon(1- \cos\left(\omega t\right) ))$ and frequency, $\omega = \frac{n \pi}{L_0}$, is given by: 
\begin{equation}
R_0(z) \rightarrow \frac{2(l + 1) }{n} + 1, \frac{2 L_0 l}{n} < z - L_0 < \frac{2 L_0 (l + 1) }{n} , l \in \mathcal{Z}
\label{eq:asymR}
\end{equation}
where $T = \frac{2 L_0}{ n}$ is the period of the drive. The case of $\omega =\frac{2 \pi}{L_0}$ and $\epsilon = 5\%$ is shown in Fig.~\ref{fig:asympt}, where one can see that the numerically integrated transformation function approaches the suggested staircase function at late times. 

In this regime, the matrices $U^\dag$ and $V^\dag$ can be analytically calculated (for details see Appendix~\ref{app:asym}):
\begin{subequations}
\begin{align}
\begin{split}
V_{\nu,\mu} =& - \frac{n}{\sqrt{\mu \nu } \pi}\sin(\frac{\mu \pi}{n}) e^{-i\frac{\pi \mu}{n}} e^{ i\pi (\nu+ \mu )\left(\frac{2 l_0}{n} - 1 + \frac{2}{n}\right)} \\ & \times \sum_{\rho = 1}^{\rho = \infty} \delta_{\mu + \nu , \rho n} ,
\end{split}\\
\begin{split}
U_{\nu,\mu} &= - \frac{n}{\sqrt{\mu \nu } \pi}\sin(\frac{\mu \pi}{n}) e^{i\frac{\pi \mu}{n}} e^{ i\pi (\nu - \mu )\left(\frac{2 l_0}{n} - 1 + \frac{2}{n}\right)}  \\ & \times \sum_{\rho = -\infty }^{\rho = \infty} \delta_{ \nu- \mu , \rho n} 
\end{split}
\end{align}
\label{eq:vulate}
\end{subequations}
which shows that asymptotically $V_{\nu,\mu}$ is 0 everywhere apart from when $\mu + \nu = \lambda n$ where $ \mu,\nu,\lambda ,n \in \mathcal{Z}$. 

Parametric resonances are expected for $\omega = \frac{n \pi}{L_0} $ at $n \geq 2$. As an illustration of the resonance at $\omega = \frac{2 \pi}{L_0}$, the element $\sum_l U_{1,l} U^\dag_{l,1}$ is calculated which contributes to the phase quantum fluctuations: 
\begin{equation}
(\underline{\underline{U}} \cdot \underline{\underline{U}}^\dag )_{1,1}
= \sum_{\rho = 0}^{\rho = \infty} \frac{4} {2 \rho + 1} \rightarrow \infty 
\label{eq:parametric}
\end{equation}
The divergence is associated with the fact that at infinite time, the first mode obtains contributions from arbitrarily high momenta. In any physical situation, there will be a momentum cut-off $\Lambda$ up to which our theory is valid. As discussed in Sec~\ref{sec:zipper} in the Luttinger liquid model this cut-off is given by the healing length, $\Lambda = \frac{1}{\xi_h}$. Since the expression in Eq.~\ref{eq:parametric} diverges logarithmically with the momentum cut-off we can find an estimate of the asymptotic values of resonant quantum fluctuations:
\begin{equation}
(\underline{\underline{U}} \cdot \underline{\underline{U}}^\dag )_{1,1} \sim 2\ln \left( \frac{L_0}{\xi_h} \right) 
\end{equation} 
Therefore, if the box length, $L_0$, is 2-3 orders of magnitude larger than the healing length, $\xi_h$, the resonant quantum fluctuations would be expected to be:
\begin{equation}
\frac{\phi_{res}^2(m=1,t \rightarrow \infty) }{\phi_{res}^2(m=1,t=0)} \sim \mathcal{O}(10)
\label{eq:asymexpe}
\end{equation}
With this figure falling off at higher momenta. This estimate is confirmed numerically with a hard cut-off as shown in Fig.~\ref{fig:parares}, where the fluctuations' matrix $ \expe{\phi_{res} (n,t) \phi_{res} (m,t)}$ as $t \rightarrow \infty$ was plotted for different resonant frequencies of the drive.

Introducing a UV cutoff in the wave equation is justified on physical grounds since at momenta $~\frac{1}{\xi_h}$ the dispersion relation is no longer linear. As a result the mode coupling mechanism is not efficient anymore and higher momenta become off-resonant in relation to lower ones.

In order, to show the range of validity of this approach, we also investigate the cross over from the perturbative state to the asymptotic one. To this end, it is more convenient to consider the phonon occupation number of the parametric resonant mode. Following the results obtained in Ref.~\cite{Dodonov93}, the late time asymptotic expansion of the $V$ matrix when driving at frequency $\omega_d = \frac{2 r \pi}{L_0} $ is given by:
\begin{subequations}
\begin{align}
V_{m,n} &= \sqrt{\frac{m}{n}} \frac{\sin\left( \frac{\pi( 2 r n \delta + m )}{2 r} \right)}{\pi (2r n \delta + m ) } \frac{\sin(\pi(m+ n)}{\sin\left( \frac{\pi(n + m) }{2 r} \right) } e^{i \pi ( n + m) (1 - \frac{1}{2r}) },\\
\delta &= \frac{e^{- r \pi \epsilon t} }{\pi r}
\end{align}
\end{subequations}
Thus for an infinite cut-off momentum the occupation number of parametrically resonant modes, increases linearly without a bound:
\begin{equation}
P(m)=  \sum_{n = 0}^{\infty} \left|V_{n,m}\right|^2 \propto t, \mbox{ for \quad} t >> \frac{1}{\epsilon \omega} 
\end{equation}  
Therefore there is a cross-over behavior between the perturbative regime and the asymptotic regime, where phonon occupation changes from a quadratic growth to a linear one. 

\begin{figure*}
    \centering
	\begin{subfigure}[b]{0.45\textwidth}
        \includegraphics[trim = {1.5cm, 1cm, 1cm, 1cm},clip,scale=0.5]{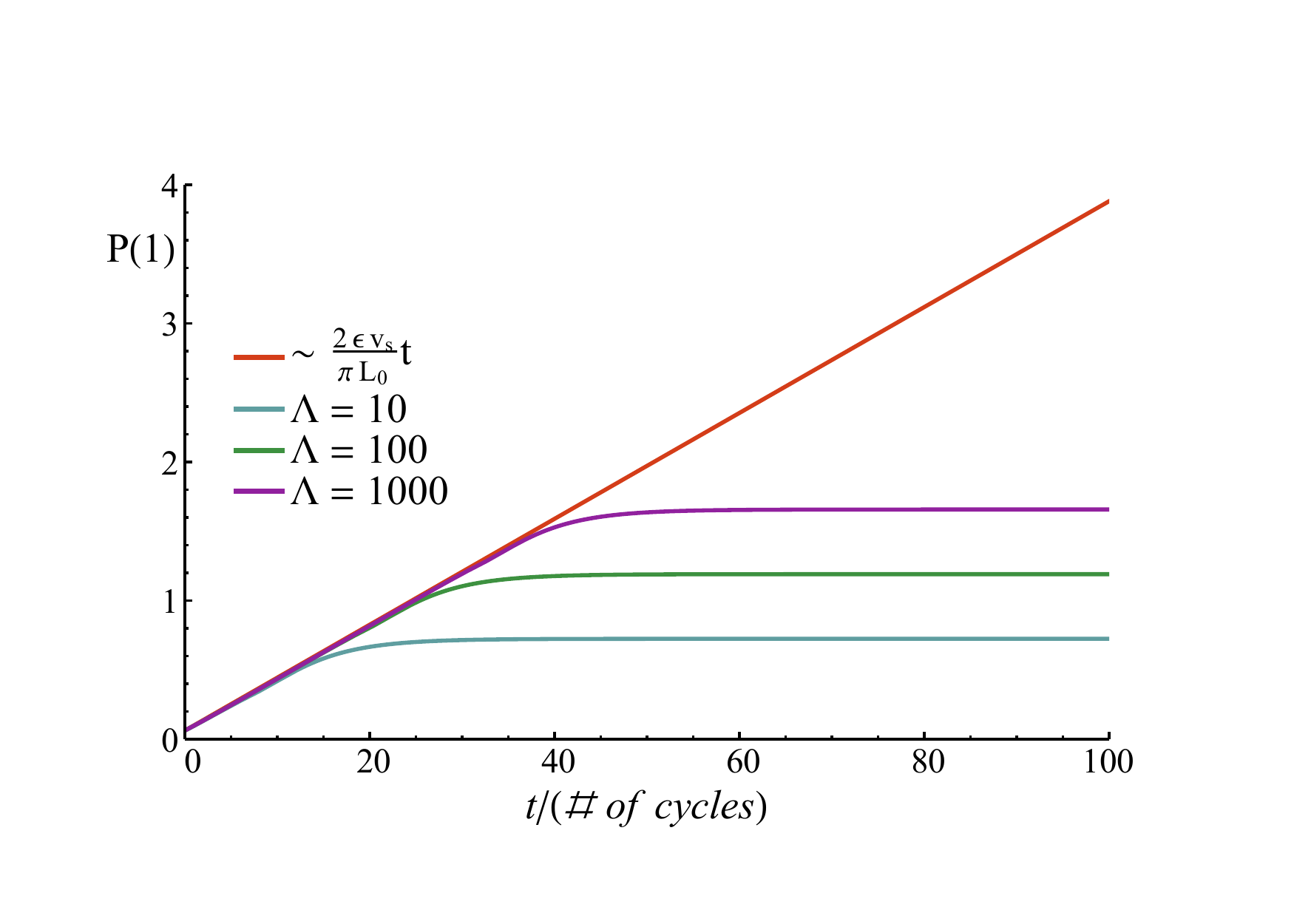}
        \caption{Asymptotic expansion with various cutoff momenta.}
        \label{fig:undriven}
    \end{subfigure}   
    \begin{subfigure}[b]{0.45\textwidth}
        \includegraphics[trim = {1cm, 1cm, 1cm, 1cm},clip,scale=0.5]{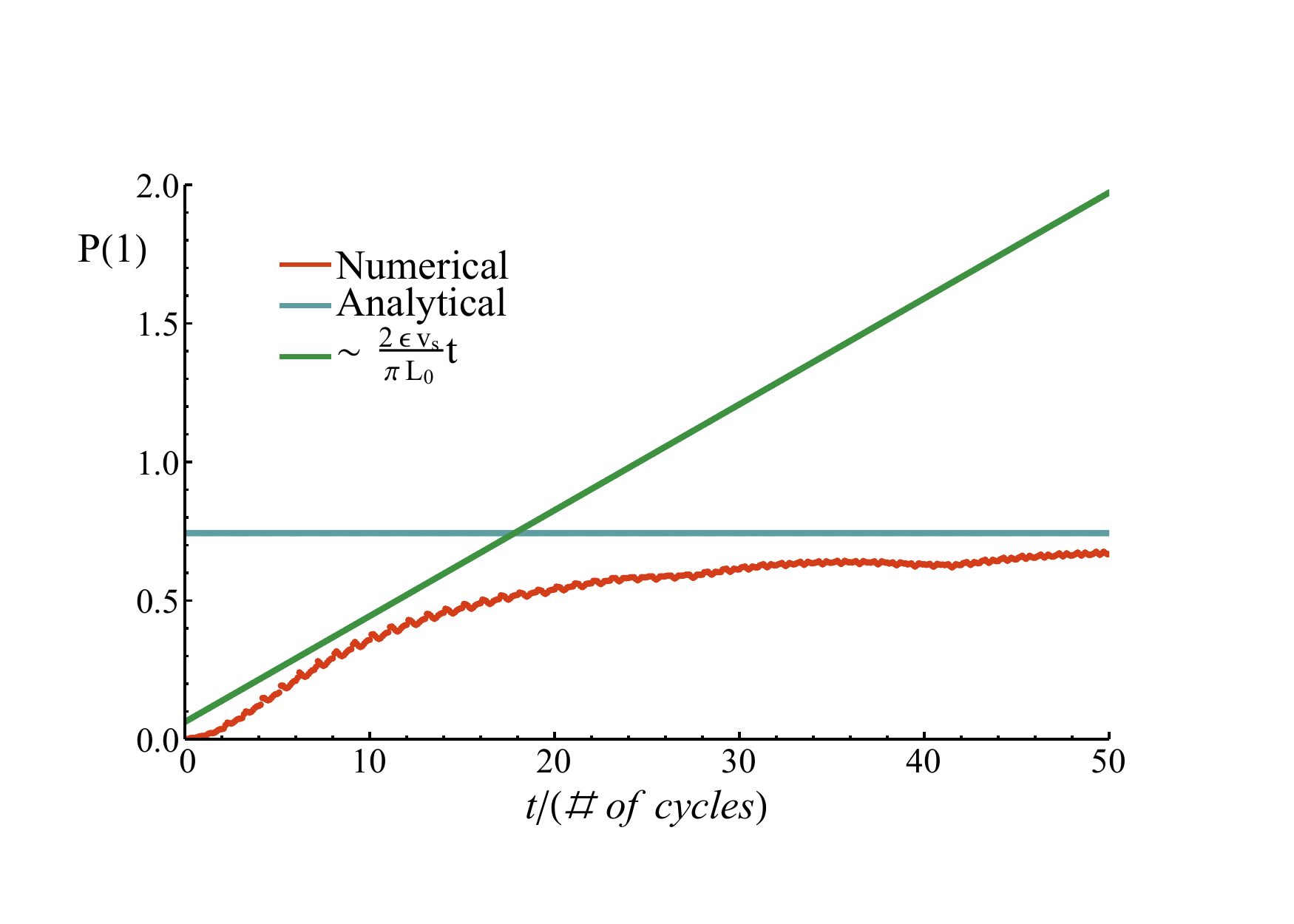}
        \caption{Numerical results with cut-off $\Lambda = 10$.}
        \label{fig:Cmn1}
    \end{subfigure}
\caption{Approach to the asymptotic state. One can see that before a time threshold is reached the phonon occupation of a single mode is independent of the cut-off and grows linearly for resonant modes. From the analytical graph one can see that both the time and the maximum occupation grow as $\log(\Lambda)$.}
\label{fig:cross}
\end{figure*}

In Fig.~\ref{fig:cross}, we present results for the system with the wall oscillating at $w_{dr} = \frac{2 \pi }{L_0}$. We compute occupation number of the first eigenmode, which is parametrically resonant with the drive. We use both the direct numerical solution and the asymptotic expansion. 
The situation is demonstrated both using the asymptotic expansion with a momentum cutoff as well as numerically in Fig.~\ref{fig:cross} for a wall oscillating at $\omega_{dr} = 2\frac{\pi}{L_0}v_s$ and focusing on the occupation of the first eigenmode which is parametrically resonant (in this example even modes are not parametrically resonant and their occupation number goes to zero at late times). The figure shows that the linear occupation growth of a particular mode is unaffected by the momentum cutoff up until the maximum occupation is reached. For this particular example the occupation grows as:
\begin{equation}
P(1) \sim \frac{2 \epsilon v_s}{\pi L_0} t 
\end{equation}
Saturation to the cut-off dependent maximum occupation value is reached at:
\begin{equation}
t = \frac{Log(\Lambda)}{\epsilon \omega} 
\end{equation}

As expected, due to the linear relationship between occupation number and time, the saturation time also grows logarithmically with the cut-off. As a result, there is a window where perturbation theory breaks down and at the same time the asymptotic state can make predictions independent of the UV physics of the theory:
\begin{equation}
1<< \epsilon \omega t_{asym} << \log(\Lambda)
\end{equation}

Numerically, from Fig.~\ref{fig:cross}, one can see the initial quadratic growth in the perturbative regime and the asymptotic approach to the saturation value.

Before concluding this section we note that a complementary way of looking at the problem is through a Floquet picture as discussed by I. Martin\cite{Martin18}. In our language the Floquet map is the Bogoliubov transformation relating two instantaneous reference frames that differ by one period, $T$. From that point of view, fixed points in the Floquet map is the cause of the staircase form of the late time asymptotic solution of the transformation function $R(z)$.

\section{Shaking Box}
\label{sec:shake}

We now turn our attention to a different setup motivated by the recent experiments in Ref.~\cite{Zoran17}. We consider a single box of a 1d quantum fluid with a moving wall. We parametrize the length dependence of the box as $L(t) = L_0(1+\epsilon(t))$. As before, we will use Luttinger liquid based wave equation to describe dynamics. We also allow for a time-dependent external potential, $V_{ext.}$ which will be useful for separating the classical and quantum parts of the response:

\begin{subequations}
\begin{align}
\partial_t \rho(x,t) &= - \frac{\hbar }{m}\partial_x \left( \rho \partial_x \phi(x,t) \right),\label{eq:continuityV} \\
\hbar \partial_t \phi(x,t) &= -U_0 \rho(x,t)- V_{ext.}(x,t),\label{eq:phieqV}\\
\left[\rho(x,t) , \phi(x',t) \right] &= i \delta(x - x')\label{eq:comrelV}
\end{align}
\label{eq:EofMV}
\end{subequations}

\textbf{Boundary conditions:} In the fixed box case boundary conditions are given by the requirement that the current, $j(x,t)  = \frac{\hbar \rho_0}{m} \partial_x \phi(x,t) $, vanishes at the edges of the box. However, when the wall is moving, the current should be finite near the wall in order for the fluid to follow the motion of the wall. To derive the new boundary conditions, we start by writing the conservation of the particle number:
\begin{equation}
\int_0^{L(t)} \rho(x,t) d x = N_0
\end{equation}
We require that
\begin{equation}
\frac{d N_0}{d t}= 0
\end{equation}
Which can be written as
\begin{equation}
\begin{split}
&\frac{\hbar \rho(L(t))}{m} \left(\partial_x \phi \left( x= L(t) \right) - \partial_x \phi \left( x = 0 \right) \right) =\\ &\frac{d L(t)}{d t} \rho\left(L(t) \right)
\end{split}
\end{equation}
This gives us
\begin{subequations}
\begin{align}
\frac{\hbar}{m} \partial_x \phi\left( x = 0 \right) &= 0 , \\
\frac{\hbar}{m} \partial_x \phi\left( x= L(t) \right) &= \frac{d L(t)}{dt}
\label{eq:fixedN}
\end{align}
\end{subequations}
Unsurprisingly, we find that the fluid velocity, $\frac{\hbar}{m} \partial_x \phi\left( x= L(t) \right)$, at the boundary should match the velocity of the wall. This causes a sloshing motion which is associated with the change of the average density due to the change of the box length. It is convenient to simplify the boundary conditions by subtracting this sloshing motion using the transformation:
\begin{subequations}
\begin{align}
\rho(x,t) &= \tilde{\rho}(x,t) - \epsilon (t)\rho_0, \\
\phi(x,t) &= \tilde{\phi}(x,t) + \frac{d\epsilon(t)}{d t} \frac{m }{2 \hbar}x^2
\end{align}
\end{subequations}
This choice of transformation simplifies the boundary conditions at the cost of introducing an effective quadratic potential perturbation:
\begin{subequations}
\begin{align}
\frac{\partial}{\partial t} \tilde{\rho}\left(x,t \right) &= -\frac{\hbar \rho_0}{m} \partial_x^2 \tilde{\phi}(x,t),\\
\hbar \frac{\partial}{\partial t} \tilde{\phi}(x,t) &= - U_0\tilde{\rho}(x,t) - V(x,t)- V_{ext}(x,t)   \\
\left[ \tilde{\rho}(x,t) , \tilde{\phi}(x',t) \right] &= i \delta(x - x'),\\
\left. \frac{d \tilde{\phi} }{dx} \right|_{x = 0} &= \left. \frac{d \tilde{\phi}}{d x}\right|_{x=L(t)} = 0,\\
V(x,t) &= \frac{d^2 e(t)}{dt^2} \frac{ m x^2}{2} - e(t) U_0 \rho_0 
\end{align}
\label{eq:tildeeq}
\end{subequations}

The new density variable, $\tilde{\rho}(x,t)$, from which the slosing motion has been subtracted, now has a constant average density:
\begin{subequations}
\begin{align}
\tilde{\rho}_{av.} &= \rho_{av.} + \epsilon(t)\rho_0 , \\
\tilde{\rho}_{av.} &= \frac{N_0}{L(t)} + \epsilon(t) \frac{N_0}{L(t)} , \\
\tilde{\rho}_{av.} &= \frac{N_0}{L_0}
\end{align}
\end{subequations}

In order to proceed, we make a change of variables to the current density, $j = \frac{\hbar \rho_0}{m} \partial_x \tilde{\phi}(x,t)$, and its conjugate momentum, $\tilde{\rho}(x,t) = \frac{\hbar \rho_0}{m} \partial_x \theta(x,t)$. Substituting these definitions in Eq.~\ref{eq:tildeeq} we find:
\begin{subequations}
\begin{align}
\frac{\hbar \rho_0}{m} \partial_t \theta &= - j,\label{eq:jcont}\\
\partial_t j &= -\hbar \frac{\rho_0^2}{m^2}\partial_x^2 \theta - \rho_0 \frac{d^2 \epsilon(t)}{dt^2} x - V_{ext}(x,t) ,\label{eq:jjoseph} \\
\left[ \tilde{j}(x,t), \tilde{\theta}(x',t) \right] &= i \delta( x - x'), \\
\tilde{j}(0,t) &= \tilde{j}(L(t),t) =0
\end{align}
\end{subequations} 
Combining equation Eq.~\ref{eq:jcont} and Eq.~\ref{eq:jjoseph}, we finally arrive at:
\begin{equation}
\left( \partial_t^2 - v_s^2 \partial_x^2 \right) j(x,t) = -\rho_0 \frac{d^3 \epsilon(t)}{dt^3} x - \partial_x V_{ext}(x,t)
\label{eq:drivenKGj}
\end{equation}
where the sound velocity, $v_s$, is given by $v_s^2 = \frac{U_0 \rho_0}{m} $. Eq.~\ref{eq:drivenKGj} with moving boundaries is what we will call inhomogeneous dynamical Casimir effect. This equation is an analogue of the dynamical Casimir effect where the current can be mapped onto the vector potential in a 1d cavity and in the absence of the effective driving term. From Eq.~\ref{eq:drivenKGj} it is clear that a compensating external potential can be applied to modulate the strength of the inhomogeneous drive, even cancel it completely, by choosing a suitably varying perturbing harmonic potential:
\begin{equation}
V_{ext}(x,t) = (\alpha- 1) \frac{d^2 e(t)}{dt^2} \frac{ m x^2}{2}
\label{eq:alpha}
\end{equation}
Eq.~\ref{eq:drivenKGj} becomes:
\begin{equation}
\left( \partial_t^2 - v_s^2 \partial_x^2 \right) j(x,t) = -\alpha \rho_0 \frac{d^3 \epsilon(t)}{dt^3} x
\end{equation}
where the inhomogeneous drive's strength is now determined by the parameter $\alpha$. The density response is found by using the continuity equation:
\begin{equation}
\partial_t \rho = - \partial_x j
\label{eq:cont}
\end{equation}
(In fact, calculating directly the density using the equations of motion may lead to results that do not take into account the boundary conditions correctly. A small discussion of this pitfall is discussed in App.~\ref{app:inhomo}.)

In solving Eq.~\ref{eq:drivenKGj}, one notices that, since the equation is linear, the solution can be decomposed into a particular solution that satisfies the inhomogeneous equation plus an arbitrary solution of the homogeneous equation.
\begin{subequations}
\begin{align}
\hat{j} &= j_{cl} + \hat{j}_q,\\
\left( \partial_t^2 - v_s^2 \partial_x^2 \right) j_{cl} (x,t) &=  -\alpha \rho_0 \frac{d^3 \epsilon(t)}{dt^3} x, \label{eq:jcldef} \\
\left( \partial_t^2 - v_s^2 \partial_x^2 \right) \hat{j}_{q} (x,t) &= 0
\end{align}
\end{subequations} 
We put hat on $\hat{j}_q$ to point out that it should be interpreted as an operator to be quantized, while $j_{cl}$ is a classical function that satisfies the inhomogeneous equation. The term, $j_{cl}$, is the classical contribution generated by the effective potential and is independent of the underlying quantum state, while the term, $j_{q}$, gives the quantum fluctuations that depend sensitively on the quantum state of the system and contain the effects of vacuum squeezing:
\begin{subequations}
\begin{align}
j_{cl} &= \expe{j}, \\
\expe{j_q j_q} &= \expe{j j } - \expe{j} \expe{j}
\end{align}
\label{eq:flucexpej}
\end{subequations}

As a result, by studying the distribution function of the current over many experiment realizations, the average value of the current gives information about the classical response of the field while the second cumulant contains information about the quantum effects. 

In the next section, the perturbative consequences of the effective driving potential are presented to built up intuition in a simple context. Then, we proceed to find the coherent evolution exactly using the formalism developed in Sec.~\ref{sec:quantum}.

Before proceeding, however, we first comment on the subtle way in which quantum fluctuations differ from the cavity QED example of dynamical Casimir effect due to the commutation relations. 

\subsection{Commutation Relations}
\label{subsec:comm}

As mentioned in the previous section, despite the seemingly straightforward identification between the shaken 1d condensate and the gauge field in a 1d cavity QED system due to the shared equations of motion and boundary conditions, the two systems are different in a subtle way. In particular, quantum variables in the former system are interchanged relative to the latter one. This is best demonstrated by the Hamiltonian formulation in the fixed length problem, where the Hamiltonian density and commutation relations for cavity QED is given by:
\begin{subequations}
\begin{align}
H =& \frac{E^2}{2} + \frac{(\partial_x A)^2}{2}, \\
\left[E(x,t) , A(x',t)\right] =& i \delta(x-x'), \\
\end{align}
\label{eq:Ahamil}
\end{subequations}
While the Hamiltonian formulation of the current in dimensionless units takes the form:
\begin{subequations}
\begin{align}
H =& \frac{j^2}{2} + \frac{(\partial_x \theta)^2}{2}, \\
\left[j(x,t) , \theta(x',t)\right] =& i \delta(x-x')
\end{align}
\label{eq:jhamil}
\end{subequations}
Comparing Eq.~\ref{eq:Ahamil} to Eq.~\ref{eq:jhamil}, we see that to preserve commutations relations, we should identify $A(x,t)$ with $\theta(x,t)$ rather than the current. 
However, in the moving box case, focusing on $\theta$ as the original field to quantize, the boundary condition is no longer conformally invariant:
\begin{equation}
\partial_t \theta(L(t),t) = 0 
\end{equation} 

For $j$, we have conformally invariant boundary conditions $\left. j \right|_{\mbox{bound}
} = 0$ but the conjugate momentum is the time integral of $j$ and not the time derivative of $j$. This follows from the equation on $\theta$ : $\partial_t \theta = - j$. If we tried to follow the formalism we presented earlier in Eq.~\ref{eq:kginner}-\ref{eq:cdef}, we would find the commutation relations
\begin{equation}
\left[A(x,t), \partial_t A(x',t) \right] = i\delta(x - x') 
\end{equation}
which do not hold for $j$.
Surprisingly, the retarded Greens function $D^R$ is unaffected by the difference in the commutation relations, as long as the function satisfies Eq.~\ref{eq:RetGreen}. This follows from the fact that $D^R$ is a purely classical object and is not affected by the commutations relations of the system. The retarded Greens function defined in Eq.~\ref{eq:Greens} is applicable for the shaking box as well.

As a result, we will limit our discussion to the coherent dynamics in the subsequent sections, while the quantum fluctuations for the shaking box will be addressed in subsequent publications.

\section{Coherent Dynamics - 1st order approximation} 
\label{sec:classical}

\begin{figure}
\begin{center}
\includegraphics[scale=0.4]{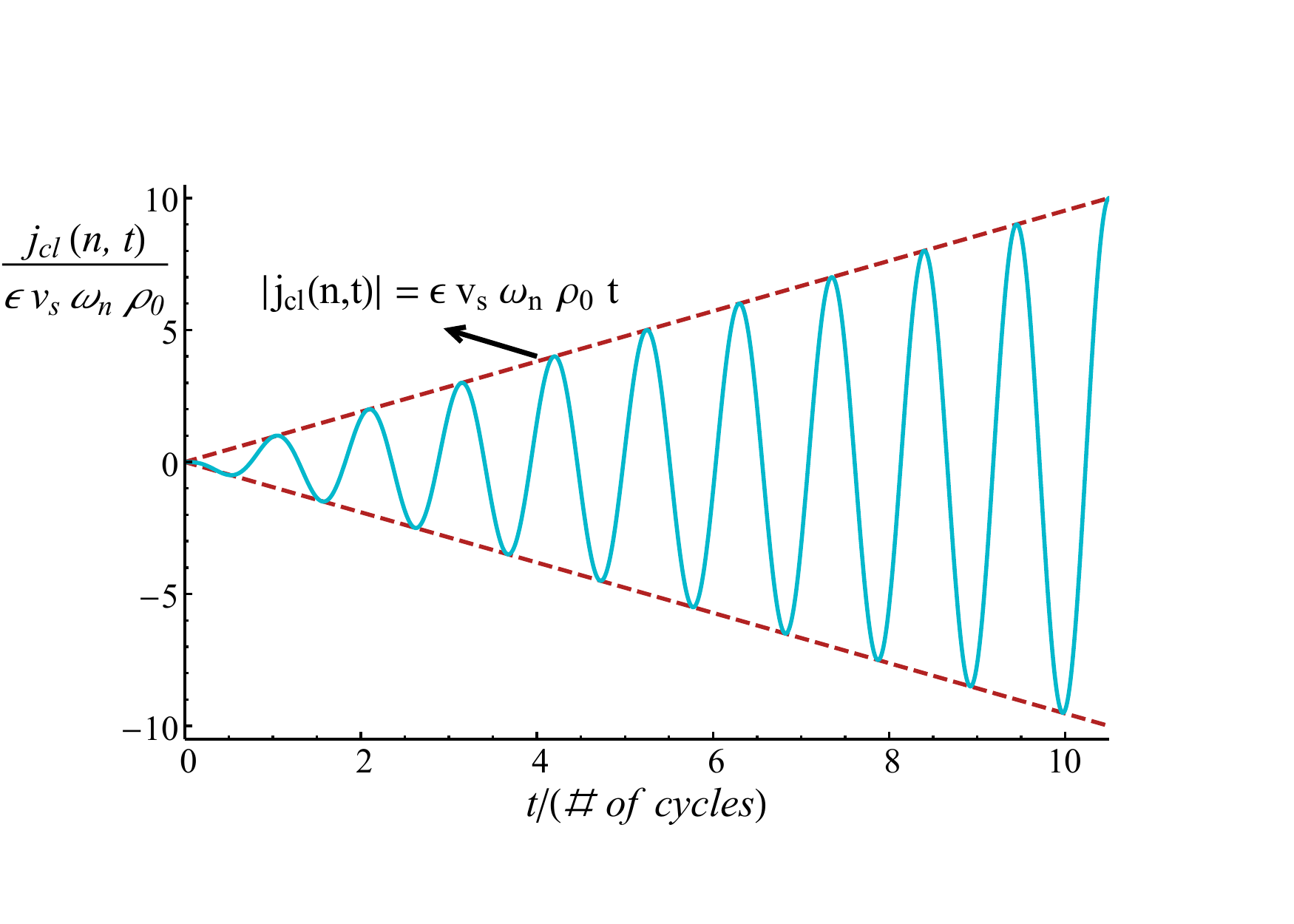}
\end{center}
\caption{In the absence of damping, driving leads to a linearly growing resonant mode for $\omega = \frac{n \pi}{L_0}v_s$. The slope of the amplitude increase of the resonant mode is proportional to the driving frequency.}
\label{fig:resnodamp}
\end{figure}

When we want to analyze driving to linear order in $\epsilon(t)$, it is sufficient to analyze Eq.~\ref{eq:tildeeq} including $V(x,t) + V_{ext.}(x,t)$ and neglecting the moving boundary conditions. The effective quadratic potential $V(x,t) + V_{ext.}$ is already linear in $\epsilon$ and boundary conditions can only modify $\expe{j}(x,t)$ and $\expe{\theta}(x,t)$ at higher order in $\epsilon$. Corrections beyond linear order due to the moving boundary will be discussed in Sec.~\ref{sec:resp}. Within linear approximation, the phonon modes become a collection of uncoupled harmonic oscillators with resonant frequencies $\omega = n \frac{\pi}{L_0} v_s $. In Fourier space, Eq.~\ref{eq:drivenKGj} is easily solved by:
\begin{subequations}
\begin{align}
\begin{split}
M(n,\omega) =& \frac{2}{L_0} \int_0^{L_0} dx \sin\left( \frac{n \pi}{L_0}x
 \right) \times \\ &\int_{-\infty}^{\infty} dt e^{i\omega t} \alpha \rho_0 \frac{d^3 \epsilon(t)}{dt^3} x,
\end{split} \\
\begin{split}
j_{cl}( n, \omega ) =&  \frac{2}{L_0} \int_0^{L_0} dx \sin\left( \frac{n \pi}{L_0} x\right) \times \\ &\int_{-\infty}^{\infty} dt e^{i\omega t} j_{cl}(x,t),
\end{split}\\
j_{cl}(n,\omega ) =& \frac{M( n,\omega)}{(\omega +i \eta )^2 - \left( \frac{n \pi}{L_0} \right)^2}
\end{align}
\end{subequations}
In the absence of damping, $\eta \rightarrow 0$ (included infinitesimally to preserve causality), the real time evolution for each mode is non-zero only for the resonant mode, $\omega = n \frac{\pi}{L_0}$, which oscillates with a linearly increasing amplitude as a function of driving time as shown in Fig.~\ref{fig:resnodamp}. The slope of the amplitude increase on resonance is proportional to the resonant frequency and the average density of the superfluid:
\begin{equation}
\mbox{Amplitude}\left[j_{cl} ( n, t) \right] = \alpha \epsilon v_s \omega_n \rho_0 t
\label{eq:ampllin}
\end{equation}
\paragraph{Effects of damping:} In this subsection we include the possibility of damping at each mode. The main mechanism with which coherent dynamics can be damped is through phonon-phonon non-linearities. Considering the standing wave a condensate in a particular phonon mode resonant with the drive, interactions would be expected to deplete this condensate. Another way to say it is that through phonon-phonon interactions a phonon mode can decay into other phonon modes through real or virtual processes. Damping in a single mode 1D models is expected to be suppressed due to integrability as argued by Tan et al.\cite{Glazman10}. However, integrability breaking contributions such as 3 body collisions and virtual hopping to higher transverse modes can still lead to damping which was investigated in earlier works both theoretically and experimentally \cite{Schmiedmayer08} \cite{Burkov07} \cite{Andreev80} \cite{Huber18} \cite{Rauer18}. Additionally, moving the boundary could break integrability such that protection from damping is not present. Here, we will simply include it as a phenomenological parameter and leave the discussion of its origin for future publications.

 Including this term, each mode becomes a damped driven harmonic oscillator with a potentially frequency and mode dependent decay rate:
\begin{equation}
\left( \partial_t^2 + 2 \gamma(n , \omega) \partial_t + \omega_n^2 \right) j_{cl} (n, t)  =  - M(n,t) 
\end{equation}
where $\omega_n = \frac{n \pi}{L_0} v_s$. We point out that decay in this context is merely a redistribution of energy from the classical expectation value to quantum noise.

\begin{figure}
\includegraphics[trim ={ 0cm 1cm 1cm 1cm},clip,scale=0.5]{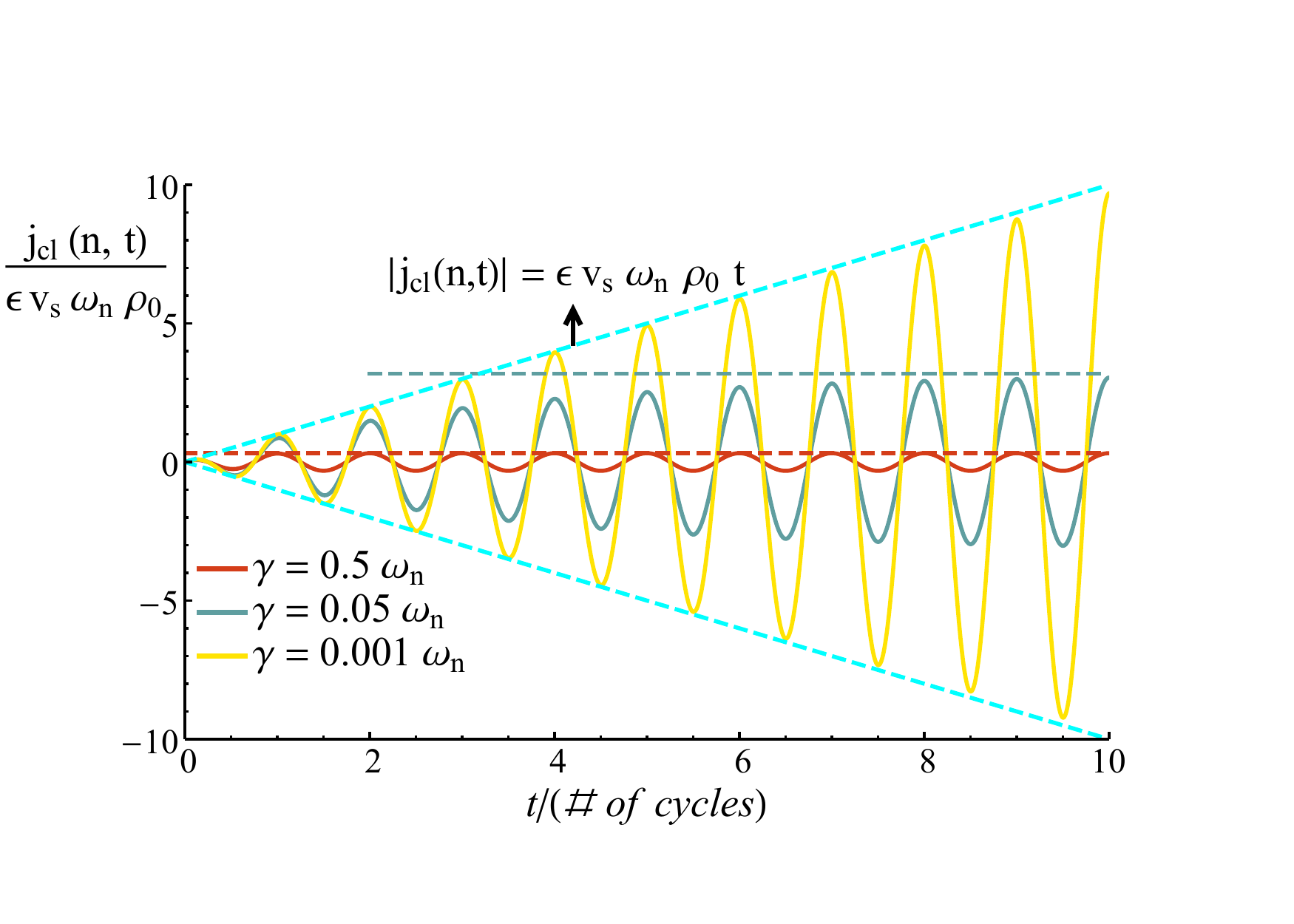}
\caption{The plot shows that in the case of resonance where $\omega = \omega_n$, the amplitude of the mode grows linearly with a rate independent of damping at early times. For times, $ t > 1/\gamma $, the current reaches a steady state with an amplitude determined by the damping. The steady state value is illustrated in the figure with dashed horizontal lines.}
\label{fig:onres}
\end{figure}

\begin{figure}
\includegraphics[trim ={ 0cm 1cm 1cm 1cm},clip,scale=0.5]{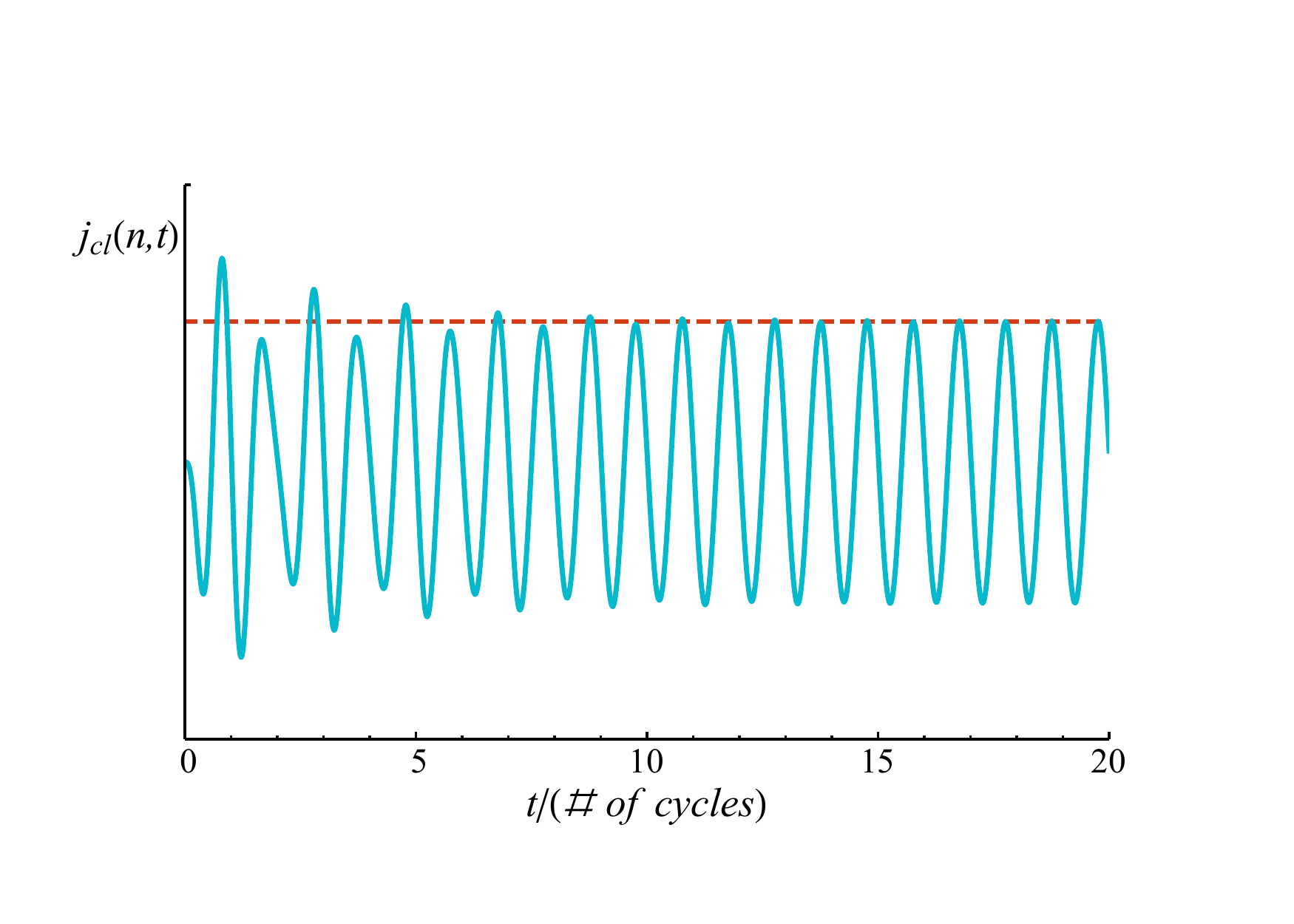}
\caption{This an example of an off-resonant response, for which $\omega = \frac{2 \pi}{L_0}$, $\omega_n = \frac{3 \pi}{L_0}$ and $\gamma(n, \omega) = 0.05\frac{2 \pi}{L_0}$. The initial response is no longer an oscillating function with a linearly increasing amplitude but for $t > 1/\gamma$ it settles into a steady state. }
\label{fig:offres}
\end{figure}

Substituting $\frac{d^3 \epsilon(t)}{dt^3} = \epsilon \omega^3 \sin(\omega t) $, the driving term has the form:
\begin{equation}
M(n,t) = \alpha \epsilon \rho_0 \frac{ L_0 \omega^3}{n \pi} \sin(\omega t) (-1)^{1+n}
\end{equation} 
where we have used $F.T.[x] = \frac{(-1)^{1+n} L_0}{n \pi}$. Defining the response function as:
\begin{equation}
\chi(\omega,n) = \frac{1}{\omega^2 + i \gamma(n, \omega) \omega - \omega_n^2}
\end{equation}
the solution is given by:
\begin{equation}
\begin{split}
&j_{cl}(n,t) = (-1)^{1+n}\alpha \epsilon \rho_0 \frac{L_0 \omega^3}{n \pi} \times \\
&\bigg( Im\{\chi(\omega,n)\}\left( \cos( \omega t) - e^{-\gamma\left(n,\omega_n^{res.}\right) t} \cos(\omega_n^{res.} t) \right)\\
&+Re\{\chi(\omega,n)\}\left( \sin( \omega t) -\frac{\omega}{\omega_n^{res.}} e^{-\gamma\left( n ,\omega_n^{res.} \right) t} \sin(\omega_n^{res.} t) \right) \bigg)
\end{split}
\end{equation}
Note that this solution satisfies the initial conditions $j_{cl}(n,t) = \partial_t j_{cl}(n,t)=0 $, $\omega_n^{res.} = \omega^2 + \gamma^2\left(n, \omega_n^{res.} \right)$.
\paragraph{On resonance:} For $\omega = \omega_n$, the bare resonance frequency, the early time behavior for $t <  \frac{1}{\gamma\left(n, \omega_n^{res.} \right)}$ is independent of damping with the amplitude growing linearly as in Eq.~\ref{eq:ampllin}:
\begin{subequations}
\begin{align}
\chi(n, \omega_n) =& \frac{1}{i \gamma(n,\omega_n) \omega_n }, \\
\begin{split}
j_{cl} (n, \omega_n) =&\frac{(-1)^{n} \alpha \epsilon \rho_0 L_0 \omega_n^2}{n \pi \gamma(n, \omega_n)}\times \\
&\left( \cos( \omega_n t) - e^{-\gamma(\omega^{res}_n,n) t} \cos( \omega_n^{res.} t) \right) ,
\end{split}\\
=& (-1)^n \alpha \epsilon \rho_0 v_s \omega_n t \cos(\omega_n t) 
\end{align}
\end{subequations}
where in the last equality the fact that $\cos(\omega_n t) \approx \cos(\omega_n^{res.} t) $, for $t < 1/\gamma(n, \omega_n)$, was used i.e. fourier broadening doesn't allow us to distinguish between the two frequencies. Moreover, the frequency dependence on $\gamma$ is assumed to be weak such that $ \gamma(n, \omega_n) \approx \gamma(n , \omega_n^{res.} ) $. Eventually, it reaches a steady state with an amplitude given by the damping:
\begin{equation}
j_{cl.}\left(n, t >  \frac{1}{\gamma (n \omega_n)}\right) = \frac{(-1)^n \alpha \epsilon \rho_0 v_s \omega_n}{\gamma(n, \omega_n) } \cos( \omega_n t) 
\end{equation}
This situation is illustrated in Fig.~\ref{fig:onres} for $n = 2$ and several values of $\gamma(n, \omega_n) $.
\paragraph{Off-resonance:} For $\omega \neq \omega_n$, as illustrated in Fig~\ref{fig:offres}, the transient response is less predictable. However one can still extract information about the damping from the steady state solution that will have an amplitude:
\begin{equation}
\mbox{Amplitude}\left[j_{cl.} (n, t> 1/\gamma) \right]  = \frac{\alpha \epsilon \rho_0 \frac{L_0 \omega^3}{n \pi}}{\sqrt{\gamma(n, \omega)^2 \omega^2 + \left(\omega^2 - \omega_0^2\right)^2 }}
\end{equation} 

Finally, note that the decay rate of the transient dynamics depends on $\gamma(n, \omega_n^{res.})$ while the decay of the steady state once the driving is switched off depends on $\gamma(n , \omega)$.
\section{Coherent Dynamics - full evolution}
\label{sec:resp}
In this section, we revisit the coherent response of the superfluid to the effective external field, $j_{cl}$., in order to derive an exact solution of Eq.~\ref{eq:drivenKGj} and go beyond the perturbative expansion. Using the tools developed in the previous sections, we notice that remarkably the retarded Greens function of the quantum homogeneous theory defined in Eq.\ref{eq:Greens}, in terms of $j_q$, constitutes the desired response function which obeys:
\begin{equation}
\left( \partial_t^2 - v_s^2 \partial_x^2 \right) G^R(x,t;x',t') = -\delta(x-x') \delta(t-t')
\end{equation}
This is shown explicitly in Appendix~\ref{app:Gr} (in dimensionless units $t \pm \frac{x}{v_s} \rightarrow t \pm x$). As a result, the classical response of the fluid obeying Eq.~\ref{eq:drivenKGj} is then given by:
\begin{equation}
\begin{split}
&j_{cl}(x,t) = \int_{-\infty}^\infty dt' \int_0^{L\left(t' \right)} dx G^R(x,t ;x',t') M(x',t'), \\
&= 2\mbox{Im} \Bigg[ \sum_n j_{n,0}(x,t) \\
&\times \int_0^t dt' \int_0^{L\left(t' \right)} dx'  j_{n,0}^*\left(x',t' \right) M(x',t') \Bigg] 
\end{split}
\label{eq:asymjcl}
\end{equation}
where $M(x,t) = \alpha \epsilon \rho_0 \frac{d^3 \epsilon(t)}{d t^3} x$ is the effective driving force. This expression gives the full non-perturbative solution of the classical response in the presence of moving boundaries and allows to study $j_{cl}$ beyond perturbative short time expansion. As before we gain intuition by moving to the Fourier basis and by decomposing $\{j_{n,0}(x,t)\}$ in terms of $\{j_{n,t}(x,t)\}$:
\begin{subequations}
\begin{align}
\begin{split}
j_{n,0}(x,t) =&\sum_m -\{j^*_{m,t}|j_{n,0} \} j_{m,t}(x,t)\\ &+ \{j_{m,t}|j_{n,0} \} j_{m,t}^*(x,t)
\end{split}\\
=& \sum_m U_{m,n}(t) j_{m,t}(x,t) + V^*_{m,n}(t)j^*_{m,t}(x,t) ,\\
\begin{split}
=& \sum_m \frac{U_{m,n}(t) e^{- i \frac{m \pi}{L(t)} t} + V^*_{m,n}(t) e^{i \frac{m \pi }{L(t)} t} } {\sqrt{m \pi}} \\  
&\times \sin\left( \frac{m \pi}{L(t)} x\right)
\end{split}  
\end{align}
\end{subequations}
Fourier transforming this result simply gives the coefficient in front of the $sin$-function:
\begin{equation}
j_{n,0} (m,t) = \frac{U_{m,n}(t) e^{- i \frac{m \pi}{L(t)} t} + V^*_{m,n}(t) e^{i \frac{m \pi }{L(t)} t} } {\sqrt{m \pi}}
\label{eq:fourierj0}
\end{equation}
Finally, using Eq.~\ref{eq:fourierj0},~\ref{eq:asymjcl} the classical response is given in Fourier space by:
\begin{equation}
\begin{split}
j_{cl}(m,t) =&2\mbox{Im} \bigg[ \sum_n  \frac{U_{m,n}(t) e^{- i \frac{m \pi}{L(t)} t}+ V^*_{m,n}(t) e^{i \frac{m \pi }{L(t)} t} } {\sqrt{m \pi}} \\
&\times \int_0^t dt' \int_0^{L\left(t' \right)} dx'  j_{n,0}^*\left(x',t' \right) M(x',t') \bigg] 
\end{split}
\label{eq:clnonpert}
\end{equation}

 \subsection{Frequency conversion}
\begin{figure*}
\captionsetup[subfigure]{justification=centering}
    \centering
    \begin{subfigure}[b]{0.45\textwidth}
        \includegraphics[trim ={ 0cm 0cm 1cm 1cm},clip, scale=0.4]{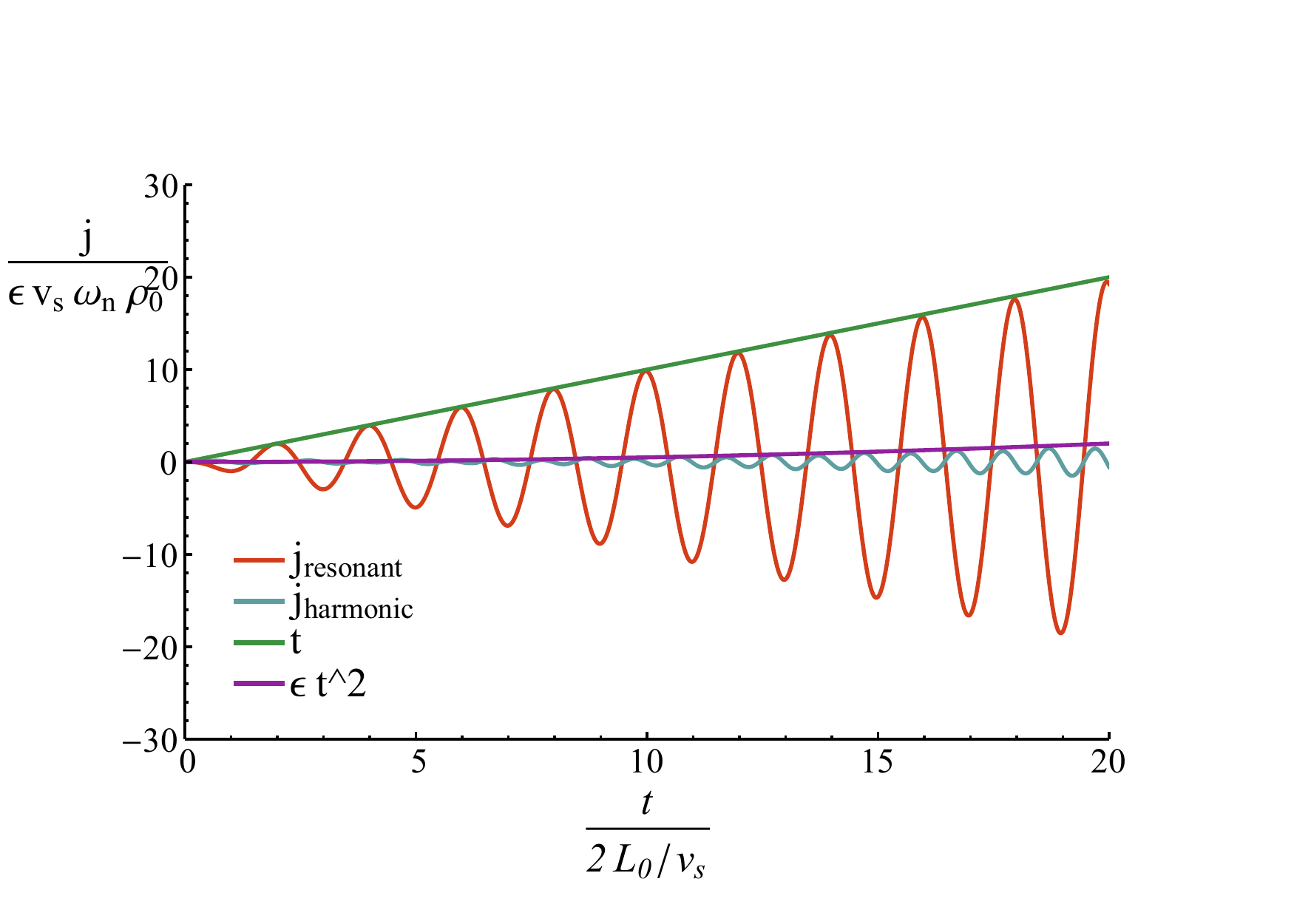}
        \caption{n = 1}
        \label{fig:pert1}
    \end{subfigure}
    \begin{subfigure}[b]{0.45\textwidth}
        \includegraphics[trim ={ 0cm 0cm 1cm 1cm},clip, scale=0.4]{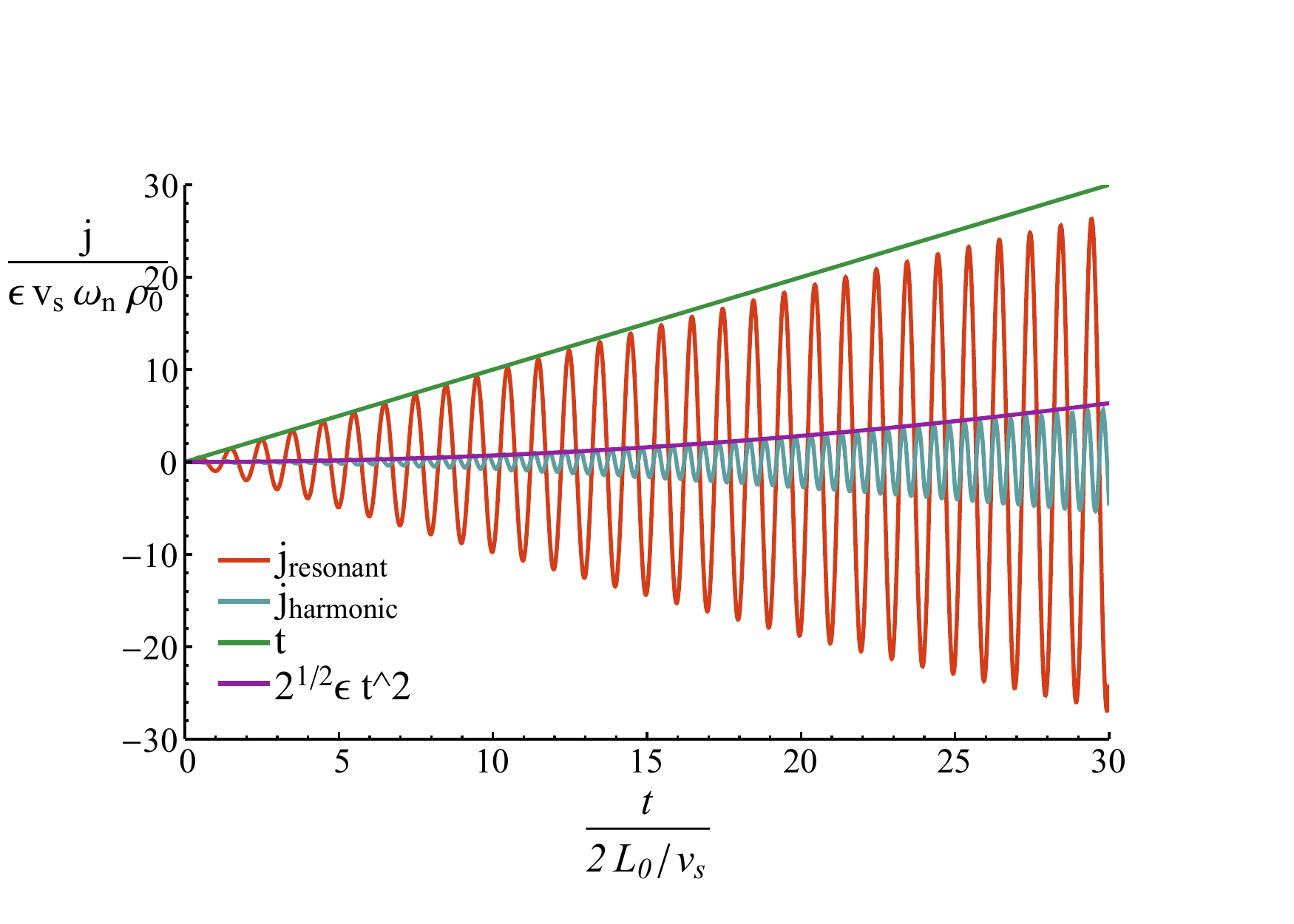}
        \caption{n=2}
        \label{fig:pert2}
    \end{subfigure}
    \begin{subfigure}[b]{0.45\textwidth}
        \includegraphics[trim ={ 0cm 0cm 1cm 1cm},clip, scale=0.4]{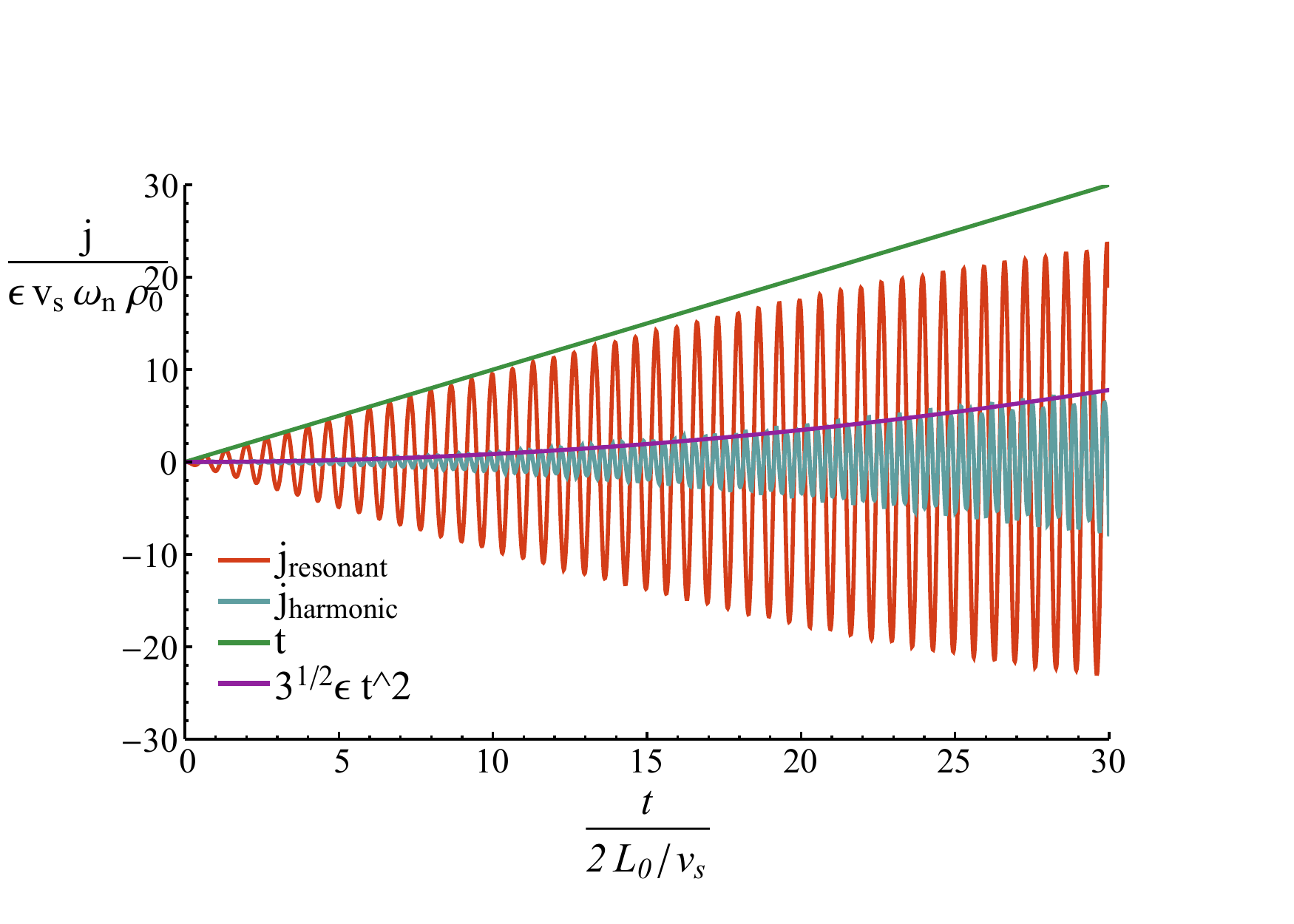}
        \caption{n=3}
        \label{fig:pert3}
    \end{subfigure}
    \begin{subfigure}[b]{0.45 \textwidth}
        \includegraphics[trim ={ 0cm 0cm 1cm 1cm},clip, scale=0.4]{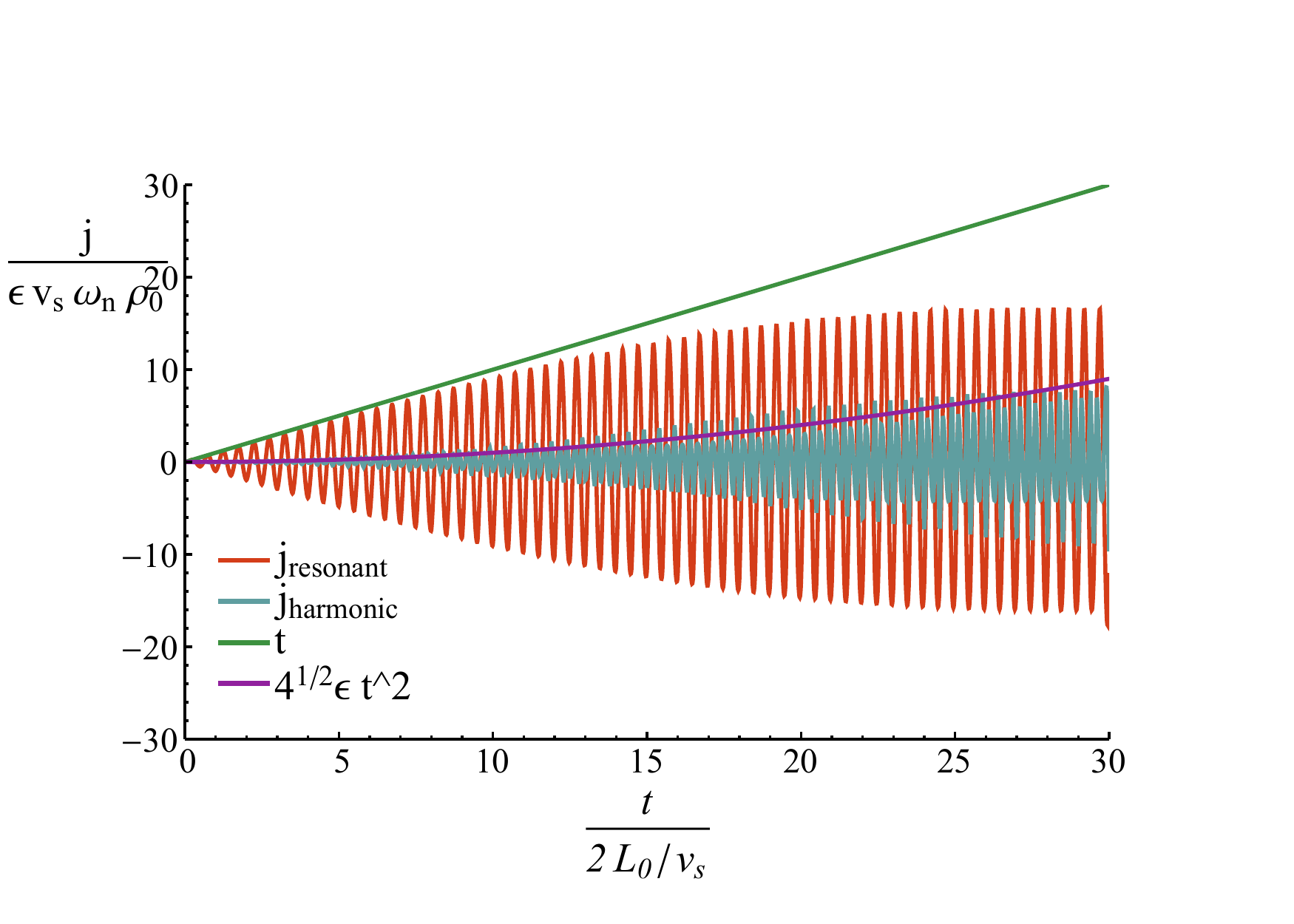}
        \caption{n=4}
        \label{fig:pert4}
    \end{subfigure}
    \begin{subfigure}[b]{0.45\textwidth}
        \includegraphics[trim ={ 0cm 0cm 1cm 1cm},clip, scale=0.4]{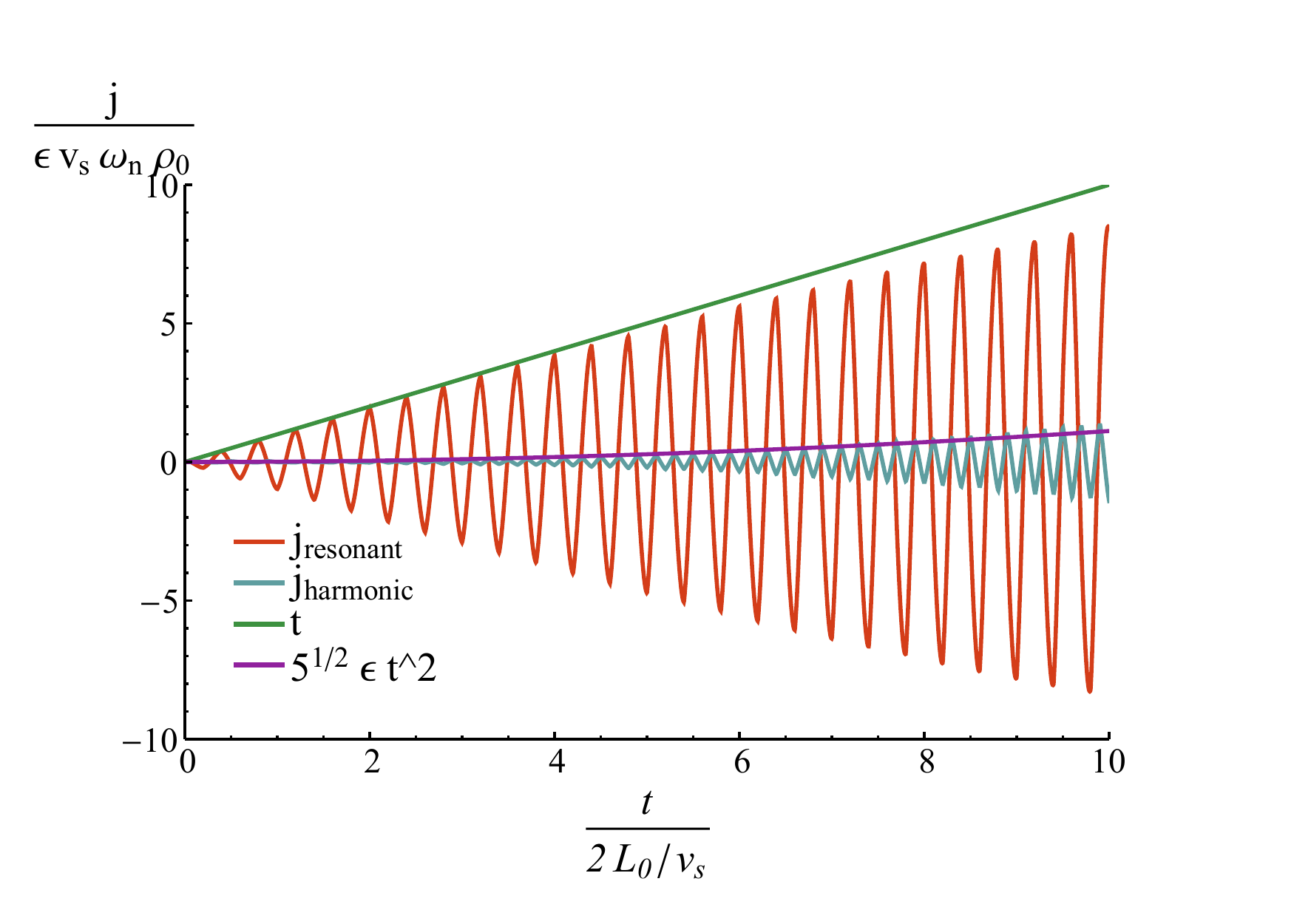}
        \caption{n=5}
        \label{fig:pert5}
    \end{subfigure}   
    \caption{The graph shows the current amplitude of the resonant mode and the second harmonic when driving at the first five resonances with an amplitude of $\epsilon  = 0.5$. At early times, there is a linear growth of the resonant mode as well as a quadratic 2nd order growth of the 2nd harmonic.}
\label{fig:freqconv}
\end{figure*}

In the fixed box case: $U_{n,m} = \delta_{n,m}$ and $V_{n,m} = 0$. As a result the system oscillates at the frequency of the drive, increasing in amplitude linearly for resonant modes as predicted in Sec.~\ref{sec:classical}. However, to 2nd order in the drive strength $\mathcal{O} \left(\epsilon^2\right)$ the resonant mode couples to twice the resonant mode, $\omega_d \rightarrow 2 \omega_d$. Remarkably, this frequency conversion, a hall mark of non-linearities, happens in the absence of any non-linearity. Perhaps, this phenomenon is unsurprising since moving the boundary breaks time-translation invariance and temporal Fourier components are no longer a good basis for the system causing them to mix. In particular, we find that when driving at the $nth$ eigenfrequency, the $nth$ eigenmode and the $2nth$ eigenmode have the following amplitude ratio:
\begin{equation}
\frac{j_{\omega_d}}{j_{2 \omega_d}} = \frac{2 L_0}{\epsilon \sqrt{n}t v_s }
\end{equation}
This relationship, should be a quantitative test of the theory, the situation is illustrated in Fig.~\ref{fig:freqconv}

Perturbation theory breaks down at $t\sim \frac{1}{\epsilon \omega} $ as mentioned earlier at which time the above ratio no longer holds. 

\subsection{Late time suppression}

The linear growth of the resonant mode at perturbative times:
\begin{subequations}
\begin{align}
j_{res}(t) \approx \alpha v_s \rho_0 \omega t, \\
\rho_{res}(t) \approx \alpha \rho_0 \omega t
\end{align}
\end{subequations}  
together with the time at which perturbation theory breaks down:
\begin{equation}
t = \frac{1}{\epsilon \omega}
\end{equation}
conspire so that the resonant mode stops increasing at:
\begin{subequations}
\begin{align}
j_{res.} &\sim \alpha v_s \rho_0, \\
\rho_{res.} &\sim \alpha \rho_0
\end{align}
\end{subequations}
By introducing an external potential that interferes with the effective potential created by the shaking we have effectively introduced a control parameter $\alpha$, that determines the overall magnitude of coherent dynamics.

Surprisingly, following the saturation at times where perturbation breaks down is a suppression of the resonant mode at late times even as the box is continuously driven at resonance with this very mode. This counter intuitive effect can be explained by first noting that a significant mode coupling causes the mode to "transfer" amplitude to higher modes, at late times these modes couple back to the resonant distractively interfering with it causing a suppression even as the drive stays turned on. This process starts as soon as perturbation theory breaks down and it is demonstrated for resonance with 2nd mode at different driving amplitudes in Fig.~\ref{fig:supp}. 

Mathematically, this is also seen in the asymptotic behaviour of $U$ and $V$ where as shown in Appendix~\ref{app:asym} at stroboscopic times both $U$ and $V$ decouple from the resonant mode and their multiples:
\begin{equation}
U , V \propto \sin\left( \frac{\mu \pi}{n} \right)
\end{equation}
where $n$ is the resonant mode and $\mu$ is one of the indices of the two matrices. While this appears as a checkered pattern in the quantum fluctuations of the quantum zipper, in the classical response it appears as a suppression of the resonant mode.   
\begin{figure}
\includegraphics[scale=0.5]{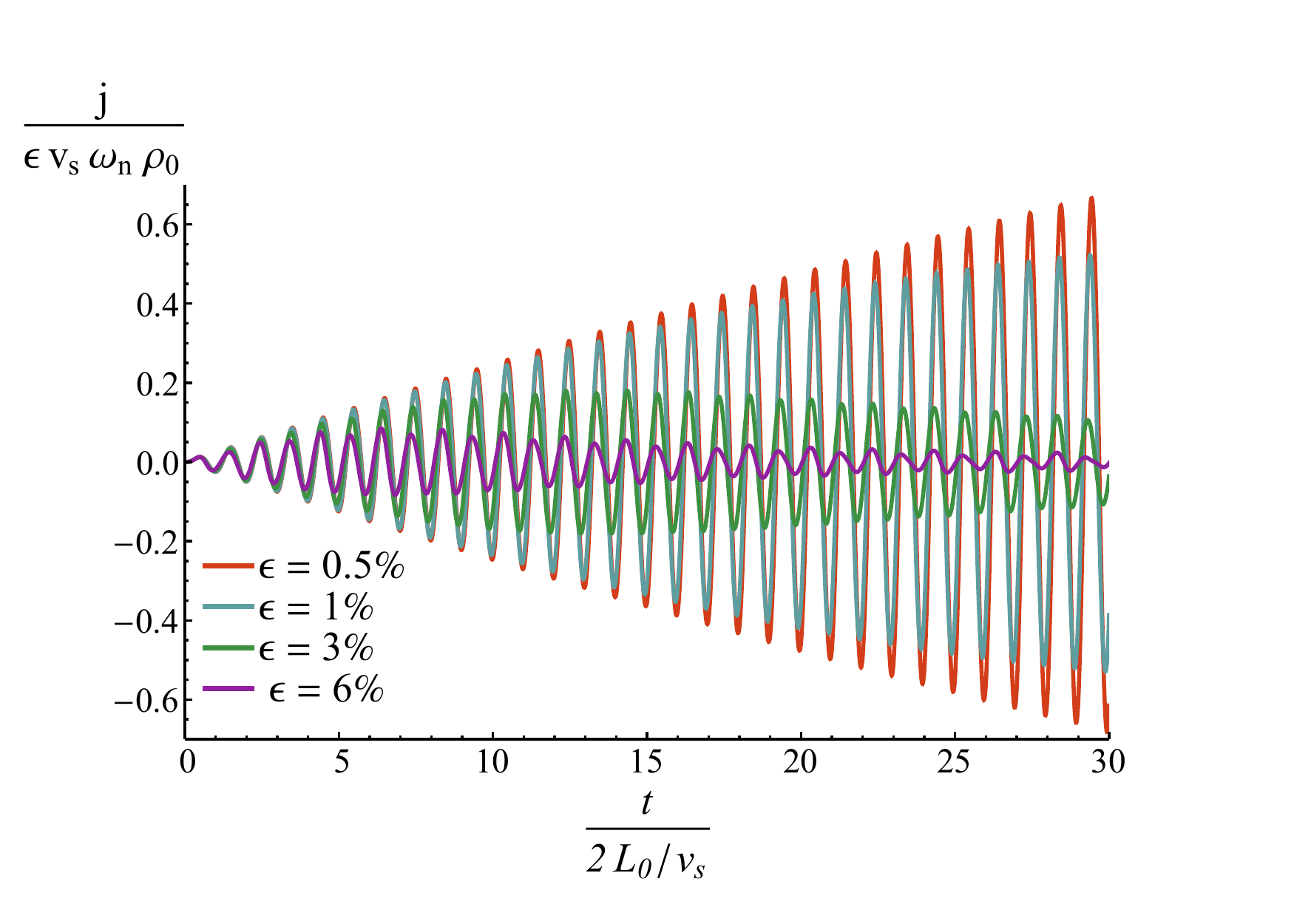}
\caption{Response of the resonant mode when driving at the 2nd harmonic for 4 different driving amplitudes. For large amplitudes perturbation theory breaks relatively quickly and starts becoming suppressed.}
\label{fig:supp}
\end{figure}

\subsection{Density Response}
\label{subsec:dens}
For convenience, we used current as our main variable. To translate back to density we use Eq.~\ref{eq:cont}, repeated here:
\begin{equation}
\partial_t \rho  = - \partial_x j
\end{equation}

An experimental protocol to relate the current amplitude of fourier modes to density amplitude is to switch off the drive at stroboscopic times and let the fluid evolve freely in a box of constant size. During the free evolution, current and density modes oscillate at their natural frequency out of phase with each other and their amplitude is related by:
\begin{equation}
\rho_{ampl}(n) = \frac{j_{ampl}(n)}{v_s}
\end{equation}

\section{Range of validity and extensions}
\label{sec:val}

In this section we provide a discussion of several contributions to dynamics beyond the $T = 0$ Luttinger liquid model that we discussed so far. We give a parameter region where our theory is expected to work well and more importantly how control parameters of the experiment can be used to access this region. 
The effects that go beyond our analysis include non-linearities, temperature and non-linear dispersion. When these effects become important, we give a discussion on how they might affect the results and propose ways of how one could take these in two account.

In the case of the shaking box, one might further object that the walls are not perfectly steep. This aspect is discussed in Appendix~\ref{app:realsqueeze}.

\subsection{Non-linearities}

There are several sources of non-linear corrections to the Luttinger liquid formalism. The full continuity equation in Eq.~\ref{eq:continuity} is non-linear (speed of sound depends on density) although so far we analyzed its linearized version. The exact fluid derivative in the Navier-Stokes equation is intrinsically non-linear $D_t v(x,t) = \partial_t v(x,t) + v(x,t) \partial_x v(x,t)$, and we only used its linear part in the Luttinger liquid formalism. Finally, when relating pressure to changes in the density one generally expects terms beyond linear ones, which should offer equally important sources of nonlinearities.

Non-linearities in the system can affect dynamic evolution in two qualitatively different ways.  Classically, this is the realm of nonlinear hydrodynamics where interactions can lead to typical nonlinear effects such as shock waves and soliton formation. On the other hand, these interacting terms couple fluctuations to expectation values. For the coherent dynamics, this can appear as a damping term that we took into account only phenomenologically. Physically, the process where the classical expectation values can be converted into quantum fluctuations through the interaction appears as dissipation in the classical equations of motion.

While the latter was addressed in Sec.~\ref{sec:classical}, the former and its relative importance to the effects presented up to now is discussed below.

\subsubsection{Classical nonlinear hydrodynamics}

In the nonlinear acoustics community, a lot of attention was given to the problem of resonant oscillations in a 1d closed tube with a moving piston\cite{Enflo02}. This represents the classical limit of our theory where a classical fluid in a tube is driven by a moving boundary.

The magnitude of non-linear effects can be deduced by the so-called Kuznetsov's equation, which represents the fluid equations up to 2nd order in small parameters (see Ref.\cite{Enflo02} for details) such as velocity amplitude and damping for an irrotational flow(for an ideal adiabatic gas):
\begin{equation}
\partial_t^2 \phi - v_s^2 \partial_x^2 \phi = \partial_t \left[ \left( \partial_x \phi \right)^2 + \frac{1}{2v_s^2} \left(\gamma-1\right) \left( \partial_t \phi \right)^2  +\frac{b}{\rho_0} \partial_x^2 \phi \right]
\end{equation} 
where $b$ the damping coefficient and $\phi$ in this equation plays the same role as the variable $\phi$ defined earlier in this paper. For the classical limit of a BEC a similar equation can be derived which is given by:
\begin{equation}
\partial_t^2 \phi - v_s^2 \partial_x^2 \phi + v_s^2 \xi_h^2 \partial_x^4 \phi = \partial_t \left[ \left( \partial_x \phi \right)^2 + \frac{1}{2v_s^2}  \left( \partial_t \phi \right)^2  +\frac{b}{\rho_0} \partial_x^2 \phi \right]
\label{eq:kutzenovbec}
\end{equation}
where the main difference is the inclusion of a non-linear dispersion whose effects are addressed in in Subsec.~\ref{subsec:disp}. 

In both equations the left hand side represents linear hydrodynamics and is accounted for simply by a dispersion relation. The right hand side contains non-linear terms. The relative magnitude of the non-linear terms compared to the linear once is given by:
\begin{equation}
\frac{\mbox{Linear terms}}{\mbox{Non-linear terms}} = \frac{v_s}{v(x,t)}
\end{equation}

Which shows that linear hydrodynamics break down only when the fluid velocity approaches the speed of sound.

From Sec.~\ref{sec:resp}, it was shown that under the influence of wall oscillations the fluid velocity grows linearly until perturbation theory breaks down where it starts getting suppressed. The maximum velocity achieved during this evolution is therefore given by:
\begin{equation}
v(x,t) \sim \alpha v_s
\end{equation} 
Applying a compensating external potential as described in Sec.~\ref{sec:shake} tunes $\alpha$ and can adjust the overall amplitude of the classical dynamics. The corresponding time scales at which the moving boundaries and nonlinearities affect the system are given by
\begin{subequations}
\begin{align}
t_{mov.bound.} &= \frac{1}{\epsilon \omega} ,\\
t_{non-lin.} &= \frac{\alpha}{\epsilon \omega}
\end{align}
\end{subequations}

This implies in the regime: 
\begin{equation}
\alpha << 1
\end{equation}
the non-linearity effectively plays no role in the problem, throughout the duration of the experiment.

It is useful to compare our solution of the linearized version of the moving piston problem to what has been discussed in the acoustics community before. Several earlier papers (\cite{Rudenko09} and citations therein) attempted to solve the problem with linear hydrodynamics and moving boundaries by introducing an approximate non-linear equation which takes the moving boundaries into account. In their analysis they found a formation of cusps that increase in height logarithmically in time\cite{Rudenko09}. 

\begin{figure}
\includegraphics[scale=0.5]{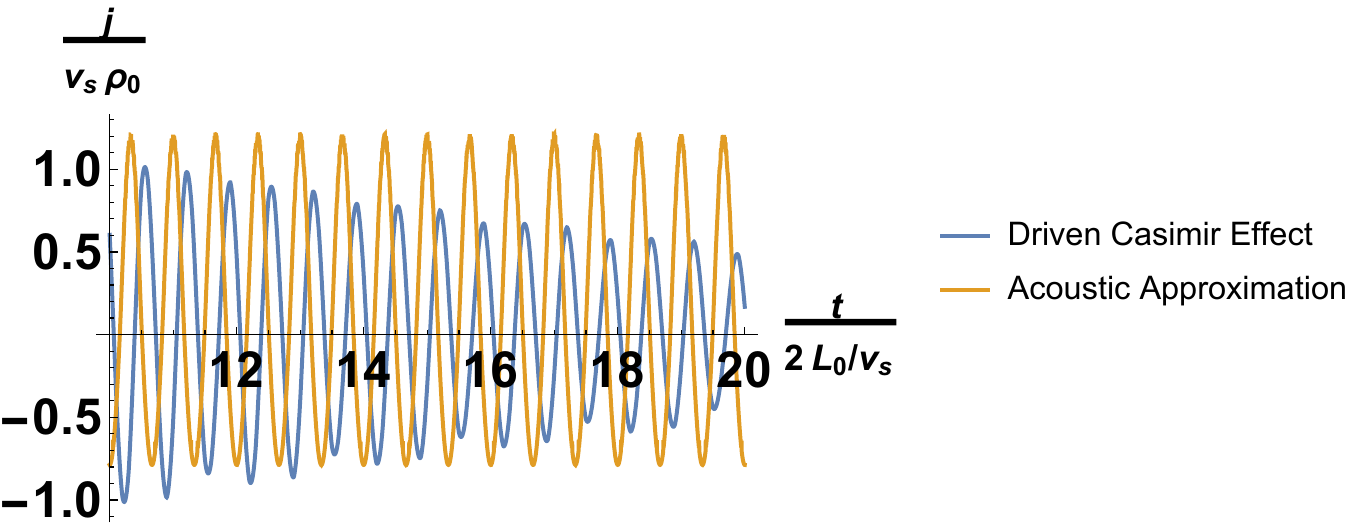}
\caption{Comparison between the resonant mode behaviour predicted by the dynamical Casimir approach and the acoustic approach. While in the acoustic approximation the resonant mode seems to saturate under continuous driving, we argue that the correct behaviour is a suppression that is somehow neglected.}
\label{fig:acvsDCE} 
\end{figure}

Our treatment is exact and takes into account the boundary conditions fully within linear hydrodynamics. Our answer is expressed as infinite sum of modes, hence the only approximation in our numerical calculations was the choice of the mode cut-off. We have argued however in Sec.~\ref{sec:quantum} that there is a window in the evolution of the system that is unaffected by this cutoff. Our theory does not predict cusps and predicts a late time suppression in the fourier component of the resonant mode that is neglected in the approximate solution as seen in Fig.~\ref{fig:acvsDCE}.

In the opposite limit:
\begin{equation}
\alpha >> 1
\end{equation}
nonlinear effects develop at time-scales much shorter than effects from the moving boundary. Therefore, the system can be treated as a nonlinear hydrodynamic system with fixed boundaries but inhomogeneous boundary conditions. While the details involve a series of simplifying approximations found in Ref.~\cite{Rudenko09} it is useful to present here the final equation used to model these systems for right moving and left moving waves:
\begin{subequations}
\begin{align}
\begin{split}
&\partial_T U(x,t) + \Delta \partial_\xi U(x,t) - \pi \epsilon U(x,t) \partial_\xi U(x,t)\\ &- \Delta \partial_\xi^2U(x,t) = -\frac{M}{2} \sin\left( \omega \xi \right),
\end{split}\\
v(x,t) &= v_s \left( U \left( \omega t - k x \right) - U \left( \omega t + k x\right) \right), \\
\xi &= \omega\left( t \pm \frac{x}{v_s} \right)
\end{align}
\end{subequations}
Where $\xi$ is the light cone coordinate depending on whether we are talking about the right moving or left moving wave and $T$ is a time coordinate of the slowly changing profile of these waves.

At late times, $T \rightarrow \infty $, the system reaches a steady state with a constants profile $\partial_T \rightarrow 0$ and the equation is solved by:
\begin{subequations}
\begin{align}
F(\chi) &=  \frac{b \omega_d }{v_s^2\rho_0 g}\frac{d}{d\chi} \log(ce_0 \left( \frac{\chi}{2}, q = \frac{2 g \epsilon v_s^4 \rho_0^2}{b^2} \right) ,\\
v(x,t) &= v_s \left(U\left(\omega\left( t - \frac{x}{v_s} \right)\right) - U\left(\omega\left( t + \frac{x}{v_s} \right)\right) \right)
\end{align}
\label{eq:shocks}
\end{subequations}
where $ce_0$ is the even Mathieu function and the small damping limit is $q \rightarrow \infty$. The behaviour of the profile in Eq.~\ref{eq:shocks} is in general very complicated but qualitatively, it includes the formation of sharp shocks and on parametric resonance the solution blows up which is qualitatively different than the results predicted by our approach. 

Interestingly, in the spirit of expanding the equation of motion to 2nd order in small parameters in this regime, $\alpha >> 1$, the conformal coordinate transformation presented in the previous sections can be used to include the effects of the moving boundary perturbatively. The coordinate transformation for times $t < \frac{1}{\epsilon \omega}$ will leave the equation of motion unchanged while the effects of breaking conformal invariance via the interaction will contribute only to higher than 2nd order terms in the equations of motion. As a result, one can replace the lightcone coordinates $u, v$ with the conformal transformed ones to obtain an answer that both obeys the boundary conditions on the moving boundaries and satisfies the equations of motion to the same level of approximation:
\begin{equation}
v(x,t) = v_s \left(U\left( \frac{\omega v_s}{L_0} R \left( t - \frac{x}{v_s} \right)\right) - U\left(\frac{\omega v_s}{L_0}R \left( t + \frac{x}{v_s} \right)\right) \right)
\end{equation}
where $R(z)$ is the transformation function defined in Sec.~\ref{sec:quantum}

\subsection{Temperature}
\label{subsec:thermal}
\begin{figure*}
\captionsetup[subfigure]{justification=centering}
    \centering
    \begin{subfigure}[b]{0.3\textwidth}
        \includegraphics[trim = {1cm, 1cm, 2cm, 2cm},clip,scale=0.3]{"Graphs/asym5".pdf}
        \caption{T = 0 , driven}
        \label{fig:thermal1}
    \end{subfigure}
    \begin{subfigure}[b]{0.3\textwidth}
        \includegraphics[trim = {1cm, 1cm, 2cm, 2cm},clip,scale = 0.3]{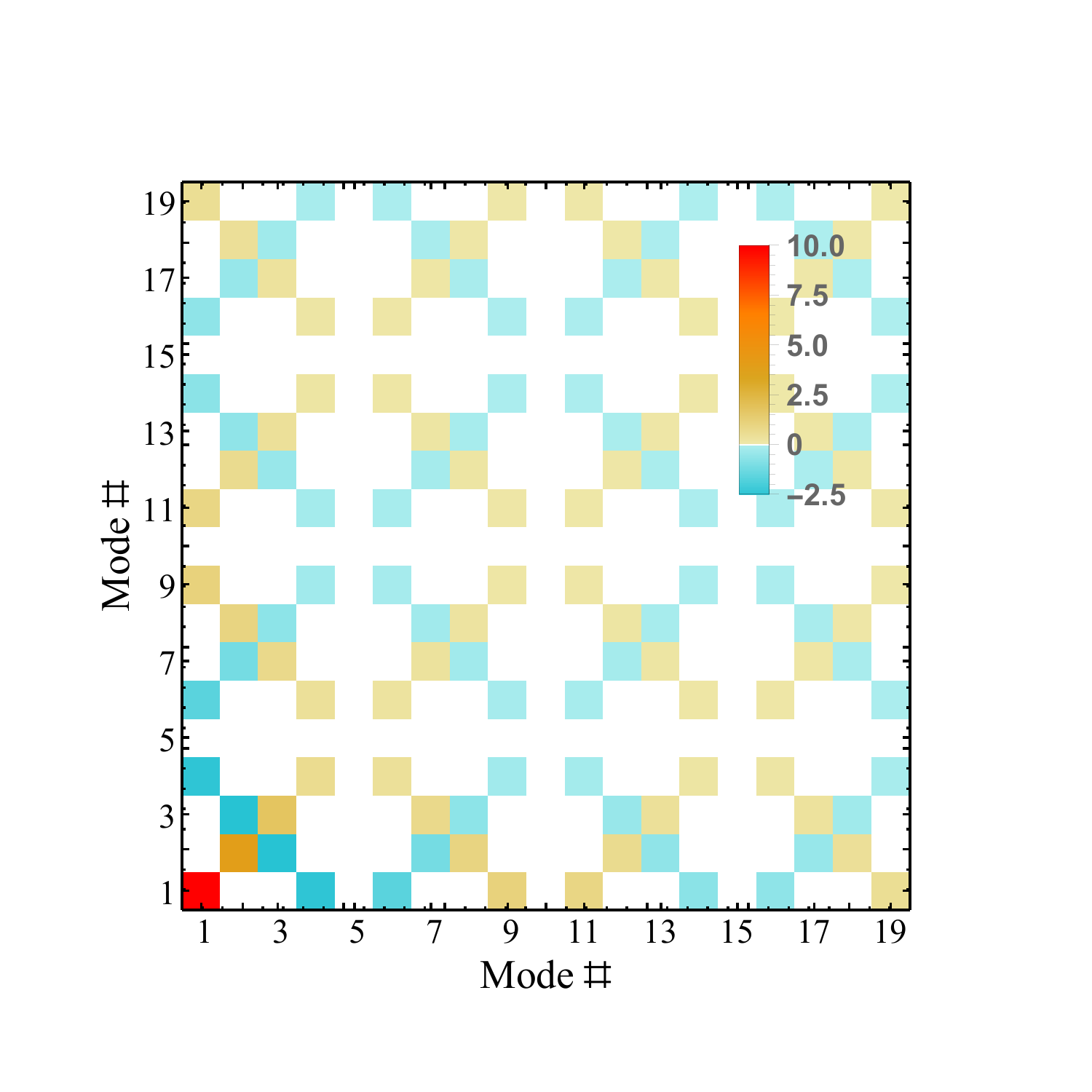}
        \caption{$T = v_s \hbar \frac{\pi}{L_0}$ , driven }
        \label{fig:thermal2}
    \end{subfigure}
    \begin{subfigure}[b]{0.3\textwidth}
        \includegraphics[trim = {1cm, 1cm, 2cm, 2cm},clip,scale=0.3]{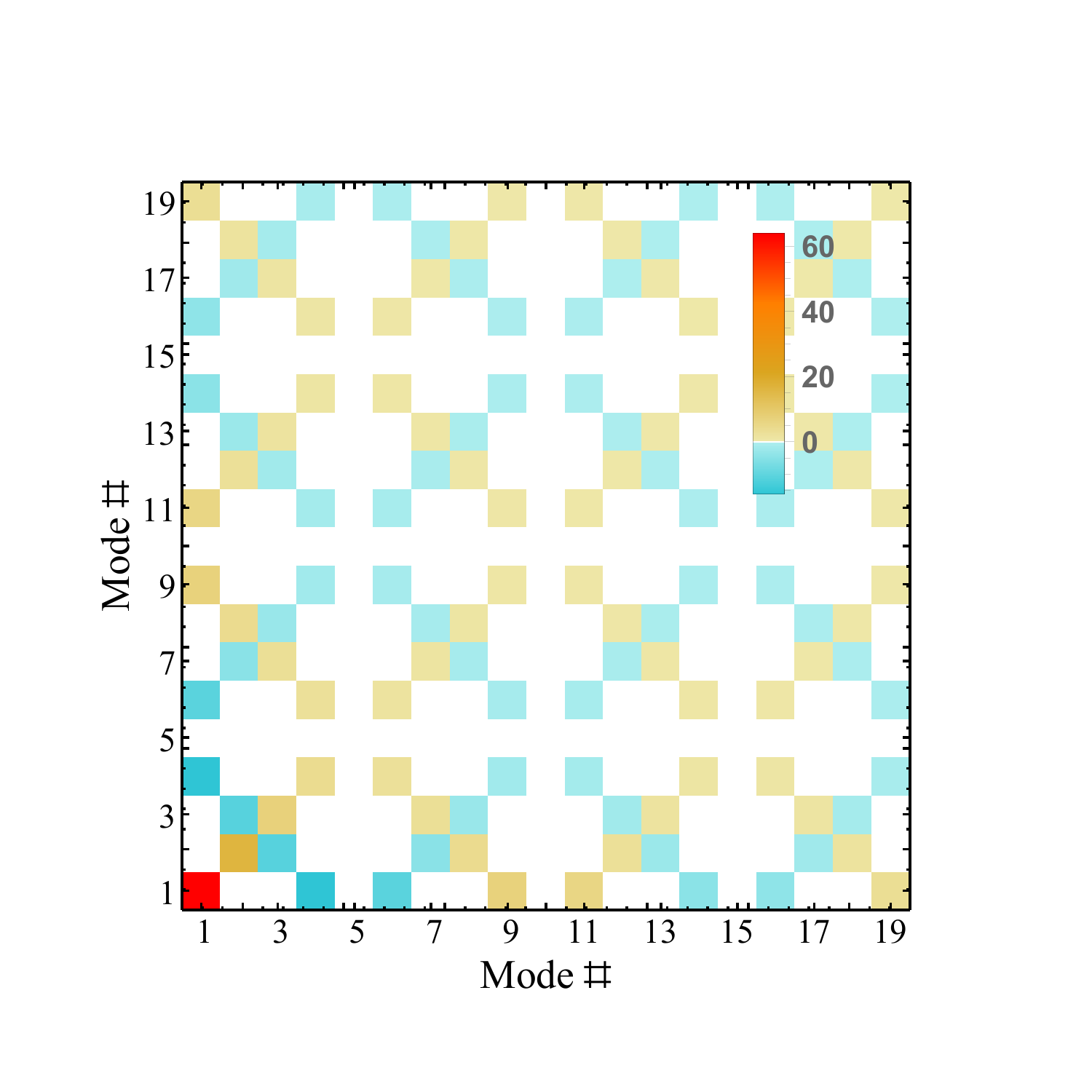}
        \caption{$T = 10 v_s \hbar \frac{\pi}{L_0}$ , driven }
        \label{fig:thermal3}
    \end{subfigure}
    \begin{subfigure}[b]{0.3 \textwidth}
        \includegraphics[trim = {1cm, 1cm, 2cm, 2cm},clip,scale=0.3]{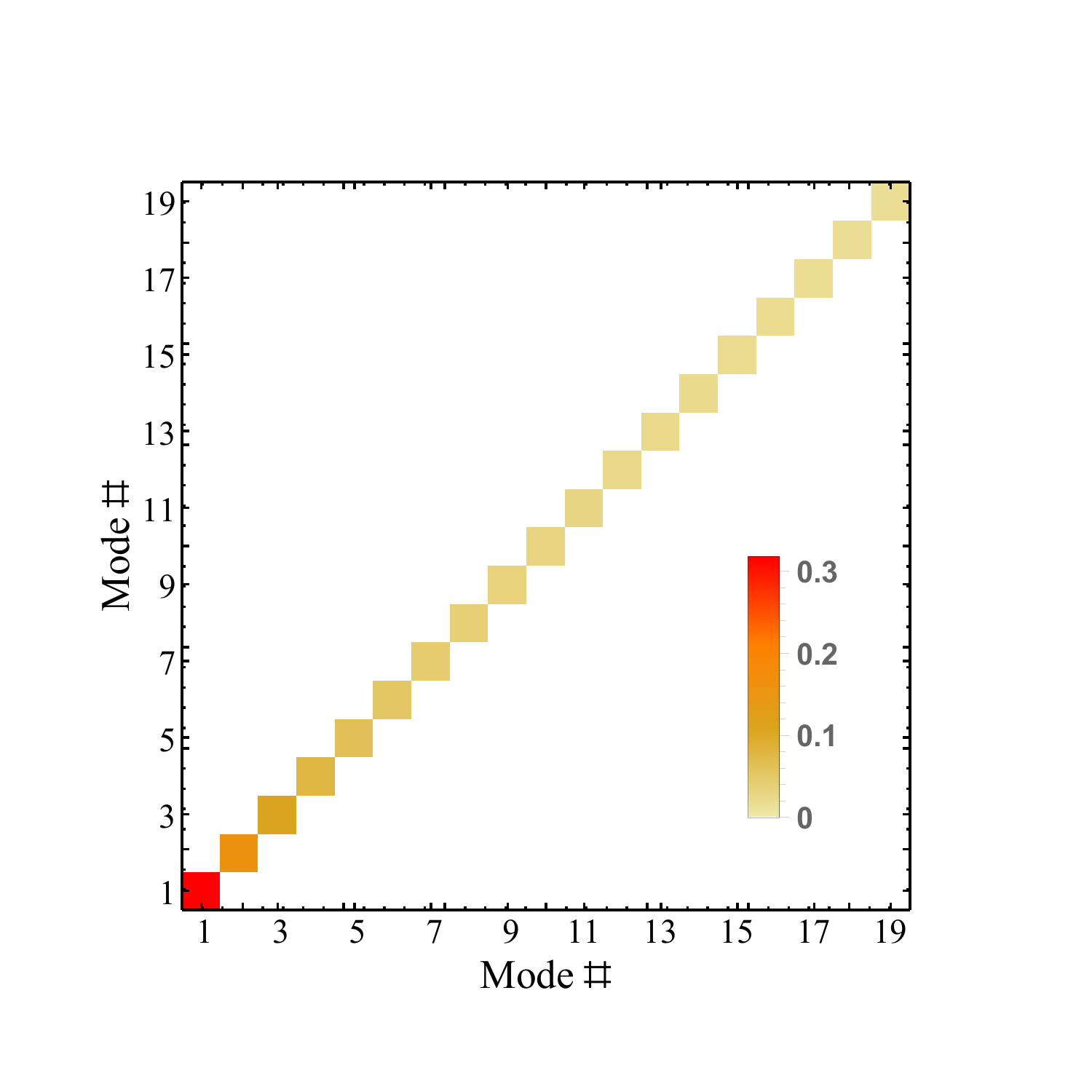}
        \caption{T = 0 , undriven}
        \label{fig:thermal4}
    \end{subfigure}
    \begin{subfigure}[b]{0.3 \textwidth}
        \includegraphics[trim = {1cm, 1cm, 2cm, 2cm},clip,scale=0.3]{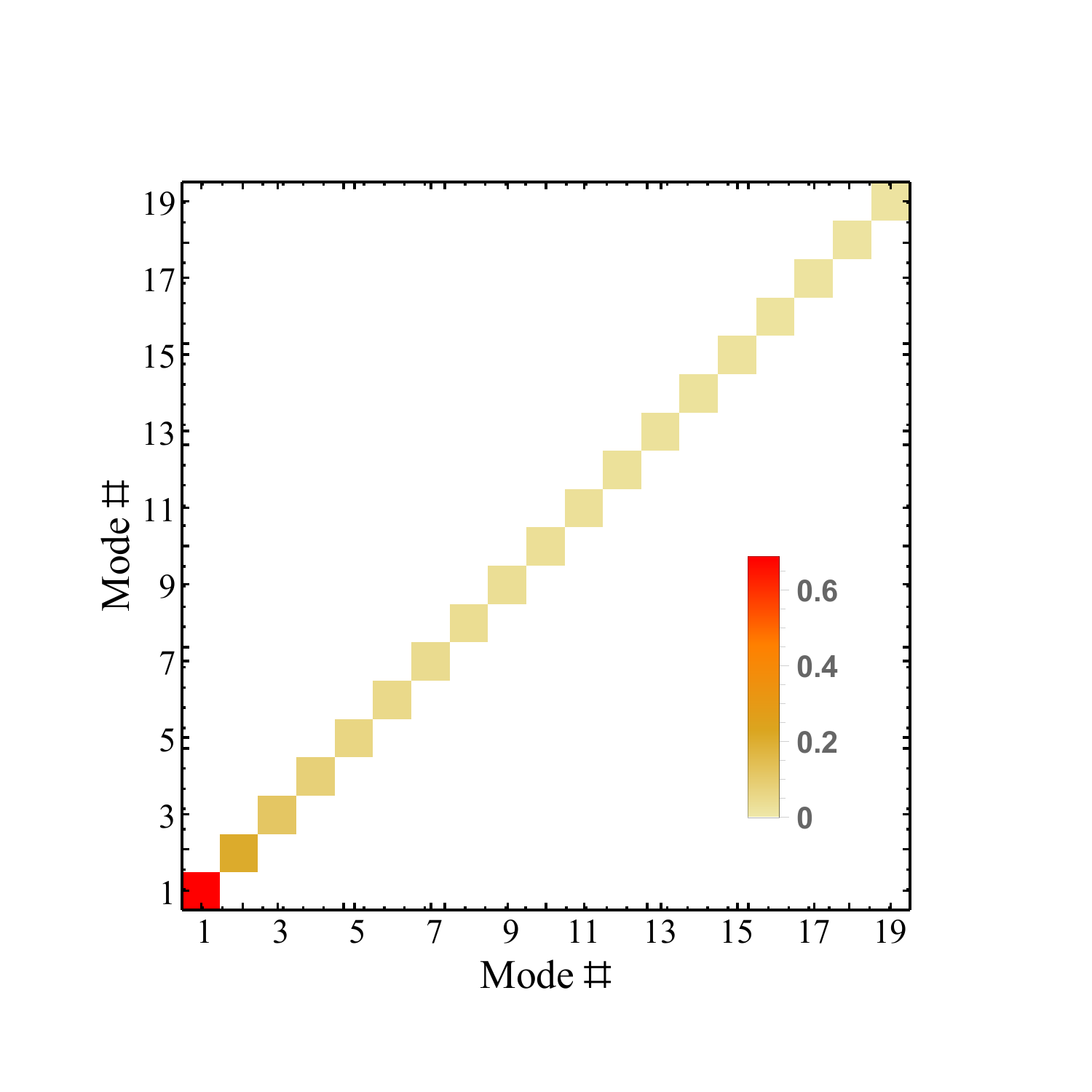}
        \caption{$T = v_s \hbar \frac{\pi}{L_0}$ , undriven }
        \label{fig:thermal5}
    \end{subfigure}   
    \begin{subfigure}[b]{0.3 \textwidth}
        \includegraphics[trim = {1cm, 1cm, 2cm, 2cm},clip,scale=0.3]{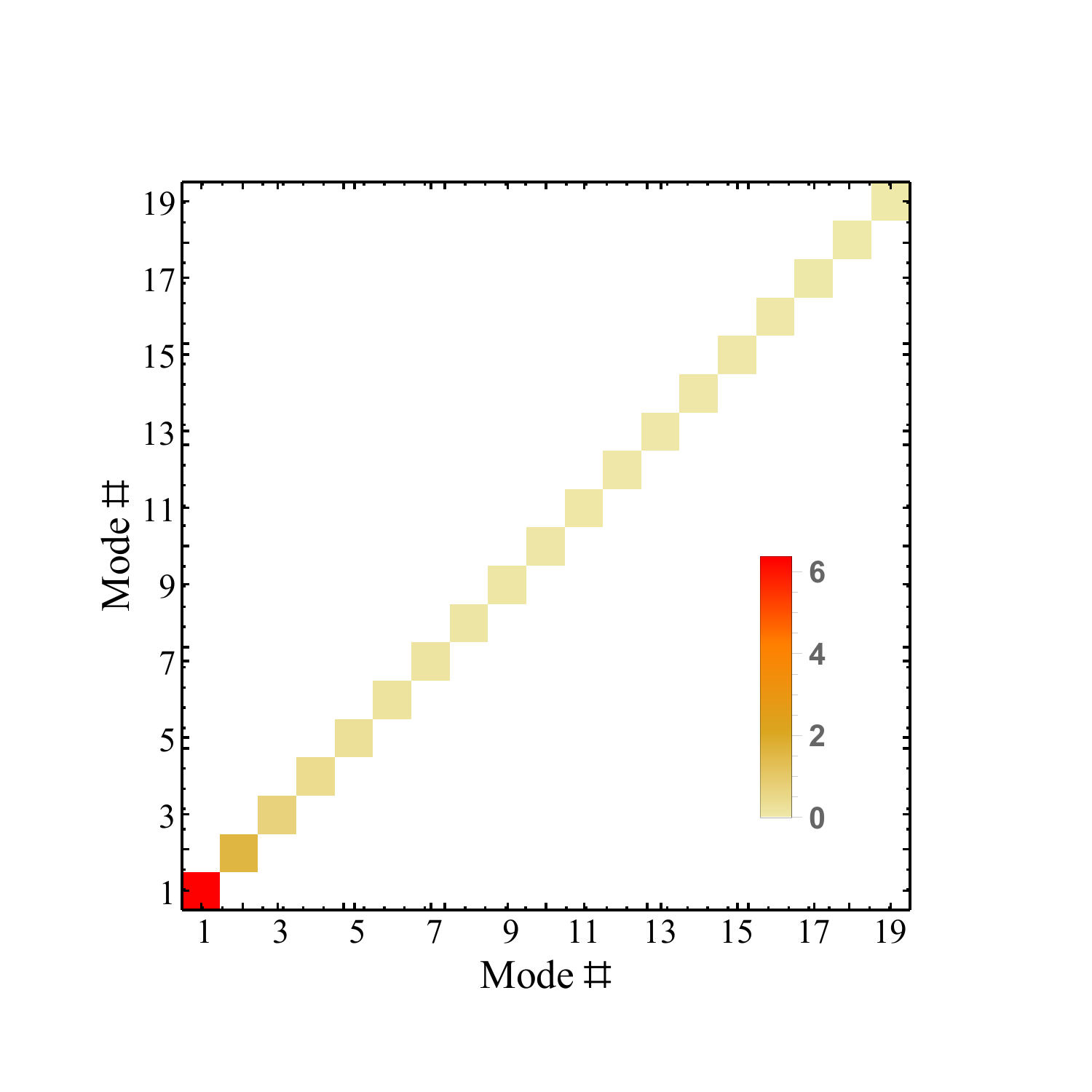}
        \caption{$T = 10 v_s \hbar \frac{\pi}{L_0}$ , undriven }
        \label{fig:thermal6}
    \end{subfigure}   
    \caption{Fluctuation matrix plotted for a driven system at $\omega = v_s \frac{ 5 \pi}{L_0}$ at various temperatures and compared against the undriven fluctuations. The pattern of resonances is not affected by the temperature, in fact temperature makes the peaks more prominent and correlations of the driven state are more greatly enhanced by temperature compared to the undriven state. The undriven state at each temperature can be thought of as the initial thermal state before driving takes place.}
\label{fig:thermal}
\end{figure*}

Temperature in many cases acts in a way that broadens and washes out features of the spectrum. This is not the case for the quantum noise predictions made in Sec.~\ref{sec:fluc}. In particular the parametric resonance pattern of the correlation matrix, $\expe{\phi_{rel}(m,t) \phi_{rel}(n,t)}$, defined in Eq.~\ref{eq:fluctT0} remains the same while the effect of the temperature is simply to enhance noise across the board, but only for the non-zero values of the correlation matrix. Remarkably, this only serves to improve the signal to noise ratio between resonant and non-resonant correlations as can be seen in Fig.~\ref{fig:thermal} where the correlation matrix has been plotted at stroboscopic late times for 3 different temperatures at driven and undriven systems (the undriven state can be thought of as the initial thermal state before driving). Details of the calculation of the correlation matrix in the presence of a thermal distribution are given in App.~\ref{app:thermal}.

For the coherent dynamics within linear hydrodynamics there is no concept of temperature since thermal fluctuations and expectation value of fields decouple. 

Temperature is expected to be important when considered in combination with interactions, that allows classical expectation value amplitude to be converted in to quantum noise by exciting quasiparticles. This process can be highly temperature dependent since pre-existing high momentum quasiparticles would make interaction with the resonant mode possible by providing more ways to satisfy energy-momentum conservation and depleting the classical expectation. This can lead to an effective highly temperature dependent damping.

\subsection{Dispersion}
\label{subsec:disp}
Working with the linearized dispersion rather than the full bogoliubov dispersion relation is a standard approximation used in order to be able to invoke the power of conformal invariance and arrive at the analytic formulas presented earlier in the paper. At high enough momenta the Luttinger-Liquid theory eventually breaks down and the non-linear dispersion relation becomes evident. This occurs at mode numbers :
\begin{equation}
k_{h} = \frac{L_0}{\xi_h}
\end{equation}

In Sec.~\ref{sec:fluc}, we showed that low energy modes of the system will not be affected by the cut-off up to times:

\begin{equation}
t \propto \frac{\log(k_h)}{\epsilon \omega}
\end{equation}

This suggests that taking a quantum fluid in a box large enough we can make this time to be longer than the duration of the experiment.
If one wishes to examine the behaviour after that time this term should be included, which would probably offer corrections to our prediction of a complete leveling off of the energy absorbed. Modeling the system by a hard cut-off is physically motivated by arguing that once higher orders in the dispersion become important the spacing between energy levels starts to increasing making those higher energy modes off-resonant with the drive.

The equation that includes the dispersion has the form:
\begin{subequations}
\begin{align}
\left( \partial_t^2 - \partial_x^2 + \xi_h^2 \partial_x^4\right) \phi(x,t) = 0,\\
\phi(0,t) = \phi(L(t),t) =0 
\end{align}
\label{eq:disp}
\end{subequations}

To investigate approximately the effects of dispersion, corrections to the fixed box eigenstates from the nonlinear dispersion could be used and the effects of the boundaries included perturbatively:
\begin{subequations}
\begin{align}
\begin{split}
f_n(w,s) =& \frac{i}{2\sqrt{ \left( n +  n^3\frac{\xi_h^2}{2 L_0^2} \right) \pi}} \Bigg( e^{- i \pi \left( \left( n +  n^3\frac{\xi_h^2}{2 L_0^2} \right) s + n w \right)}\\ &- e^{- i \pi \left( \left( n +  n^3\frac{\xi_h^2}{2 L_0^2} \right) s - n w \right)} \Bigg)
\end{split}\\
\begin{split}
f_n(x,t) =& \frac{i}{2\sqrt{ \left( n +  n^3\frac{\xi_h^2}{2 L_0^2} \right) \pi}} \left( e^{- i \pi n R(t +x)} - e^{- i \pi n R(t-x)} \right)\\
&\times e^{-i \pi  n^3\frac{\xi_h^2}{4 L_0^2} \left( R(t-x) + R(t +x) \right)}
\end{split}
\end{align}
\label{eq:dispmodes}
\end{subequations}
where $s$ and $w$ are the conformal coordinates as usual.

This is a good set of eigenstates in the perturbative regime $ t < \frac{1}{\epsilon \omega} $ because it solves Eq.~\ref{eq:disp} to order $\mathcal{O}( \frac{\xi_h^2 }{L_0^2}\epsilon) $ which is an improvement of order $\mathcal{O}\left(\epsilon\right)$. Moreover, it is an orthonormal set which can be easily verified by direct calculation of the KG inner product in the conformal frame, and ofcourse it satisfies the boundary conditions.



\section{Beyond Luttinger Liquid}
\label{sec:beyond}
Our analysis relied heavily on the emergent conformal symmetry of the low energy dynamics of 1 dimensional systems. This, however, is not true for the entire spectrum. It is then important to consider generalizations of the space-time transformations which allow to map systems with time-dependent geometry to more familiar classes of problems. In this section we discuss the scaling transformation which allows to map the system of atoms in a time-dependent box into a system in a fixed box but with time-dependent parameters. In a different context such transformations have been discussed earlier in Ref.\cite{Barmettler10},\cite{Martino13}. We note that transforming the problem into a more familiar setting of the Hamiltonian with time varying parameters makes it easier to connect to the general ideas of linear response formalism. We will see however that time dependent perturbations that appear in this case are very non-local: they involve changes in the mass of the particles, the interaction strength, and the overall quadratic potential.

In the absence of shaking, the system is described by the Hamiltonian:
\begin{equation}
H_0 =  \sum_i \frac{p_i^2}{2 m} - \frac{U_0}{2} \sum_{i,j}  \delta^D (x_i - x_j)
\end{equation}
where $m$ is the mass of the atoms and $U_0$ the strength of the point interaction. Upon shaking we can recover the constant length of the box by performing a time-dependent dilation transformation:
\begin{subequations}
\begin{align}
U_{dil.} &= e^{- i  \alpha(t) \sum_i x_i p_i },\\
U^\dag_{dil.} x_j U_{dil.} &= e^{\alpha(t)} x_j, \\
U^\dag_{dil.} p_j U_{dil.} &= e^{-\alpha(t)} p_j
\end{align}
\end{subequations}
where we need to take $\alpha(t) = - \ln\left( \frac{L(t)}{L_0} \right) $, so that the length of the box is restored to its original value, $L(t) \rightarrow L_0$. After the scaling transformation, the new Hamiltonian will contain terms that couple position and momentum, $ H_{new} \supset x_i p_i$. These can be simplified by an additional rotating frame transformation:
\begin{equation}
U_{rot}  = e^{- i F(t) \sum_i \frac{\hat{x_i}^2}{2}}
\end{equation}
with $F(t) = \frac{ m \partial_t L(t)}{2 L(t)}$. The full transformation:
\begin{equation}
U = U_{dil.} U_{rot.}
\end{equation}
leads to the following Schr\"odinger equation:
\begin{subequations}
\begin{align}
	i \frac{\partial }{\partial t} \ket{\Phi( \{x_i\},t )} =& \hat{H}' \ket{\Phi( \{x_i\},t)}, \\
\begin{split}
\hat{H}' = U \hat{H} U^\dag - i U \partial_t U^\dag =& \frac{1}{2 m(t)} \sum_i p_i^2 + \frac{a(t)}{2} \sum_i x_i^2 + \\
&\frac{g(t)}{2} \sum_{i,j,i \neq j} \delta^D \left( x_i - x_j \right), 
\end{split}\\
\ket{\Phi( \{x_i = 0 \},t )} =& \ket{\Phi( \{x_i = L_0 \},t )}
\end{align}
\end{subequations}
the time-dependent parameters are given by:
\begin{subequations}
\begin{align}
m(t) =& m \frac{L^2(t)}{L_0^2} ,\\
a(t) =& m \frac{L(t)}{L_0^3} \partial^2_t L(t), \\
g(t) =& \frac{g L_0^D}{L(t)^D}
\end{align}
\end{subequations}\
If the model requires a high energy cut-off $\Lambda$ (for example in higher dimensions), it would also be affected by the transformation as $\Lambda(t) = \Lambda \frac{L(t)}{L_0}$. 
Therefore, we can replace the effects of the moving boundaries by introducing time-dependence into both the mass and the interaction strength, while adding a time-varying quadratic external potential. In particular, note that to first order this quadratic potential is the same as the one found within the Luttinger liquid formalism in Eq.~\ref{eq:tildeeq}.

Experimentally, this can be an indispensable tool for probing the system in new ways, potentially capable of exciting collective excitation that were elusive in experiments with previous probes.

\section{Summary and outlook}
\label{sec:conc}

In this paper we considered two types of geometric parametric driving of 1d systems and provided a detailed theoretical analysis for both. The first set-up that we discussed is based on an atomic interferometer of variable length. It provides an acoustic resonator analogue of the dynamical Casimir effect in cavity QED. The second system that we analyzed is a 1d condensate in a box with periodically modulated length.  The unique feature of this system is that it combines classical driving with dynamical Casimir like parametric mode squeezing. In both systems, we focused on analyzing dynamics of the low energy modes which can be described within the Luttinger liquid framework and thus exhibit effective conformal invariance. The key advantage of cold atoms analogues of the dynamical Casimir effect is that relativistic dynamics require velocities comparable to the speed of sound, in contrast to the speed of light in the case of QED. Hence, the systems that we discuss should make it possible to achieve inter-mode coherence and not only the squeezing of individual modes observed in earlier experiments. We used the intricate connection between correlation functions and Greens functions to determine the classical evolution of expectation values of a single BEC in a box with a moving wall. A moving boundary through Doppler shift gives rise to mode coupling. Mode coupling appears, in this context, as higher harmonic generation, where on resonance the resonant mode increases linearly, while the 2nd harmonic grows quadratically in time. Interestingly, at late times the amplitude of the resonant mode becomes suppressed even under continuous driving, an effect we ascribe to higher modes coupling back to the resonant one and destructively interfering.

Our results are expected to be accurate when the following conditions are satisfied:
\begin{subequations}
\begin{align}
\frac{\xi_h}{L_0} &<< 1,\\
\alpha &\rightarrow 0,\\
T &<< \mu, \omega_L
\end{align}
\end{subequations}
where the ratio of the healing length over the size of the box  $\frac{\xi_h}{L_0}$ is indicative of how important effects from the true non-linear dispersion are in both the zipper and the shaken box. The parameter $\alpha$, defined in Eq.~\ref{eq:alpha}, controls the overall scale of the classical response by including a compensating quadratic potential in the shaken box experiment. A small $\alpha$ suppresses non-linear dynamics of the classical evolution without influencing the mode squeezing effects from the boundaries. Finally, $T$ is the temperature which has to be low enough so that the system is truly 1d and thermal excitations are within the linear dispersion regime.

Before concluding the paper, we mention several interesting directions in which our work can be extended. The main part of our discussion relied on the Luttinger Liquid formalism with a high energy cut-off on the scale of the healing length. We pointed out however that one can define a general dilation transformation which eliminates the time-dependence of the system size at the expense of introducing time dependent interactions and masses and adding a time-varying parabolic potential. The non-perturbative effects of the strong potential perturbation at the edges are swapped with a perturbation of a more complicated operator that can be analyzed within conventional linear response theory. This procedure is applicable in any dimension and can be used as a probe in more complicated systems to detect collective excitations of many body systems that do not couple to density perturbations, e.g. the Higgs mode in Fermi superfluids. Another future direction is to extend our analysis from 1d systems to 2d superfluids in box potentials, such as realized in experiments by Chomaz et al. \cite{Dalibard15}. In this case one can consider protocols that involve moving the walls of the box in a way that excites modes in both directions. Systems that we discuss offer interesting questions about dynamics of nearly-integrable models. For particles in a fixed 1d box integrability can be understood as arising from the fact that momenta are exactly conserved during both collisions of particles and reflections from static walls. In the case of a shaken box, particles can change their momenta due to Doppler shifts on moving walls. This should act as integrability breaking, however its role and efficiency are not immediately clear. Finally, we point out that in addition to exploring the dynamical Casimir effect, cold atoms quantum resonators can be used to explore other analogues of non-equilibrium phenomena in field thoeries in non-stationary space-time. For example, by moving the wall of the box or the connection point in the case of the zipper with relativistic acceleration one can realize the Unruh effect \cite{FagnocchiUn}, or even simulate inflation of the universe of a rapidly expanding gas\cite{Eckel18}.

\section*{Acknowledgments}
We are grateful for useful discussions with S. Fagnocchi, E. Altman, D. Siels, J. Dalibard, I. Bloch, I. Martin, M. Zwierlein and  
R. N. Joaquin. We acknowledge support from AFOSR-MURI Quantum Phases of Matter (grant FA9550-14-1-0035) and DARPA DRINQS program (award D18AC00014).

\appendix

\section{Retarded Greens function}
\label{app:Gr}
In this section, we show explicitly that the retarded Greens function, defined as the commutator of the quantum component of the field in Eq.~\ref{eq:Greens}, is indeed a Greens function of the equations of motion and satisfies:
\begin{equation}
\left( -\partial\\
_t^2 + \partial_x^2 \right) D^R \left( x,t;x',t' \right) = \delta(x- x') \delta(t - t')  
\label{eq:eqofgreens}
\end{equation}
Re-writing the definition of $G^R$ here for convenience:
\begin{equation}
D^R(x,t;x',t')  = 2 \theta(t -t') \mbox{Im} \left[ \sum_n f_{n,t''} (x,t) f^*_{n,t''}(x',t') \right]
\label{eq:retard}
\end{equation}
note that the subscript $t''$ implies that any valid set of modes can be used in the summation since the retarded Greens function depends only on the commutator of the fields and as a result is independent of the underlying quantum state.

Explicit substitution of Eq.~\ref{eq:retard} into the left hand side of Eq.~\ref{eq:eqofgreens} gives:
\begin{equation}
\begin{split}
&\left( -\partial
_t^2 + \partial_x^2 \right) D^R \left( x,t;x',t' \right) = \\
=& - \partial_t \delta(t-t')2 \mbox{Im} \left[ \sum_n \partial_t \left( f_{n,t''} (x,t) \right)f^*_{n,t''}(x',t') \right] \\
&- 2 \delta(t - t')   2 \mbox{Im} \left[ \sum_n \partial_t \left( f_{n,t''} (x,t) \right)f^*_{n,t''}(x',t') \right]
\end{split}
\end{equation}
where the identity $\left(-\partial_t^2 + \partial_x^2 \right) f_{n,t''}(x,t)  = 0$ was used. The delta function is only defined under integration and as a result we use the property:
\begin{equation}
-\partial_t \left( \delta(t-t') \right) f(t) = \delta(t-t') \partial_t f(t)
\end{equation}
This leads to the equation:
\begin{equation}
\begin{split}
&\left( -\partial
_t^2 + \partial_x^2 \right) D^R \left( x,t;x',t' \right) = \\
=& - \delta(t - t')   2 \mbox{Im} \left[ \sum_n \partial_t \left( f_{n,t''} (x,t) \right)f^*_{n,t''}(x',t) \right]
\end{split}
\end{equation}
In order to calculate the sum we can now use the freedom of choosing any modes we like and in particular we can choose $t'' = t$. Under this choice we have:
\begin{subequations}
\begin{align}
f_{n,t''= t} (x,t) &= \frac{e^{- i \frac{n \pi}{L(t)} t} \sin\left( \frac{n \pi}{L(t)} x \right) }{\sqrt{n \pi}} \\
\partial_t f_{n,t'' = t} (x,t) &= - \frac{ i n \pi}{L(t)} j_{n,t'' = t}(x,t)
\end{align}
\end{subequations}
Finally, putting everything together we have:
\begin{subequations}
\begin{align}
\begin{split}
&\left( -\partial
_t^2 + \partial_x^2 \right) D_R \left( x,t;x',t' \right) = \\
=& - \delta(t - t')   \frac{2}{L(t)} \sum_n \left(- n \pi \right) \frac{ \sin\left( \frac{n \pi}{L(t)}x \right) \sin\left( \frac{n \pi}{L(t)} x'\right)}{n \pi}
\end{split}, \\
=& \delta(t-t') \frac{2}{L(t)} \sum_n \sin\left( \frac{n \pi}{L(t)}x \right) \sin\left( \frac{n \pi}{L(t)} x'\right), \\
=& \delta(t-t') \delta(x-x') 
\end{align}
\end{subequations}
Completing the proof.

\section{Instantaneously at rest modes}
\label{app:inst}
In this Appendix, it is shown that as long as the wall is moving with subsonic speeds, it is always possible to construct a basis of eigenmodes that have the fixed box eigenmodes' form instantaneously at $t = t'$. To achieve that we need to be able to assign the following boundary conditions to the transformation function $R(z)$:
\begin{equation}
R(z) = \frac{z}{L(t)} , \quad z\in \left( t - L(t) , t+ L(t) \right) 
\end{equation}
the rest of the function can then be determined using the recursion relation, Eq.~\ref{eq:recR}. This is possible as long as the recursion relation does not couple any two points with in the interval $\left(t + L(t) , t - L(t) \right)$. In particular, we need to show that
\begin{itemize}
\item for any $t'$ such that $ t - L(t) < t' + L(t') < t + L(t) $, the quantity $t' - L(t')$ is outside this interval, $ t' - L(t') < t - L(t)$.
\item similarly for any $t'$ such that $ t - L(t) <  t' - L(t') < t+ L(t)$, the quantity $t' + L(t')$ is again outside this interval, $t + L(t) < t' + L(t')$.
\end{itemize}

Consider two times, $t_0 ,t_1$ such that $ t_0 + L(t_0) > t_1 + =  L(t_1)$ $$\Rightarrow t_0 - t_1 = \Delta t   > - \Delta x = - (L(t_0) - L(t_1)) $$ and also require that the speed of the wall at any point is smaller than the sound velocity i.e. $ \left| \frac{\Delta x}{\Delta t} \right| < v_s = 1$. In this situation we have 4 scenarios:
\begin{enumerate}
\item For $ \begin{matrix*}[r] \Delta t >& 0\\ \Delta x >&0 \end{matrix*} $ , trivially satisfies $\Delta t > - \Delta x$ 
\item For $ \begin{matrix*}[r] \Delta t <& 0\\ \Delta x >&0 \end{matrix*} \Rightarrow |\Delta t| <  |\Delta x| $ and hence it is not allowed.
\item For $ \begin{matrix*}[r] \Delta t >& 0\\ \Delta x <&0 \end{matrix*} \Rightarrow |\Delta t| >  |\Delta x| $ and hence it is allowed.
\item For $ \begin{matrix*}[r] \Delta t <& 0\\ \Delta x <&0 \end{matrix*}$, $\Delta t > - \Delta x$ is not satisfied and hence it is not allowed. 
\end{enumerate}
Now consider the quantity $\Delta s  = t_0 - L(t_0) - (t_1 - L(t_1)) = \Delta t - \Delta x$ from the two allowed scenario:
\begin{itemize}
\item For  $ \begin{matrix*}[r] \Delta t >& 0\\ \Delta x >&0 \end{matrix*} $ , $\Delta s =  |\Delta t| - |\Delta x|  > 0$ since $|\Delta t| > |\Delta x|$.
\item For $ \begin{matrix*}[r] \Delta t >& 0\\ \Delta x <&0 \end{matrix*} $ , $\Delta s =  |\Delta t| + |\Delta x|  > 0$, trivially.
\end{itemize}
From the above analysis we deduce that as long as $\left| \frac{d L(t)}{dt} \right| <1$, if $t_1 + L(t_1) \in  ( t_0 - L(t_0) , t_0 + L(t_0) )$ then $t_1 - L(t_1) \notin  ( t_0 - L(t_0) , t_0 + L(t_0) ) $. A similar analysis shows that if if $t_1 - L(t_1) \in  ( t_0 - L(t_0) , t_0 + L(t_0) )$ then $t_1 + L(t_1) \notin  ( t_0 - L(t_0) , t_0 + L(t_0) ) $.

$\Rightarrow$ No two points are coupled in a chosen interval and we are free to choose the value of $R(z)$, through out such an interval. This choice uniquely specifies the transformation.

\section{Computing overlaps}
\label{app:comp}
For general two sets of solutions we have:
\begin{subequations}
\begin{align}
g_n(x,t) &= \frac{i}{\sqrt{n \pi}} \frac{e^{-i n \pi R(t +x) } - e^{-i n \pi R(t - x) }}{2}, \\
f_n(x,t) &= \frac{i}{\sqrt{n \pi}} \frac{e^{ - i n \pi R'(t+x)} - e^{- i n \pi R'(t -x) }}{2}
\end{align}
\end{subequations}
It is convenient to define complementary functions:
\begin{subequations}
\begin{align}
\tilde{g}_n(x,t) &= \frac{i}{\sqrt{n \pi}} \frac{e^{-i n \pi R(t +x) } + e^{-i n \pi R(t - x) }}{2}, \\
\tilde{f}_n(x,t) &= \frac{i}{\sqrt{n \pi}} \frac{e^{ - i n \pi R'(t+x)} + e^{- i n \pi R'(t -x) }}{2}
\end{align}
\end{subequations}
for which we have the following relations:
\begin{equation}
\partial_t f_n = \partial_x \tilde{f}_n
\end{equation}
and similarly for $g_n$.
The inner product between two modes takes the form:
\begin{subequations}
\begin{align}
\{ f_m| g_n \} &= -i \int dx \left( f_m \partial_t g_n - g_n \partial_t f_m \right) ,\\
&= -i \int dx \left( f_m \partial_t g_n + \tilde{f}_n \partial_x g_n \right)
\end{align}
\end{subequations}
This inner product is itself time-independent. As a result, we are free to choose to evaluate it at any time. In particular, it is convenient to choose the time for which $g_n$ looks like the stationary solution, Eq.~\ref{eq:eigenfixed}:
\begin{subequations}
\begin{align}
g_n(x,t) &= \frac{i}{\sqrt{n \pi}} \frac{e^{- i n \pi \frac{t+x}{L(t)} } - e^{- i n \pi \frac{t - x}{L(t)}} }{2} , \\
\partial_t g_n  &= - \frac{ i n \pi}{L(t)} g_n, \\
\partial_x g_n &= - \frac{ i n\pi}{L(t)} \tilde{g}_n, \\
\Rightarrow \{f_m |g_n \} &= - \frac{n \pi}{L(t)} \int dx \left( f_m g_m + \tilde{f}_m \tilde{g}_n \right) ,\\
&= \sqrt{\frac{n}{m} } \frac{1}{2 L(t)} \times \\
&\int_0^{L(t)} dx \bigg( e^{- i \pi \left( n\frac{t+x}{L(t)} +m R'(t+x) \right)}\\&+ e^{- i \pi \left( n \frac{t-x}{L(t)} + m R'(t-x) \right)} \bigg)
\end{align}
\end{subequations}
The two terms in the integrand of the final expression can be combined by changing the variables of the second term, $x \rightarrow -x$, leading to:
\begin{equation}
\begin{split}
&\{f_m |g_n \} =\\ &\sqrt{\frac{n}{m} } \frac{1}{2 L(t)} \times\int_{-L(t)}^{L(t)} dx e^{- i \pi \left( n\frac{t+x}{L(t)} +m R'(t+x) \right)}
\end{split}
\end{equation}
Similarly, one can derive the formulas shown in the text for the bogoliubov coefficients.

\section{Pertubartive expansion of $R(z)$} 
\label{app:pert}
A pertubative expansion of $R(z)$ is found using the recursion relation and initial conditions of $R(z)$:
\begin{subequations}
\begin{align}
&R(z +L(z) ) - R(z -  L(z) ) = 2 ,\\
&R(z) = \frac{z}{L_0}, \mbox{  for } z \in (-L_0, L_0)
\end{align}
\end{subequations}

The next step is to expand $R(z)$ in power series of the perturbating function, $\epsilon(t)$:
\begin{subequations}
\begin{align}
L(t) &= L_0( 1+ e(t)), \\
R(z) &= \sum_n R^{(n)} (z) e(z)^n
\end{align}
\end{subequations}
Expanding terms in the recursion relation and matching term by term gives us the perturbative expansion. The recursion relation becomes:
\begin{equation}
\begin{split}
\sum_n &\frac{1}{n!}\Bigg( \left. \frac{ d^n R(x)}{d x^n} \right|_{z = z + L_0}(L_0 e(z) )^n  \\
&-  (-1)^n \left. \frac{d^n R(x)}{d x^n} \right|_{x= z -L_0} (L_0 e(z))^n \Bigg) = 2 
\end{split}
\end{equation}

Using the equation above the 0th order term is:
\begin{subequations}
\begin{align}
&R^{(0)}(z + L_0)  - R^{(0)} (z - L_0)  =  2, \\
&R^{(0)} (z) = \frac{z}{L_0} , \mbox{  for } z \in ( -L_0, L_0), \\
\Rightarrow &R^{(0)}(z)= \frac{z}{L_0}
\end{align}
\end{subequations}

To 1st order we have:
\begin{subequations}
\begin{align}
&R^{(1)}(z + L_0 )e(z)  -  e(z) R^{(1)} (z - L_0) ) =  - 2 e(z), \\
&R^{(1)}(z) = 0 , \mbox{  for } z \in ( -L_0, L_0), \\
\begin{split}
\Rightarrow &R^{(1)}(z)  = - 2 n, \\
&\mbox{  for } z \in ( -L_0 + 2 n L_0, L_0 + 2 n L_0)
\end{split}
\end{align}
\end{subequations}

The 2nd order term becomes:
\begin{subequations}
\begin{align}
\begin{split}
&R^{(2)}(z + L_0 )e^2(z)  -   R^{(2)} (z - L_0) )e^2(z) = \\ 
&- R^{(1)}(z + L_0) \left( \left. \frac{d e(x)}{dx}\right|_{x = z +L_0 } \right) ( L_0 e(z)) \\
&- R^{(1)}(z - L_0) \left( \left. \frac{d e(x)}{dx}\right|_{x = z - L_0 } \right) ( L_0 e(z)),
\end{split} \\
&R^{(2)}(z) = 0 , \mbox{  for } z \in ( -L_0, L_0), \\
\end{align}
\end{subequations}
for resonant wall motion we have $e(z \pm L_0) = e(z) $, hence the 2nd order term obtains the form:
\begin{subequations}
\begin{align}
\begin{split}
e^2(z) R^{(2)}(z)  =& e^2(z) R^{(2)}(z - 2 L_0) \\
&+ (2n - 1) L_0 \frac{d e(z)^2}{dz} ,
\end{split}\\
\begin{split}
e^2(z) R^{(2)}(z) =& n^2 L_0\frac{d e^2(z)}{dz} ,\\
 \mbox{  for } z \in& ( -L_0 + 2 n L_0, L_0 + 2 n L_0)
\end{split}
\end{align}
\end{subequations}
Finally, joining the pieces together up to 2nd-order, our expansion takes the form:
\begin{subequations}
\begin{align}
R(z) =& \frac{z}{L_0} - 2n e(z) + n^2 L_0 \frac{d e^2(z)}{dz},\\
&z \in ( -L_0 + 2 n L_0, L_0 + 2 n L_0) 
\end{align}
\end{subequations}
As claimed in the text, given Eq.~\ref{eq:pertR}, one can estimate where the perturbative expansion breaks down. The nth order scales a $~ (t e)^n \omega^{(n-1)}$. As a result this perturbation is valid up to times:
\begin{equation}
t_{pert.} < \frac{1}{e \omega} 
\end{equation}

\section{Late time behavior of Bogoliubov matrices}
\label{app:asym}
The asymptotic form of the Bogoliubov matrices can be calculated using Eq.~\ref{eq:bogmat} and the asymptotic form of $R(z)$, Eq.~\ref{eq:asymR}.

Here we will concentrate on calculating the bogoliubov matrices at stroboscopic times i.e. integer multiples of the period of the drive, $t = l_0 2 L_0 / n $, $l_0 \in \mathcal{Z}$ 
\begin{subequations}
\begin{align}
\begin{split}&V_{\nu,\mu} = -\sqrt{\frac{\mu}{\nu}}\frac{e^{i \frac{2 \mu \pi l_0}{n} }}{2 L_0} \int_{-L_0}^{L_0} dx e^{i \frac{\mu \pi x}{L_0}} e^{i \nu \pi R( l_0 T + x )}\\
&= - \sqrt{\frac{\mu}{\nu}} \frac{ e^{i \frac{2 \mu \pi l_0}{n}}}{2 L_0}\int_{-2 L_0}^0 d u e^{i \frac{\mu\pi}{L_0} ( u + 1 )} e^{ i \nu \pi R\left(\frac{2 l_0 L_0}{n} + u + L_0\right)}\\
 &= -\sqrt{\frac{\mu}{\nu}}\frac{e^{i \frac{2 \mu \pi l_0}{n} }}{2 L_0} \sum_{k=0}^{n-1} \int_{- 2 L_0 + \frac{2 L_0 k }{n}}^{- 2 L_0 +\frac{2 L_0 (k+1)}{n}} du e^{\frac{i \mu \pi (u + L_0)}{L_0}} \\ 
& \times e^{i \nu \pi \left( \frac{2 l_0}{n} + \frac{-2 L_0 + 2 L_0 (k+1)/n}{L_0}+1 \right)},\end{split} \\
\begin{split} 
&\int_{-2 L_0 + \frac{2 L_0 k }{n}}^{- 2 L_0 +\frac{2 L_0 (k+1)}{n}} dx e^{\frac{i \mu \pi x}{L_0}} = \\  
&=\frac{L_0}{i \mu \pi} \left( e^{i\mu \pi \left( -2 + \frac{2(k+1)}{n} \right)} - e^{i\mu \pi \left( -2 + \frac{2k}{n} \right)} \right)\\
&= \frac{2 L_0}{\mu \pi } e^{i \mu \pi \left( \frac{2 k}{n} + \frac{1}{n} \right)} \sin \left( \frac{\mu \pi}{n} \right),
\end{split} \\
&\sum_{k =0}^{n-1} e^{ i 2 \pi k \frac{ \mu + \nu }{n}} = n \delta_{\mu +\nu, \rho n},  \mbox{    where   } \rho \in \mathcal{Z},\\
\begin{split} 
&\Rightarrow V_{\nu,\mu} = \\
&-\frac{e^{ i\frac{2 \mu \pi l_0}{n}}}{\sqrt{\mu \nu } \pi}\sin(\frac{\mu \pi}{n}) n e^{i \mu \pi \left( -1 + \frac{1}{n}\right)} e^{ i \nu \pi\left(\frac{2 l_0}{n} - 1 + \frac{2}{n}\right)} \\  & \times \sum_{\rho = 1}^{\rho = \infty} \delta_{\mu + \nu , \rho n}  
\end{split}\\
\begin{split} &V_{\nu,\mu} = - \frac{n}{\sqrt{\mu \nu } \pi }\sin(\frac{\mu \pi}{n}) e^{-i\frac{\pi \mu}{n}} e^{ i\pi (\nu+ \mu )\left(\frac{2 l_0}{n} - 1 + \frac{2}{n}\right)}  \\
&\times \sum_{\rho = 1}^{\rho = \infty} \delta_{\mu + \nu , \rho n} 
\end{split}
\end{align}
\label{eq:vmatlate}
\end{subequations}
As required. $U_{\nu,\mu}$ is found similarly.

\section{Inhomogeneous Differential Equations}
\label{app:inhomo}
The possible pitfalls of dealing with inhomogeneous differential equations are presented here by way of a simplified example similar to the equations of motion found in the main text. To present the core issue it is enough to work with an inhomogeneous differential equation with fixed boundaries and then discuss the consequences for the moving boundary case. We wish to solve the following equation:
\begin{subequations}
\begin{align}
\left( \partial_t^2 -\partial_x^2 \right) j &= 0, \\
j(x = L_0) &= a(t), \quad j(x = 0 ) = 0 
\end{align}
\end{subequations}
The most straightforward way to deal with this equation is to shift $j$ by a linear function: 
\begin{equation}
j = \tilde{j} + x \frac{a(t)}{L_0} 
\end{equation}
Now, the variable $\tilde{j}$ satisfies homogeneous boundary conditions and in particular, can be expanded in terms of a sin-series: 
\begin{subequations}
\begin{align}
\tilde{j}(x=0) &= \tilde{j}(x=L_0) =0 , \\
\Rightarrow \tilde{j}(x,t) &= \sum_n a_n(t) \sin\left( \frac{n \pi}{L_0} x \right) 
\label{eq:sinexpa}
\end{align}
\end{subequations}
The cost is introducing an inhomogeneous term in the wave equation:
\begin{equation}
\left( \partial_t^2 - \partial_x^2 \right) j = - x \frac{a''(t)}{L_0}
\label{eq:inhomodiff} 
\end{equation}
where $a''$ denotes 2nd derivative with respect to time. This equation is solved by substituting the expansion of Eq.~\ref{eq:sinexpa} in Eq.~\ref{eq:inhomodiff} and take the Fourier transform on both sides:
\begin{equation}
\left( \partial_t^2 + \left( \frac{ n \pi}{L_0}\right)^2\right) a(t) = \frac{2}{L_0} \int_0^{L_0} dx \left( - x \frac{a''(t)}{L_0} \right) \sin\left( \frac{n \pi}{L_0} \right) 
\label{eq:modifinhom}
\end{equation}
However, in arriving at Eq.~\ref{eq:modifinhom} we have effectively replaced the linear potential by its sin-series expansion:
\begin{equation}
x \rightarrow \sum_n b_n \sin\left( \frac{n \pi}{L_0} x \right) 
\end{equation}
This is a drastic step, effectively eliminating all the cosine contributions from the expansion. This is justified by the fact that even though there is an external potential, the boundaries exert normal forces effectively cancelling all cosine contributions in order to preserve the boundary conditions.
The apparent paradox comes from trying to calculate the density in this model, which involves taking the spatial derivative of current, $h(x,t) = \partial_x \tilde{j}(x,t)$. As before, $h(x,t)$, can be expanded in cosine-series:
\begin{equation}
h(x,t) = \sum_n c_n \cos\left( \frac{n\pi}{L_0}x \right) 
\end{equation}
Naively, the equations of motion for $h$ would be:
\begin{equation}
\left( \partial_t^2 - \partial_x^2 \right) h(x,t) = -\frac{a''(t)}{L_0}
\label{eq:eomforh}
\end{equation}
However, the operation of taking a spatial derivative of the inhomogeneous term and expanding that term in the suitable cosine or sine series do not commute.
\begin{subequations}
\begin{align}
x &\rightarrow \sum_n b_n \sin\left(\frac{n\pi}{L_0}x \right)  , \qquad b_n \rightarrow \frac{1}{n}, \\
\partial_x x &\rightarrow \sum_n d_n \cos\left(\frac{n\pi}{L_0}x \right), \qquad d_n \rightarrow 1  , \label{eq:derexp} \\
\mbox{or} \quad \partial_x x = 1 &\rightarrow \sum_n g_n \cos\left(\frac{n\pi}{L_0}x \right), \qquad g_n \rightarrow 0  ,\label{eq:expder}  
\end{align}
\end{subequations}
Eq.~\ref{eq:derexp} represents expressing $x$ in a sine-series first and then taking a spatial derivative, while Eq.~\ref{eq:derexp} represents taking the derivative of $x$ and then expanding in a cosine-series. It is clear however, that the correct procedure is the one that gives consistent results of the solution of $h$ and $j$ in Eq.~\ref{eq:derexp}. Upon re-summing the expansion one can show that: 
\begin{equation}
\partial_x x \rightarrow \sum_n d_n \cos\left(\frac{n \pi}{L_0} x \right) = 1 - \delta(x-L_0) 
\end{equation} 
As a consequence, the naive guess Eq.~\ref{eq:eomforh}, derived by taking a spatial derivative of the equation of motion for $j$ to find the equation of motion for $h$ is wrong. Instead, we can find the correct result by replacing $x \rightarrow \theta(x) x \theta( 1-x) $. 
In the moving wall case, relevant for this paper, the eigenmodes are not simple sines and cosines in space and it becomes unclear what the equations of motion for the density are imposed by the boundary conditions of the current. Instead, to avoid this pitfall one should calculate the response of the current first and from there find the density using the continuity equation.

\section{Thermal fluctuations}
\label{app:thermal}
As mentioned in Sec.~\ref{sec:quantum}, we are working in the Heisenberg picture where the state is time-independent but the operators like the current operator, $\hat{j}$, evolve in time. However, using the relativistic formalism we where able to define time-independent creation and annihilation operators, $\{ c_m(t), c_m^\dag(t) \}$. The variable $t$ denotes the time the corresponding orthonormal basis functions have the fixed box basis functions' form. The generalization of the state of the system being the ground state of the $\{c_m(0)\}$ annihilation operators at all times, for finite temperature, T, is the following thermal density matrix:
\begin{equation}
\rho(T) = \frac{e^{- \beta \sum_l \frac{ l \pi }{L_0}c_l^\dag(0) c_l(0)}}{\mathcal{Z}}   
\end{equation}
As a result the Keldysh function, using $\expe{c^\dag_n(0) c_m (0) } = \delta_{n,m} n_b\left(\frac{n \pi}{L_0} \right)$, ($n_b$ is the bose distribution function) takes the form: 
\begin{subequations}
\begin{align}
&G^K(x,t;x',t') = - i \expe{\left\{j_q(x,t) j_q(x',t') \right\} } \\ \begin{split}
=& - 2 i\sum_n \mbox{Re} \Bigg[ j_{n,0}(x,t) j_{n,0}^*(x',t') \\ &\times \left(1 + 2 n_b\left( \frac{ n \pi}{L_0}\right) \right) \Bigg]
\end{split}
\end{align}
\label{eq:keldyshish}
\end{subequations}
This equation is the finite temperature generalization of Eq.~\ref{eq:fluct0temp}. In the absence of interactions thermal fluctuations leave the classical response unaffected. However, quantum/thermal fluctuations are included in the self-energy in the presence of interactions leading to temperature depended damping factors.

In order to understand the effect of temperature, it is helpful writing out the equivalent expression of Eq.~\ref{eq:fluctT0} for finite temperature, $T$:
\begin{equation}
\begin{split}
&\expe{ j_q(n,t) j_q(m,t) } = \frac{1}{\pi \sqrt{n m}}\times \\
&\sum_l \bigg(V^\dag_{n,l}V_{l,m}e^{i (n - m) \pi t/L(t)} +  \\
&U_{n,l} U^\dag_{l,m} e^{ - i ( n - m ) \pi t/L(t)} + \\
&V^\dag_{n,l}U^\dag_{l,m}e^{i ( n + m) \pi t/L(t)} + \\
&U_{n,l}V_{l,m} e^{ - i \pi ( n + m ) t/L(t) }  \bigg)\\
&\times \left( 1 + n_b\left( \frac{ l \pi}{L_0} \right) \right)
\end{split}
\label{eq:fluctT0}
\end{equation}
It differs only, by a thermal factor on the summation over $l$. For $k_b T < \hbar v_s \frac{\pi}{L_0}$, thermal effects are exponentially suppressed. This factor enhances the significance of lower modes in the summation, but preserves the matrix pattern of resonances as it can be seen in Fig.~\ref{fig:thermal}. The mode $n =5$ was used as an example for temperatures, $k_b T = \left\{0, \hbar v_s \frac{\pi}{L_0}, 10 \hbar v_s \frac{\pi}{L_0}\right\}$ and compared to the undriven fluctuations of the system. The resonance effect is in fact enhanced by temperature, with peaks in the correlations becoming more prominent. Furthermore, fluctuations of the driven state are more greatly enhanced by temperature compared to the undriven case.

\section{Squeezing in realistic systems}
\label{app:realsqueeze}
In realistic systems, the box potential is not perfectly steep and moving of the boundary occurs by thickening the wall via amplitude modulation of the light creating the wall. An imperfect box potential can be thought of, expanding around the middle of the box, as a Taylor series where low order terms have been eliminated making it look very flat. The box potential and the actual potential have the form:
\begin{subequations}
\begin{align}
V_{Box} &= V_0 \Theta\left(|x| - \frac{L_0}{2} \right), V_0 \rightarrow \infty,\\
V_{actual} &= \left( \frac{2 x}{L_0} \right)^{2 N}, N >>2
\end{align}
\end{subequations}
The actual potential becomes the box potential in the limit of $N \rightarrow \infty$. Upon driving, intensity modulation of the laser creating the wall translates to a time-depends multiplicative factor on the potential:
\begin{subequations}
\begin{align}
V_{actual} &= A(t) \left( \frac{2 x}{L_0} \right)^{2 N} , \\
&= \left( \frac{2 x}{L(t)} \right)^{2 N}
\end{align}
\end{subequations}
where $A(t)$ is the intensity modulation and $L(t) = \frac{L_0}{A^{1/2N} (t)}$ the effective time dependent box potential. For small intensity fluctuations, $A(t) = 1 + \epsilon f(t) $ , $\Rightarrow L(t) \approx L_0( 1 + \frac{\epsilon}{2 N} f(t))$. The box limit $N \rightarrow \infty$ must be associated with $\epsilon \rightarrow \infty $ such that $\frac{\epsilon}{N} \rightarrow$ const. As a result for sufficiently steep boxes we expect our analysis to be valid.

\bibliography{references}{}

\end{document}